\documentclass{article}

\usepackage{epsf}
\usepackage{longtable}

\usepackage{epubtk}

\usepackage{graphicx}

\def\tr{\hbox{Tr}}
\def\Tr{\hbox{Tr}}
\def\r{\rho}
\def\be{\begin{eqnarray}}
\def\ee{\end{eqnarray}}
\def\lb{\label}
\def\o{\over}

\begin{document}

\title{ Entanglement entropy of black holes}

\author{\epubtkAuthorData{Sergey Solodukhin}{%
Laboratoire de Math\'ematiques et Physique Th\'eorique\\
Universit\'e Fran\c cois-Rabelais Tours F\'ed\'eration Denis Poisson - CNRS,\\
Parc de Grandmont, 37200 Tours, France}{%
Sergey.Solodukhin@lmpt.univ-tours.fr}
{%
}%
}
\date{}
\maketitle

\begin{abstract}
The entanglement entropy is a fundamental quantity which characterizes the correlations between sub-systems in a larger 
quantum-mechanical  system. For two sub-systems separated by a surface the entanglement entropy is proportional to the area of the surface
and depends on the UV cutoff which regulates the short-distance correlations. The geometrical nature of the entanglement entropy calculation is particularly intriguing
when applied to black holes  when the entangling surface is the  black hole horizon. I review a variety of  aspects of this calculation: the useful mathematical tools such as
the geometry of spaces with conical singularities and the heat kernel method, the UV divergences in the entropy and their renormalization, the logarithmic terms  in the entanglement entropy in 4 and 6 dimensions and  
their relation to the conformal anomalies. The focus in the review is  on the systematic use of the conical singularity method. The relations to other known approaches such as 't Hooft's brick wall model and the Euclidean path integral in the optical metric  are discussed in detail. The puzzling behavior of the entanglement entropy due to fields which  non-minimally couple to gravity is emphasized. The holographic description of the entanglement entropy of the black hole horizon is illustrated on the two- and four-dimensional examples.  Finally, I examine the possibility to interpret  the Bekenstein-Hawking entropy entirely as  the entanglement entropy.

\end{abstract}

\epubtkKeywords{entanglement entropy,  Bekenstein-Hawking entropy, black holes}

\newpage
    \tableofcontents
\pagebreak

\newpage

\section{Introduction}
\label{section:introduction}

One of the mysteries in  modern physics is why  black holes have an entropy. This entropy, known as the Bekenstein-Hawking entropy, was first introduced by Bekenstein   
\cite{Bekenstein:1972tm}, \cite{Bekenstein:1973ur}, \cite{Bekenstein:1974ax} rather as a useful analogy. Soon after that, this idea   was put on a firm ground by Hawking  \cite{Hawking:1974sw} who showed that black holes thermally radiate and calculated the black hole temperature. 
The main feature of the Bekenstein-Hawking entropy  is its proportionality to the area of the black hole horizon. This property makes it rather different from the usual entropy, for example the entropy of a thermal gas in a box, which is proportional to the volume.

In 1986 Bombelli, Koul, Lee and Sorkin \cite{Bombelli:1986rw} published a paper where they considered  the so-called  reduced density matrix, obtained by 
tracing over the degrees of freedom of 
a quantum field that are inside the horizon. This procedure appears to be very natural for black holes, since the  black hole horizon plays the role of a causal boundary, which does not allow  anyone outside the black hole to have access to the events which take place inside the horizon. Another attempt to understand the entropy of black holes was made by
't Hooft in 1985 \cite{'tHooft:1984re}. His idea  was to calculate the entropy of the thermal gas of Hawking particles which propagate just outside the horizon. This calculation has uncovered two remarkable features: the entropy does turn out to be proportional to the horizon area, however, in order to regularize the density of states very close to the horizon, 
it was necessary to introduce the so-called ``brick wall'', a boundary  which is placed at a small distance from the actual horizon. This small distance plays the role of a regulator in the 't Hooft's calculation. Thus, the first indications that entropy may grow as area were found.

An important step in the development of these ideas was made in 1993 when a  paper of Srednicki \cite{Srednicki:1993im} appeared. In this very inspiring paper 
Srednicki calculated the reduced density and the corresponding entropy directly in flat spacetime by tracing over the degrees of freedom residing inside  an imaginary surface. 
The entropy defined in this calculation has became known as the {\it entanglement entropy}. Sometimes the term {\it geometric entropy} is used as well. The entanglement entropy, as was shown by Srednicki, is proportional to the area of the entangling surface.  This fact  is naturally explained by observing that the entanglement entropy is non-vanishing due to the
short-distance correlations present in the system. Thus only modes which are located in a small region close to the surface contribute to the entropy. By virtue of this fact one finds that the size of this region plays the role of the UV regulator so that the entanglement entropy is a UV sensitive quantity. A surprising feature of Srednicki's calculation is that no black hole is actually needed: the entanglement entropy of a quantum field in flat spacetime  already establishes the area law. In an independent paper Frolov and Novikov \cite{Frolov:1993ym} applied a  similar approach directly to a black hole.  
These results have sparked the interest in the entanglement entropy. In particular, it was realized that the `` brick wall'' model of t'Hooft studies a similar entropy and that the two entropies are in fact related.  On the technical side of the problem,  a very efficient method was developed  to calculate the entanglement entropy.
This method, first considered by Susskind \cite{Susskind:1993ws}, is based on a simple replica trick in which one first introduces a small conical singularity at the entangling surface, evaluates  the effective action of a quantum field on the background of metric with a conical singularity and then differentiates the action with respect to the deficit angle. By means of this method one has developed a systematic calculation of the UV divergent terms in the geometric entropy of black holes, revealing the covariant structure of the divergences \cite{Callan:1994py}, \cite{Solodukhin:1994st}, \cite{Fursaev:1995ef}. In particular, the logarithmic UV divergent terms in the entropy were found \cite{Solodukhin:1994yz}. 
The other aspect, which was widely discussed in the literature, is whether the UV divergence in the entanglement entropy could be properly renormalized.
It was suggested by Susskind and Uglum \cite{Susskind:1994sm} that the standard renormalization of Newton's constant makes the entropy finite provided one considers the entanglement entropy as a quantum contribution to the Bekenstein-Hawking entropy.  This proposal however did not answer the question whether the  Bekenstein-Hawking entropy itself can be considered
as an entropy of entanglement. It was proposed by Jacobson \cite{Jacobson:1994iw} that, in models in which Newton's constant is induced in the spirit of Sakharov's ideas, 
the Bekenstein-Hawking entropy would also be properly induced. A concrete model to test this idea was considered in \cite{Frolov:1996aj}. 

Unfortunately, in the 90-s the study of the entanglement entropy could not compete with the booming success of the  string theory  (based on D-branes) calculations of the black hole entropy \cite{Strominger:1996sh}. The second wave of interest in the entanglement entropy has started in 2003 with works studying the entropy in condensed matter systems and in lattice models. These studies revealed the universality of the approach based on the replica trick and the efficiency of the conformal symmetry to compute the entropy in two dimensions. The black holes again became  in the focus of  study in 2006 after work of Ryu and Takayanagi \cite{Ryu:2006bv} where a holographic interpretation of the entanglement entropy was proposed. In this proposal, in the frame of the AdS/CFT correspondence, the entanglement entropy, defined on a boundary of anti-de Sitter,  is  related to the area of a certain minimal surface in the bulk of the anti-de Sitter spacetime. This proposal opened interesting possibilities for computing, in a purely geometrical way, the entropy and for addressing in a new  setting the question of the statistical interpretation of the Bekenstein-Hawking entropy.  

The progress made in the recent years and the intensity of the on-going research indicate that the entanglement entropy is a very promising direction which in the coming years  may lead to a breakthrough in our understanding of the black holes and  Quantum Gravity. 
A number of very nice reviews appeared in the recent years that address the role of the entanglement entropy for black holes \cite{Bekenstein:1994bc},  \cite{Frolov:1998vs}, \cite{Jacobson:2003wv},  \cite{Das:2008sy}; review the calculation of the entanglement entropy in quantum field theory in flat spacetime 
\cite{Eisert:2008ur}, \cite{Casini:2009sr}  and the role of the conformal symmetry \cite{Calabrese:2009qy};   focus on the holographic aspects of the entanglement entropy \cite{Nishioka:2009un}, \cite{Barbon:2009zz}. In the present review I build on these works and focus on the study of the entanglement entropy as applied to black holes. 
The goal of this review is to collect a complete variety of results and present them in a systematic and self-consistent  way  without neglecting neither technical nor principal aspects of the problem.

\section{Entanglement entropy in Minkowski space-time}
\label{section: Minkowski} 

\subsection{Definition}
\label{subsection: Definition}

Consider a pure vacuum state $|\psi>$ of a quantum system defined inside a space-like region $\cal O$ and suppose that the degrees of freedom in the system can be considered as located inside certain  sub-regions of $\cal O$.
A simple example of this sort is a system of coupled oscillators placed in the sites of a space-like lattice.  Then for an arbitrary imaginary  surface $\Sigma$ which separates the region $\cal O$  in two complementary sub-regions $A$ and $B$,  the system in question can be represented as a union of two sub-systems.  The wave function of the global system is given by a linear combination of the product of quantum states  of each sub-system, $|\psi>=\sum_{i,a}\psi_{ia}|A>_i|B>_a$.  The states $|A>_i$ are formed by the  degrees of freedom  localized in the region $A$  while the states $|B>_a$ by those which are defined in the region $B$.  The density matrix that corresponds to a pure quantum  state $|\psi>$ 
\be
\rho_0(A,B)=|\psi><\psi|
\lb{density-pure_state}
\ee
has zero entropy. By tracing over the degrees of freedom in region $A$ we obtain a density matrix 
\be
\rho_B=\tr_A\rho_0(A,B)\, 
\lb{trace-rho}
\ee
with elements $(\rho_{B})_{ab}=(\psi\psi^\dagger)_{ab}$. The statistical entropy, defined for this density  matrix by the standard formula
\be
S_{B}=-\tr\rho_{B}\ln\rho_{B}\, 
\lb{entropy}
\ee
is by definition the \emph{entanglement entropy} associated with the surface $\Sigma$. We could  have traced over the degrees of freedom located in region $B$ and formed the density matrix $(\rho_{A})_{ij}=(\psi^T\psi^*)_{ij}$. It is clear that\epubtkFootnote{For finite matrices this property indicates that the two density matrices have the same eigenvalues.} 
$$
\tr \rho_{A}^k=\tr \rho_{B}^k
$$
 for any integer $k$.  Thus we conclude that the entropy (\ref{entropy}) is the same for both  density matrices $\rho_{A}$ and $\rho_{B}$, 
\be
S_{A}=S_{B}\, .
\lb{in=out}
\ee
This property indicates that the entanglement entropy for a system in a pure quantum state is not an extensive quantity. In particular, it does not depend on the size of each region  $A$ or $B$ and thus is only determined by the geometry of $\Sigma$.

\subsection{Short-distance correlations}

On the other hand, if  the entropy (\ref{entropy}) is non-vanishing, this shows that in the global system there exist correlations  across the surface $\Sigma$ between modes which reside on different sides of the surface. In this review we shall consider the case when the system in question is a quantum field. 
The short-distance correlations that exist in this system  have two important consequences:
\begin{itemize}
\item the \emph{entanglement entropy} becomes dependent on the UV cut-off $\epsilon$ which regularizes the short-distance (or the large-momentum) 
behavior of the field system
\item  to leading order in ${\epsilon}^{-1}$ the \emph{entanglement entropy} is proportional to the area of the surface $\Sigma$
\end{itemize}
For a free massless scalar field the 2-point correlation function in $d$ space-time dimensions has the standard form 
\be
<\phi(x),\phi(y)>=\frac{\Omega_d}{ |x-y|^{d-2}}~~,
\lb{2-pt-function}
\ee
where $\Omega_d=\frac{\Gamma (\frac{d-2}{2})}{ 4\pi^{d/ 2}}$. Correspondingly, the typical behavior of the entanglement entropy in $d$ dimensions is
\be
S\sim \frac{A(\Sigma)}{ \epsilon^{d-2}}\, ,
\lb{UV-entropy}
\ee
where the exact pre-factor depends on the regularization scheme. Although the  similarity between (\ref{2-pt-function}) and (\ref{UV-entropy}) illustrates well the field-theoretical origin of the entanglement entropy, the exact relation between the short-distance behavior of  2-point correlation functions in the field theory and the UV divergence of the entropy is more subtle, as we shall discuss later in the paper. 

\subsection{Thermal entropy}
Instead of a pure state one could have started with a mixed thermal state at temperature $T$ with density matrix $\rho_0(A,B)=e^{-T^{-1}H(A,B)}$, where $H(A,B)$ is the Hamiltonian of the global system. In this case  the relation
(\ref{in=out}) is no more valid and the entropy depends on the size of the total system as well as on the size of each  sub-system. By rather general arguments,  in the limit of large volume  the reduced density matrix approaches the thermal density matrix. So that  in this limit the entanglement entropy (\ref{entropy})  reproduces the thermal entropy.  For  further references we give here the expression 
\be
S_{thermal}=\frac{d}{\pi^{d/2}}\Gamma (\frac{d}{2})\zeta(d)\ T^{d-1}V_{d-1}\, 
\lb{Sthermal}
\ee
for the thermal entropy of a massless field residing inside  a spatial $(d-1)$-volume $V_{d-1}$ at temperature $T$.

\subsection{Entropy of a system of finite size  at finite temperature}

In more general situation one starts with a system of finite size $L$ in a mixed thermal state at temperature $T$. This system   is divided by the entangling surface $\Sigma$
in two sub-systems of characteristic size $l$. Then the entanglement entropy is a function of several parameters (if the field in question is massive then mass $m$ should be added to the parameters on which the entropy should depend on)
\be
S=S(T,\, L,\, l,\, \epsilon)\, ,
\lb{Sgeneral}
\ee
where $\epsilon$ is a UV cut-off. Clearly, the entanglement entropy in this general case is due to a combination of different factors: the entanglement between two sub-systems and the thermal nature of the initial  mixed state. In $d$ dimensions even for simple geometries this function of 4 variables is not known explicitly.
However, in two space-time dimensions, in some particular cases, the explicit form of this function is known.

\subsection{Entropy in (1+1)-dimensional space-time}

The state of a quantum field in two dimensions is defined on a union of intersecting intervals $A\cup B$.
The 2-point correlation functions behave logarithmically in the limit of coincident points. Correspondingly, the leading UV divergence of the entanglement entropy in two dimensions is logarithmic. For example, for a 2d massless conformal field theory, characterized by a central charge $c$, the entropy is \cite{Srednicki:1993im}, \cite{Callan:1994py}, \cite{Holzhey:1994we}
\be
S_{2d}=\frac{cn}{ 6}\ln \frac{l_A}{\epsilon}+s(l_A/l_B)\, ,
\lb{S2d}
\ee
where $n$ is the number of intersections of intervals $A$ and $B$ where the sub-systems are defined, $l_A$ ($l_B$) is the length of the interval $A$ ($B$). The second term in (\ref{S2d}) is a UV finite term. In some cases the  conformal symmetry in two dimensions can be used  to calculate 
not only the UV divergent term in 
the entanglement entropy but also the UV finite term, thus obtaining the complete answer for the entropy, as was shown by Holzhey, Larsen and Wilczek \cite{Holzhey:1994we} (see \cite{Korepin:2004zz}, \cite{Calabrese:2004eu} for more recent developments).  There are two different limiting cases when the conformal symmetry is helpful. In the first case one considers a pure state of the conformal field theory on a circle of circumference $L$, the subsystem is defined on a segment of size $l$ of the circle.  In the second situation the system is defined on an infinite line, the subsystem lives on interval of length $l$ of the line and the global system is in a thermal mixed state with temperature $T$. In Euclidean signature both geometries represent a cylinder. For a thermal state the compact direction on the cylinder corresponds to Euclidean time $\tau$ compactified to form a circle of circumference $\beta=1/T$.  In both cases the cylinder  can be further conformally mapped to a plane. The invariance of the entanglement entropy under conformal transformation
can be  used to obtain 
\be
S=\frac{c}{3}\ln \left(\frac{L}{\pi\epsilon}\sin(\frac{\pi l}{L}\right)
\lb{entropy-circle}
\ee
in the case of a pure state on a circle and
\be
S=\frac{c}{3}\ln \left(\frac{\beta}{\pi\epsilon}\sinh(\frac{\pi l}{\beta}\right)
\lb{entropy-T}
\ee
for a thermal mixed state on an infinite line. In the limit of large $l$  the entropy (\ref{entropy-T})  approaches 
\be
S=\frac{c}{3}\pi l T+\frac{c}{3}\ln (\frac{l}{\pi\epsilon})+{c\over 3}\ln{\beta\over l}\, ,
\lb{entropy-Tl}
\ee
where the first term represents the entropy of the thermal gas (\ref{Sthermal}) in a cavity of size $l$ while the second term represent the purely entanglement contribution (note that  the intersection of  $A$ and $B$  contains two points in this case so that  $n=2$).
The third term is an intermediate term due to the interaction of both factors,  thermality and entanglement.
This example clearly shows that  for a generic thermal state the entanglement entropy is due to the combination  of two factors: the entanglement between two subsystems and the  thermal nature of the mixed state of the  global system.

\subsection{The Euclidean path integral representation and the replica method}
\lb{subsection: replica method}
A technical method very useful for the calculation of the entanglement
entropy in a field theory is the so-called {\it the replica trick}, see
ref.\cite{Callan:1994py}. Here we illustrate this method for a field theory described by a second order Laplace type operator.
 One considers a quantum field $\psi(X)$ in a
$d$-dimensional spacetime and  chooses the Cartesian coordinates
$X^\mu=(\tau,x, z^i,\, i=1,..,d{-}2)$, where $\tau$ is Euclidean time,
such that the surface $\Sigma$ is defined by the condition $x=0$ and
$(z^i,\, i=1,..,d{-}2)$ are the coordinates on $\Sigma$. In the
subspace $(\tau,x)$ it will be  convenient to choose the polar
coordinate system $\tau=r\sin(\phi)$ and $x=r\cos(\phi)$, where the
angular coordinate $\phi$ varies between $0$ and $2\pi$. We note that if the field theory in question is relativistic then the field operator is invariant under  the shifts
$\phi\rightarrow \phi +w$, where $w$ is an arbitrary constant.

One first defines the vacuum state of the quantum field in question
by the path integral over a half of the total Euclidean spacetime
defined as $\tau\leq 0$ such that the quantum field satisfies the
fixed boundary condition $\psi(\tau=0,x,z)=\psi_0(x,z)$ on the
boundary of the half-space,
 \begin{equation}
  \Psi[\psi_0(x,z)]=\int\limits_{\psi(X)|_{\tau=0}=\psi_0(x,z)}
   {\cal D}\psi\; e^{-W[\psi]}~,
  \lb{W1}
  \end{equation}
where $W[\psi]$ is the action
of the field. The surface $\Sigma$ in our case is a plane and the
Cartesian coordinate $x$ is orthogonal to $\Sigma$. 
The co-dimension 2 surface $\Sigma$ defined by the conditions $x=0$ and $\tau=0$ 
naturally separates   the hypersurface $\tau=0$ in two parts:
$x<0$ and $x>0$. These are the two sub-regions $A$ and $B$ discussed in section \ref{subsection: Definition}.

The boundary data $\psi(x,z)$ is also
separated into $\psi_-(x,z)=\psi_0(x,z),\; x<0$ and
$\psi_+=\psi_0(x,z),x>0$. By tracing over $\psi_-(x,z)$ one defines a reduced
density matrix
 \begin{equation}
  \rho(\psi_+^1,\psi_+^2)=\int {\cal D}\psi_-\Psi(\psi_+^1,\psi_-)\Psi(\psi_+^2,\psi_-)~,
 \lb{W2}
 \end{equation}
where the path integral goes over fields defined on the whole
Euclidean spacetime except a cut $(\tau=0,x>0)$. In the path
integral the field $\psi(X)$ takes the boundary value $\psi_+^2$
above the cut and $\psi_+^1$ below the cut. The trace of the $n$-th
power of the density matrix (\ref{W2})  is then  given by the
Euclidean path integral over fields defined on an $n$-sheeted
covering  of the cut spacetime. In the polar coordinates
$(r,\phi)$ the cut corresponds to values $\phi=2\pi k,$
$k=1,2,..,n$. When one passes across the cut from one sheet to
another, the fields are glued analytically. Geometrically this
$n$-fold space is a flat cone $C_n$ with angle deficit $2\pi(1-n)$
at the surface $\Sigma$. Thus we have
 \begin{equation}
  \Tr\rho^n=Z[C_n]~,
 \lb{W3}
 \end{equation}
where $Z[C_n]$ is the Euclidean path integral over the $n$-fold cover of the Euclidean space, i.e. over the cone $C_n$.
Assuming that in (\ref{W3}) one can analytically continue to
non-integer values of $n$, one  observes that 
$$-\Tr \hat{\rho} \ln
\hat{\rho}=-(\alpha\partial_\alpha-1)\ln\Tr
\rho^\alpha|_{\alpha=1}\, ,
$$
 where $\hat{\rho}=\frac{\rho}{ \Tr\rho}$ is
the renormalized matrix density. Introduce the effective action
$W[\alpha]=-\ln Z(\alpha)$, where $Z(\alpha)=Z[C_\alpha]$ is the
partition function of the field system in question on a Euclidean
space with conical singularity at the surface $\Sigma$. In the
polar coordinates $(r,\phi)$ the conical space $C_\alpha$  is
defined by making the coordinate $\phi$ periodic with  period
$2\pi\alpha$, where $(1-\alpha)$ is very small. The invariance under the abelian isometry $\phi\rightarrow \phi+w$
helps to construct without any problem the correlation functions  with the required periodicity $2\pi\alpha$ starting from the $2\pi$-periodic correlation functions. The analytic continuation of $\Tr \rho^\alpha$ to $\alpha$ different from 1 in the relativistic case is naturally provided by the path integral $Z(\alpha)$ over the conical space $C_\alpha$.
 The entropy  is then calculated by the replica trick
 \begin{equation}  \lb{SS}
  S=(\alpha\partial_\alpha-1)W(\alpha)|_{\alpha=1}\, .
 \end{equation}
One of the advantages of this method is that we do not need to care about the normalization of the
reduced density matrix and can deal with a matrix which is not properly normalized.

\subsection{Uniqueness of analytic continuation}
\label{section: analytic continuation}

The uniqueness of the analytic continuation of $\Tr \rho^n$ to
non-integer $n$ may not seem  obvious, especially if the field
system in question is not relativistic so that there is no
isometry in the polar angle $\phi$ which would allow us without
any trouble to glue together pieces of the Euclidean space to
form a   path integral  over a conical space $C_\alpha$. However,
some arguments can be  given that the analytic continuation to
non-integer $n$ is in fact unique. 

Consider a renormalized density matrix $\hat{\rho}=\frac{\rho}{ \Tr
\rho}$. The eigenvalues of $\hat{\rho}$ lie in the interval
$0<\lambda<1$. If this matrix were a  finite matrix we could use
the triangle inequality to show that
 \[ |\Tr \hat{\rho}^\alpha|<|(\Tr\hat{\rho})^\alpha|=1~
   \quad {\rm if}\quad Re(\alpha)>1\;. \]
For infinite size matrices  the trace is usually infinite so that
a regularization is needed.  Suppose that $\epsilon$
is the regularization parameter and $\Tr_\epsilon$ is the
regularized trace. Then
 \be
  |\Tr_\epsilon\hat{\rho}^\alpha|<1~\quad {\rm if}\quad Re(\alpha)>1\;.
 \lb{W5}
 \ee
Thus $\Tr\hat{\rho}^\alpha$ is a bounded function in
the complex half-plane, $Re(\alpha)>1$.  Now suppose that we know that $\Tr_\epsilon
\rho^\alpha|_{\alpha=n}=Z_0(n)$ for integer values of $\alpha=n,\;
n=1,2,3,..$. Then, in the region $Re(\alpha)>1$, we can represent
$Z(\alpha)=\Tr_\epsilon\rho^\alpha$ in the form
 \begin{equation} \lb{W7}
  Z(\alpha)=Z_0(\alpha)+\sin(\pi\alpha)g(\alpha) \,,
 \end{equation}
where the function $g(\alpha)$ is  analytic (for
$Re(\alpha)>1$). Since by condition (\ref{W5}) the function
$Z(\alpha)$ is bounded, we obtain that, in order
to compensate for the growth of the sine in (\ref{W7}) for complex
values of $\alpha$, the function $g(\alpha)$ should satisfy the
condition
 \begin{equation} \lb{W8}
  |g(\alpha=x+iy)|<e^{-\pi |y| }~.
 \end{equation}
By Carlson's theorem \cite{Carlson} an analytic function, which is   bounded in the region
$Re(\alpha)>1$ and  which satisfies condition (\ref{W8}),
vanishes identically. Thus we conclude that $g(\alpha)\equiv0$
and there is only one analytic continuation to
non-integer $n$, namely the one given by function $Z_0(\alpha)$.

\subsection{Heat kernel and the Sommerfeld formula}
\lb{Heat kernel}
Consider for concreteness a quantum bosonic field  described by a field operator $\cal D$ so that the partition function is $Z=\det^{-1/2}{\cal D}$. Then the effective action defined as
 \begin{equation}
  W=-\frac{1}{ 2}\int_{\epsilon^2}^\infty \frac{ds}{ s}\Tr K (s)\lb{W}\,,
 \end{equation}
where parameter $\epsilon$ is a UV cutoff, is expressed in terms of the trace of the heat kernel $K(s,X,X')=<X|e^{-s{\cal D}}|X'>$. The latter is defined as a solution  to the heat  equation
 \begin{equation}\label{K}
  \left\{
  \begin{array}{l}
    (\partial_s+{\cal D}) \,K(s,X,X')=0 \,, \\
    K(s{=}0,X,X')=\delta(X,X') \,.
  \end{array}
    \right.
 \end{equation}
In order to calculate the effective action $W(\alpha)$ we use the heat kernel method. In the context of  manifolds with conical singularities this method was  developed in great detail in \cite{Dowker:1977zj}, \cite{Fursaev:1994in}. In the Lorentz invariant case  the invariance under the abelian symmetry $\phi\rightarrow \phi +w$ plays an important role. The heat kernel $K(s,\phi,\phi')$ (where we omit the coordinates other than the angle $\phi$) on regular flat space  then depends on the difference $(\phi-\phi')$. This function is $2\pi$ periodic with respect to $(\phi-\phi')$. The heat kernel $K_\alpha (s,\phi,\phi')$ on a space with a conical singularity is supposed to be $2\pi\alpha$ periodic. It is  constructed from the $2\pi$ periodic  quantity by applying the Sommerfeld formula
\cite{Sommerfeld}
  \begin{equation}
   K_\alpha(s,\phi,\phi')
    = K(s,\phi-\phi')  + \frac{ i}{4\pi\alpha}\int_\Gamma \cot \frac{w}{2\alpha} K(s,\phi-\phi'+w)dw \,.
  \label{Sommerfeld}
  \end{equation}
That this quantity still satisfies the heat kernel equation is a consequence of the invariance under the abelian isometry $\phi\rightarrow \phi +w$.
The contour $\Gamma$
consists of two vertical lines, going from $(-\pi+ i \infty )$
to $(-\pi- i \infty )$ and from $(\pi- i \infty )$ to
$(\pi-+ i \infty )$ and intersecting the real axis between the
poles of the $\cot \frac{w}{2\alpha}$: $-2\pi\alpha$, $0$ and $0,$
$+2\pi\alpha$, respectively. For $\alpha=1$ the integrand in
(\ref{Sommerfeld}) is a $2\pi$-periodic function and the
contributions of these two vertical lines cancel each other. Thus,
for a small angle deficit the contribution of the integral in
(\ref{Sommerfeld}) is proportional to $(1-\alpha)$.

\subsection{An explicit calculation}
\lb{Explicit-calculation}
Consider an infinite $(d-2)$-plane in $d$-dimensional space-time. The calculation of the entanglement entropy for this plane can be done explicitly
by means of the  heat kernel method.
In flat space-time, if the operator $\cal D$ is the Laplace operator,
$$
{\cal D}=-\nabla^2\, ,
$$
 one can use the Fourier transform  in order to solve the heat equation. In $d$ spacetime dimensions one has
 \begin{equation}
  K(s,X,X')={1\over (2\pi)^d}\int d^dp \,e^{ip_\mu(X^\mu{-}X'^\mu)}~e^{-sF(p^2)}\,.
 \lb{heat-kernel}
 \end{equation}
Putting $z^i=z'^i,\, i=1,..,d-2$  and choosing in the polar coordinate system $(r,\phi)$, that $\phi=\phi'+w$ we have that $p_\mu(X{-}X')^\mu=2pr\sin{w\o 2}\cos\theta$, where  $p^2=p^\mu p_\mu$ and $\theta$ is the angle between the $d$-vectors $p^\mu$ and $(X^\mu{-}X'^\mu)$. The radial momentum $p$ and angle $\theta$, together with  the  other $(d{-}2)$ angles  form a spherical coordinate system in the space of momenta $p^\mu$. Thus one has for the integration measure $\int d^dp=\Omega_{d-2}\int_0^\infty dp\,p^{d-1}\int_0^\pi d\theta\,\sin^{d-2}\theta\,$, where 
$\Omega_{d-2}={2\, \pi^{(d-1)/2} \over\Gamma((d-1)/2)}$ is the area of a unit radius sphere in $d{-}1$ dimensions. Performing the integration in (\ref{heat-kernel}) in this coordinate system we find
\be
  K(s,w,r)=\frac{\Omega_{d-2}\sqrt{\pi}}{ (2\pi)^d} \frac{\Gamma({d-1\over 2})}{(r\sin{w\over 2})^{(d-2)/2}}
   \int_0^\infty dp \, 
  p^{\frac{d}{ 2}}J_{\frac{d-2}{2}} \big(2rp\sin\frac{w}{2}\big)\,e^{-sp^2} \,.
 \lb{Kw}
\ee
For the trace one finds
 \begin{equation}
  \Tr K(s,w) = \frac{s}{(4\pi s)^{\frac{d}{2}}} {\pi \alpha\o \sin^2{w\o 2}} A(\Sigma) ~,
 \lb{TrK}
 \end{equation}
where $A(\Sigma )=\int d^{d-2}z$ is the area of the surface $\Sigma$.
One uses the integral $\int_0^\infty dx x^{1-\nu}
J_\nu(x)={2^{1-\nu}\over \Gamma(\nu)}$ for the derivation of (\ref{TrK}). 
The integral over the contour $\Gamma$ in the Sommerfeld formula (\ref{Sommerfeld})
is calculated via residues (\cite{Dowker:1977zj}, \cite{Fursaev:1994in})
 \begin{equation} \lb{C}
  C_2(\alpha)\equiv{ i \over 8\pi\alpha}\int_\Gamma \cot{w\o 2\alpha}\, {dw\o \sin^2{w\o 2}}={1\o 6\alpha^2}(1-\alpha^2)~.
 \end{equation}
Collecting everything together one finds that in flat Minkowski space-time
 \begin{equation}
  \Tr K_\alpha(s) =\frac1{(4\pi s)^{d/2}} \big( \alpha V 
   \, + \,2\pi \alpha C_2(\alpha)\, s\, A(\Sigma)\big)\, ,
 \lb{P}
 \end{equation}
where $V=\int d\tau d^{d-1}x$ is the volume of space-time and $A(\Sigma)=\int d^{d-2}x$ is the area of the surface $\Sigma$. Substituting (\ref{P}) into equation (\ref{W}) we obtain that the effective action contains two terms. The one proportional to the volume $V$ reproduces the vacuum energy in the effective action. The second term proportional to the area $A(\Sigma)$ is responsible for the entropy. Applying the formula (\ref{SS}) we obtain the entanglement entropy
 \be
  S=\frac{A(\Sigma)}{6 (d-2) (4\pi)^{(d-2)/2}\epsilon^{d-2}}\,
 \lb{Sent}
 \ee
 of an infinite plane $\Sigma$ in $d$ space-time dimensions.  Since any surface, locally, looks like a plane and a curved spacetime, locally, is approximated by Minkowski space, this result gives the leading 
 contribution to the entanglement entropy of any surface $\Sigma$ in flat or curved space-time.
 
\subsection{Entropy of massive fields}  
The heat kernel of a massive field described by the wave operator ${\cal D}=-\nabla^2 +m^2$ is expressed in terms of the heat kernel of massless field,
$$
K_{(m\neq 0)}(x,x',s)=K_{(m=0)}(x,x',s)\cdot e^{-m^2 s}\, .
$$
 Thus one finds
\be
\Tr K^{(m\neq 0)}_\alpha(s)=\Tr K^{(m=0)}_\alpha(s) \cdot e^{-m^2s}\, ,
\lb{trace massive field}
\ee
where the trace of the heat kernel for vanishing mass is given by (\ref{P}).
Therefore the entanglement entropy of a massive field is
\be
S_{m\neq 0}={A(\Sigma)\o  12 (4\pi)^{(d-2)/2}}\int_{\epsilon^2}^\infty {ds\o s^{d/2}}e^{-m^2s}\, .
\lb{massive field entropy}
\ee
In particular, if $d=4$, one finds that
\be
S_{m\neq 0}= {A(\Sigma)\o  12 (4\pi)^{}}({1\o \epsilon^2}+2m^2\ln\epsilon+m^2\ln m^2+m^2(\gamma-1) +O(\epsilon m))\, .
\lb{d=4 massive field entropy}
\ee
The logarithmic term in the entropy that is due to the mass of the field appears in any even dimension $d$. The presence of a UV finite term proportional to the $(d-2)$-th power of mass is the other general feature of (\ref{massive field entropy}), (\ref{d=4 massive field entropy}).

\subsection{An expression in terms of the determinant of the Laplacian on the surface}

Even though the entanglement entropy is determined by the geometry of the surface $\Sigma$, in general this can be not only its intrinsic geometry but also
how the surface is embedded in the larger space-time. The embedding is determined by the extrinsic curvature. 
The curvature of the larger spacetime enters through the Gauss-Cadazzi relations.
But in some particularly simple cases the entropy can be given a purely intrinsic interpretation. To see this for the case when $\Sigma$ is a plane we note that the
entropy (\ref{Sent}) or  (\ref{massive field entropy}) originates from the surface term in the trace of the  heat kernel (\ref{P}) (or (\ref{d=4 massive field entropy})). To leading order in $(1-\alpha)$, the surface term in the case of a massive scalar field is
$$
(1-\alpha) \cdot {1\o 6} \cdot \Tr K_\Sigma (s)\, ,\,\, 
$$
where 
$$
\Tr K_\Sigma(s)=\frac{A(\Sigma)}{(4\pi s)^{{d-2}\over 2}}\cdot e^{-m^2 s}\,
$$
can be interpreted as  the trace of the heat kernel of operator $-\Delta(\Sigma)+m^2$, where $\Delta(\Sigma)$ is the intrinsic Laplace operator  defined on $(d-2)$-plane $\Sigma$.
The determinant of the operator $-\Delta(\Sigma)+m^2$ is determined by 
$$
\ln\det(-\Delta(\Sigma)+m^2)=-\int_{\epsilon^2}^\infty {ds\o s}\Tr K_\Sigma (s)\, .
$$ 
Thus we obtain an interesting expression for the entanglement entropy 
\be
S=-{1\o 12}\ln\det(-\Delta(\Sigma)+m^2)\,
\lb{Sdeterminant}
\ee
in terms of geometric objects defined intrinsically on the  surface $\Sigma$. A similar expression in the case of an ultra-extreme black hole was
obtained in \cite{Mann:1997hm} and for a generic black hole with horizon approximated by a plane was obtained in \cite{Frolov:1998ea}.

\subsection{Entropy in theories with a modified propagator}
\label{section: theories with a modified propagator}
In certain physically interesting situations the propagator of a quantum field is different from the standard $1/(p^2+m^2)$ and is described by some 
function  as $1/F(p^2)$.  The quantum field in question then satisfies a modified Lorentz invariant field equation 
\be
{\cal D}\psi= F(\nabla^2)\psi=0\, .
\lb{F-operator}
\ee
Theories of this type naturally arise in models with extra dimensions. The deviations from the standard form of propagator may be
both in the UV regime (large values of $p$) or in the IR regime (small values of $p$). If the function $F(p^2)$ for large values of $p$ grows 
faster than $p^2$ this theory is characterized by improved UV behavior.

The calculation of the entanglement entropy performed in section \ref{Explicit-calculation} can be generalized to include theories with 
operator (\ref{F-operator}).  This example is instructive  since, in particular,  it illuminates the exact relation between the structure of 2-point function
(the Green's function in the case of free fields) and the entanglement entropy  \cite{Nesterov:2010jh}.

In $d$ spacetime dimensions one has
 \begin{equation}
  K(s,X,Y)={1\over (2\pi)^d}\int d^dp \,e^{ip_\mu(X^\mu{-}Y^\mu)}~e^{-sF(p^2)}\,.
 \lb{2-K}
 \end{equation}
 Note that we consider Euclidean theory so that $p^2\geq 0$.
The Green's function 
\be
G(X,Y)=<\psi(X), \psi(Y)>
\lb{Green's}
\ee 
is a solution to the field equation with a delta-like source
\be
{\cal D} \, G(X,Y)=\delta(X,Y)
\lb{G}
\ee
and can be expressed in terms of the heat kernel as follows
\be
G(X,Y)=\int_0^\infty ds \, K(s,X,Y)~~.
\lb{GK}
\ee
Obviously, the Green's function can be represented in terms of the Fourier transform in a manner similar to (\ref{2-K}),
\be
G(X,Y)={1\over (2\pi)^d}\int d^dp ~e^{ip_\mu(X^\mu{-}Y^\mu)}~ G(p^2)\,,\,\, G(p^2)=1/F(p^2)\, .
 \lb{2-2}
 \ee
The calculation of the trace of the heat kernel for operator (\ref{F-operator}) on  a space with a conical singularity 
goes along the same lines as in section \ref{Explicit-calculation}. This was performed in \cite{Nesterov:2010yi} and the result is
\begin{equation}
  \Tr K_\alpha(s) =\frac1{(4\pi)^{d/2}} \big( \alpha V P_{d}(s)
   \, + \,2\pi \alpha C_2(\alpha)A(\Sigma)P_{d-2}(s)\big)\, ,
 \lb{K-general-F}
 \end{equation}
where the functions $P_n(s)$ are defined as
\begin{equation}
  P_{n}(s)=\frac2{\Gamma({n\o 2} )}\int_0^\infty dp\, p^{n-1}\, e^{-s F(p^2)}\, .
 \label{functions-Pn}
 \end{equation}
 The entanglement entropy  takes the form (we remind the reader that for simplicity we take the surface $\Sigma$ to be 
a $(d-2)$-dimensional plane) \cite{Nesterov:2010yi}
\be
  S=\frac{A(\Sigma)}{12\cdot (4\pi)^{(d{-}2)/2}}
  \int_{\epsilon^2}^\infty {ds\o s}P_{d-2}(s)\,,
 \lb{Sent-general-F}
 \ee
 It is important to note that \cite{Nesterov:2010yi},  \cite{Nesterov:2010jh}
 
 \medskip

\noindent (i) the area law in the entanglement entropy is universal and is valid for any function $F(p^2)$;

\medskip
 
\noindent (ii) the entanglement entropy is UV divergent independently of the function $F(p^2)$, with the degree of divergence 
depending on the particular function $F(p^2)$;

\medskip

\noindent (iii) in the coincidence limit, $X=Y$, the Green's function (\ref{2-2})
\be
G(X,X)={2\o \Gamma({d\o 2})}{1\o (4\pi)^{d\o 2}}\int_0^\infty dp~p^{d-1}G(p^2)
\lb{X=Y}
\ee
may take a finite value if $G(p^2)=1/F(p^2)$ is decaying faster than $1/p^d$. However, even for this function $F(p^2)$ the entanglement entropy is UV divergent.

\medskip

As an example, consider a function which grows for large values of $p$ as $F(p^2)\sim p^{2k}$. 
The   2-point correlation function in this theory behaves as
\be
<\phi(X),\phi(Y)>\sim {1\o |X-Y|^{d-2k}}
\lb{2-point-function}
\ee
 and for $k>d/2$ it is regular in the coincidence limit. On the other hand, the entanglement entropy  scales as
 \be
 S\sim {A(\Sigma)\over \epsilon^{d-2\o k}}
 \lb{entropy-general-k}
 \ee
 and remains divergent for any positive value of $k$.  Comparison of (\ref{2-point-function}) and (\ref{entropy-general-k}) shows that only for $k=1$ (the standard form of the wave operator and the propagator) the short-distance behavior of the 2-point function is similar to the UV divergence of the entanglement entropy.

\subsection{Entanglement entropy in   non-Lorentz invariant theories}
\label{subsection: non-Lorentz invariant}

Non-Lorentz invariant theories are characterized by a modified dispersion relation, $\omega^2+F({\bf p}^2)=0$, between the energy $\omega$ and the 3-momentum $\bf p$.
These theories can be described by a wave operator of the following type
\be
{\cal D}=-\partial_t^2+F(-\Delta_x)\, ,
\lb{non-Lorentz operator}
\ee
where $\Delta_x=\sum_i^{d-1}\partial_i^2$ is the spatial Laplace operator. Clearly, the symmetry with respect to the Lorentz boosts is broken in operator (\ref{non-Lorentz operator}) if $F(q)\neq q$. 

As in the Lorentz invariant case to compute the entanglement entropy associated with a surface $\Sigma$ we choose  $(d-1)$ spatial coordinates $\{x^i,~i=1,..,d-1\}=\{x,z^a, ~a=1,..,d-2\}$, where $x$ is the coordinate orthogonal to the surface $\Sigma$ and $z^a$ are the coordinates on the surface $\Sigma$. Then, after going to Euclidean time $\tau=i t$, we switch to the polar coordinates, $\tau=r\sin(\phi)$, $x=r\cos(\phi)$.  In the Lorentz invariant case  the conical space which is needed for calculation of the entanglement entropy is obtained by making the angular coordinate $\phi$ periodic with  period $2\pi \alpha$ by applying the Sommerfeld formula (\ref{Sommerfeld}) to the heat kernel.
If  Lorentz invariance is broken, as it is for the operator (\ref{non-Lorentz operator}) there are  certain difficulties in applying the method of the conical singularity when one computes the entanglement entropy.
The difficulties come from the fact that the wave operator $\cal D$, if written in terms of the polar coordinates $r$ and $\phi$, becomes an explicit function of the angular coordinate $\phi$. As a result of this, the operator $\cal D$ is not invariant under shifts of $\phi$ to arbitrary  $\phi+w$.  Only shifts with $w=2\pi n$, where $n$ is an integer are allowed. Thus, in this case one cannot apply the Sommerfeld formula since it explicitly uses the symmetry of the differential operator under shifts of angle $\phi$.
On the other hand,  a conical space with angle deficit $2\pi (1-n)$ is exactly what we need to compute $\Tr \rho^n$ for the reduced density matrix. In ref.\cite{Nesterov:2010yi},
by using some  scaling arguments it was shown that the trace of the heat kernel $K(s)=e^{-s{\cal D}}$  on a conical space with $2\pi n$ periodicity, is
 \begin{equation} \lb{trace  non-Lorentz heat kernel}
  \Tr K_n(s)=n\Tr K_{n=1}(s)+\frac{1}{(4\pi)^{d/2}} 2\pi n\,C_2(n)\,A(\Sigma)\,P_{d-2}(s) \, ,
 \end{equation}
where $n\Tr K_{n=1}(s)$ is the bulk contribution. By the arguments presented in Section \ref{section: analytic continuation} there is a unique analytic extension of this formula to non-integer $n$. A simple comparison with the surface term in the heat kernel  of the
Lorentz invariant operator, which was obtained in the previous section, shows that the surface terms of  the two kernels are identical. We thus conclude that the entanglement entropy is given by the same formula
 \begin{equation} \lb{non-Lorentz entropy}
  S=\frac{A(\Sigma)}{12\cdot (4\pi)^{(d{-}2)/2}}
  \int_{\epsilon^2}^\infty {ds\o s}P_{d-2}(s)\, ,
 \end{equation}
where $P_n(s)$ is defined in (\ref{functions-Pn}), as in the Lorentz invariant case  (\ref{Sent-general-F}). A similar property of the entanglement entropy was observed for a non-relativistic theory described by the Schr\"{o}dinger operator \cite{Solodukhin:2009sk} (see also \cite{deBoer:2011wk} for a holographic derivation). For polynomial operators, $F(q)\sim q^k$, some scaling arguments can be  used \cite{Solodukhin:2009sk} to get the form of the entropy  that follows from (\ref{non-Lorentz entropy}).

\medskip

In the rest of the review we shall mostly focus on the study of Lorentz invariant theories with  field operator quadratic in derivatives, of the Laplace type, ${\cal D}=-(\nabla^2+X)$.

\subsection{Arbitrary surface in curved space-time: general structure of UV divergences}
The definition of the entanglement entropy and the  procedure for its calculation   generalize  to curved spacetime.
The surface $\Sigma$ can then be any smooth closed co-dimension two surface\epubtkFootnote{If the boundary of $\Sigma$ is not empty there could be extra terms in the entropy proportional to the ``area'' of the boundary $\partial \Sigma$  as was shown in \cite{Fursaev:2006ng}. We do not consider this case here.}  which divides the space in two sub-regions. In the next section we will consider in detail the case where this surface is a black hole horizon. 
Before proceeding to the  black hole case  we would like to specify the general structure of UV divergent terms in the entanglement entropy.
In $d$-dimensional curved  spacetime  entanglement entropy is presented in the form of a Laurent series with respect to the
UV cutoff $\epsilon$ (for $d=4$ see  	\cite{Solodukhin:2008dh})
\be
S={s_{d-2}\o \epsilon^{d-2}}+{s_{d-4}\o \epsilon^{d-4}}+.. +{s_{d-2-2n}\o \epsilon^{d-2-2n}}+..+s_0\ln\epsilon+s(g)\; ,
\lb{2.1}
\ee
where $s_{d-2}$ is proportional to the area of the surface $\Sigma$. All other terms in the expansion 
(\ref{2.1}) can be presented as integrals over  $\Sigma$ of local quantities constructed in terms of the Riemann curvature of the spacetime and the extrinsic curvature of the surface $\Sigma$. The intrinsic curvature of the surface $\Sigma$ of course can be expressed in terms of $\cal R$ and $k$ using the Gauss-Codazzi equations.
Since nothing should depend on the direction of  vectors normal to $\Sigma$, the integrands in expansion (\ref{2.1}) should be even powers of extrinsic curvature. The general form of the $s_{d-2-2n}$ term can be symbolically presented in the form
\be
s_{d-2-2n}=\sum_{l+p=n}\int_\Sigma {\cal R}^l\, k^{2p}\, ,
\lb{general-form}
\ee
where $\cal R$ stands for components of the Riemann tensor  and their projections onto the sub-space orthogonal to  $\Sigma$ and $ k$  labels  the components of the extrinsic curvature. 
Thus, since the integrands are even in derivatives, only terms $\epsilon^{d-2n-2}$, $n=0, 1, 2, ..$ appear in (\ref{2.1}). If  $d$ is even then there also may appear a logarithmic term $s_0$.  The term $s(g)$ in (\ref{2.1}) is a UV finite term, which also may depend on the geometry of the surface $\Sigma$, as well as on the geometry of the space-time itself.

\section{Entanglement entropy of non-degenerate  Killing horizons}

\subsection{The geometric setting of black hole spacetimes}
The notion of entanglement entropy is naturally applicable to a black hole.  In fact, probably the only way to  separate a system in two sub-systems is to place  one of them inside  a black hole horizon. The important feature that, in fact, defines the black hole is the existence of a horizon. Many useful definitions of a horizon are known.  In the present paper we shall consider only the case of the so-called {\it eternal} black holes for which different definitions  of the horizon coincide. The corresponding spacetime then admits a maximal analytic extension which we shall use in our construction. The simplest example is the Schwarzschild black hole,
the maximal extension of which is demonstrated on the well known Penrose diagram.    The horizon of the Schwarzschild black hole is an example of a so-called {\it Killing horizon}. 
The spacetime in this case possesses a global Killing vector, $\xi_t=\partial_t$, which generates  time translations. The Killing horizon is defined as a null hypersurface on which the Killing vector $\xi_t$ is null, $\xi_t^2=0$.  The null surface in the maximal extension of an eternal black 
consists of two parts: the future horizon and the past horizon. The two intersect on a co-dimension two compact surface $\Sigma$ called the {\it bifurcation} surface. In the maximally extended spacetime a hypersurface ${\cal H}_t$ of constant time $t$ is  a Cauchy  surface. 
The bifurcation surface $\Sigma$ naturally splits the Cauchy surface in two parts, ${\cal H}_-$ and ${\cal H}_+$, respectively {\it inside} and {\it outside} the black hole. For asymptotically flat spacetime, a such as the Schwarzschild metric, the hypersurface ${\cal H}_t$ has the topology of a wormhole.
(In the case of the Schwarzschild metric it is called the Einstein-Rosen bridge.) The surface $\Sigma$ is the surface of minimal area in ${\cal H}_t$.
In fact the bifurcation surface $\Sigma$ is a minimal surface not only in the $(d-1)$-dimensional Euclidean space ${\cal H}_t$ but also in the $d$-dimensional spacetime. As a consequence, as we show below, the components of the extrinsic curvature  defined for two vectors normal to $\Sigma$, vanish
on $\Sigma$.

The space-time in question admits a Euclidean version by analytic continuation $t\rightarrow i\tau$. It is a feature of regular metrics with a Killing horizon  that the direction of Euclidean time $\tau$ is compact with  period $2\pi\beta_H$ which is determined by the condition of regularity, i.e.
the absence of a conical singularity. In the vicinity of the bifurcation surface $\Sigma$, the spacetime then is a product of a compact surface $\Sigma$ and a
two dimensional disk, the time coordinate $\tau$ playing the role of the angular coordinate on the disk. The latter can be made more precise by introducing a new angular variable $\phi=\beta^{-1}_H\tau$ which varies from $0$ to $2\pi$.
We consider the static space-time with  Euclidean metric of the general type
\be
ds^2={\beta^2_H}g(\rho)d\phi^2+d\rho^2+\gamma_{ij}(\rho,\theta)d\theta^id\theta^j\, .
\lb{metric}
\ee
The radial coordinate $\rho$ is such that the surface  $\Sigma$ is defined by the condition $\rho=0$. Near this point the functions $g(\rho)$ and $\gamma_{ij}(\rho,\theta)$
can be expended as
\be
g(\rho)={\rho^2\over \beta_H^2}+O(\rho^4)\, , \,\, \gamma_{ij}(\rho,\theta)=\gamma_{ij}^{(0)}(\theta)+O(\rho^2)\, ,
\lb{g-and-r}
\ee
where $\gamma_{ij}^{(0)}(\theta)$ is the metric on the bifurcation surface $\Sigma$ equipped  with coordinates $\{\theta^i\, , \, i=1,..,d-2\}$. This metric describes what is called a non-degenerate horizon. The Hawking temperature of the horizon is finite in this case and equal to $T_H=1/(2\pi\beta_H)$.

It is important to note that  the metric (\ref{metric}) does not have to satisfy any field equations.  The entanglement entropy can be defined for any metric which possesses a Killing type horizon. In this sense the entanglement entropy is an {\it off-shell} quantity. It is useful to keep this in mind when one compares the entanglement entropy with some other approaches in which an entropy is assigned to a black hole horizon. Even though the metric (\ref{metric})
with (\ref{g-and-r}) does not have to satisfy the Einstein equations we shall still call the complete space described by the Euclidean metric
(\ref{metric}) the Euclidean black hole {\it instanton} and will denote it by $E$.

\subsection{Extrinsic curvature of horizon, horizon as a minimal surface}

The horizon surface $\Sigma$ defined by the condition $\rho=0$ in the metric (\ref{metric}) is a co-dimension 2 surface. It has two normal vectors: a spacelike vector $n^1$ with the only non-vanishing component $n^1_\rho=1$
and a timelike vector $n^2$ with the non-vanishing component $n^2_\phi=1/\rho$. With respect to each normal vector 
one defines an extrinsic curvature, $k_{ij}^a=-\gamma_{i}^{\ l}\gamma_j^{\ p}\nabla_l n^a_p$, $a=1,2$. The extrinsic curvature $k^2_{ij}$ identically vanishes. It is a consequence of the fact that $n^2$ is a Killing vector which generates the time translations. Indeed, the extrinsic curvature can be also written as a Lie derivative, $k_{\mu\nu}=-{1\o 2}{\cal  L}_n g_{\mu\nu}$, so that it vanishes if $n$ is a Killing vector. The extrinsic curvature associated to the vector $n^1$, 
\be
k^1_{ij}=-{1\o 2}\gamma_{i}^{\ l}\gamma_j^{\ p}\partial_\rho\gamma_{kn}\, ,
\lb{extrinsic curvature static}
\ee
is  vanishing when restricted to the surface defined by the condition $\rho=0$. It is due  to the fact that the term linear in $\rho$ is absent in the $\rho$-expansion
for $\gamma_{ij}(\rho,\theta)$ in the metric (\ref{metric}). This is required by the regularity of the metric (\ref{metric}): in the presence of such a term  the Ricci scalar would be singular at the horizon,  $R\sim 1/\rho$.

The vanishing of the extrinsic curvature of the horizon   indicates that the horizon is necessarily a minimal surface. It has the minimal area considered as a surface in $d$-dimensional spacetime.  On the other hand, in the Lorentzian signature, the horizon $\Sigma$ has the minimal area if considered on the hypersurface of constant time $t$, ${\cal H}_t$, the latter thus has the topology of a wormhole.

\subsection{The wave function of a black hole}
Although the entanglement entropy can be defined for any co-dimension two surface in  spacetime 
when the surface is a horizon particular care is required.
 In order to apply the general prescription outlined in section \ref{subsection: Definition}, we first of all need to specify the corresponding wave function. Here we will follow the prescription proposed by Barvinsky, Frolov and Zelnikov \cite{Barvinsky:1994jca}. This prescription is a natural generalization of the one in flat spacetime discussed in section \ref{subsection: replica method}. On the other hand, it is similar to the ``no-boundary'' wave function of the Universe introduced in \cite{Hartle:1983ai}. We define the wave function of a black hole by the Euclidean path integral over field configurations on the half-period Euclidean instanton defined by the metric (\ref{metric}) with angular coordinate $\phi$ changing in the interval from $0$ to $\pi$. This half-period instanton has Cauchy surface $\cal H$ (on which we can choose  coordinates $x=(\rho,\theta)$) as a boundary where we specify the boundary conditions in the path integral,
\begin{equation}
  \Psi[\psi_-(x),\psi_+(x)]=\int\limits_{\begin{array}{ll}
                                          \psi(X)|_{\phi=0}=\psi_+(x)\\
                                           \psi(X)|_{\phi=\pi}=\psi_-(x)
                                           \end{array} }
   {\cal D}\psi\; e^{-W[\psi]}~,
  \lb{wave function}
  \end{equation}
where $W[\psi]={1\o 2}\int \psi {\hat{\cal D}}\psi$ is the action
of the quantum  field $\psi$.  The functions $\psi_-(x)$ and $\psi_+(x)$ are the boundary values defined on the  part of  the hypersurface $\cal H$ which is respectively inside (${\cal H}_-$) and  outside (${\cal H}_+$) the  horizon $\Sigma$.  As was shown in  \cite{Barvinsky:1994jca} the wave function (\ref{wave function}) corresponds to the Hartle-Hawking vacuum state \cite{Hartle:1976tp}.

\subsection{Reduced density  matrix and entropy}

The density matrix $\rho(\psi_+^1,\psi_+^2)$ defined by tracing over $\psi_-$-modes is given by the Euclidean path integral over 
field configurations on the complete instanton $ (0< \phi <2\pi )$ with a cut along the axis $\phi=0$ where the field $\psi(X)$ in the path integral takes the values $\psi_+^1(x)$ and $\psi_+^2(x)$ below and above the cut respectively. The trace $\tr \rho$ is obtained by equating
the fields across the cut and doing the unrestricted Euclidean path integral on the complete Euclidean instanton $E$.  Analogously, $\tr \rho^n$
is given by the path integral over field configurations defined on the n-fold cover $E_n$ of the complete instanton.  This space is described by the metric
(\ref{metric}) where angular coordinate $\phi$ is periodic with  period $2\pi n$. It has  a conical singularity on the surface $\Sigma$ so that
in a small vicinity of $\Sigma$ the total space $E_n$ is a direct product of $\Sigma$ and a two-dimensional cone ${\cal C}_n$ with angle deficit $\delta=2\pi (1-n)$. Due to the abelian isometry generated by the Killing vector $\partial_\phi$ this construction can be analytically continued to arbitrary (non-integer) $n\rightarrow \alpha$. So that one can define a partition function 
\be
Z(\alpha)=\tr \rho^\alpha
\lb{Z}
\ee
by the path integral over field configurations over $E_\alpha$, the $\alpha$-fold cover of the instanton $E$. 
For a bosonic field described by the field operator $\hat{\cal D}$ one has that $Z(\alpha)=\det^{-1/2}\hat{\cal D}$.
Defining the effective action as
$W(\alpha)=-\ln Z(\alpha)$, the entanglement entropy is still given by (\ref{SS}), i.e. by differentiating the effective action with respect to the angle deficit. Clearly, only the term linear in $(1-\alpha)$ contributes to the entropy. The problem thus reduces to the calculation of this term in the effective action.

\subsection{The role of the rotational symmetry} 
We emphasize  that the presence of the so called {\it rotational symmetry}
with respect to the Killing vector $\partial_\phi$, which generates rotations in the 2-plane orthogonal to the entangling surface $\Sigma$, plays an important role in our construction. Indeed, without such a symmetry it would be impossible to interpret $\tr \rho^\alpha$ for an arbitrary $\alpha$ as a partition function in some gravitational background. In general,  two points are important for this interpretation:

\medskip

\noindent i) that  the spacetime possesses, at least locally near the entangling surface, a rotational symmetry so that, after the identification $\phi\rightarrow \phi+2\pi\alpha$
we get a well-defined spacetime $E_\alpha$ with no more than just a conical singularity; this holds automatically  if the surface in question is a Killing horizon;

\medskip

\noindent ii) and that the field operator is invariant under the ``rotations'', $\phi\rightarrow \phi +w$; this is automatic if the field operator is a covariant operator. 

\medskip

In particular, the point ii) allows us to use the Sommerfeld formula (more precisely its generalization to a curved spacetime) in order to define the Green's function or the heat kernel 
on the space $E_\alpha$. 
As is shown in \cite{Nesterov:2010yi} (see also discussion in section  \ref{subsection: non-Lorentz invariant}) in the case of the non-Lorentz invariant field operators in flat Minkowski spacetime   the lack of the symmetry ii) makes the whole ``conical space'' approach rather obscure.  On the other hand, in the absence of the rotational symmetry i) 
there may appear terms in the entropy that are ``missing'' in the  naively applied conical space approach: the extrinsic curvature contributions \cite{Solodukhin:2008dh} or even  some curvature terms \cite{Hung:2011xb}.

In what  follows we consider  the entanglement entropy  of the Killing horizons and deal with the covariant operators so that we do not have to worry about i) or ii).

\subsection{Thermality of the reduced density matrix of a Killing horizon}
\label{section:formal proof}
The quantum state defined by equation (\ref{wave function}) is the Hartle-Hawking vacuum \cite{Hartle:1976tp}. The Green's function in this state is defined by analytic continuation from the Euclidean Green's function. The periodicity $t\rightarrow t+i\beta_H$ is thus inherent in this state. This periodicity indicates that the correlation functions computed in this state  are in fact thermal correlation functions when continued to the Lorentzian section. This fact generalizes to an arbitrary interacting quantum field as shown in \cite{Gibbons:1976pt}. On the other hand, being globally defined, the Hartle-Hawking state is a pure state which involves correlations between modes localized on different sides of the horizon. This state however is described by a thermal density matrix if reduced to modes defined on one side of the horizon as was shown by Israel \cite{Israel:1976ur}. 
That the reduced density matrix obtained by tracing over modes inside the horizon is thermal  can be formally seen by using angular quantization. Introducing the Euclidean Hamiltonian $H_E$ which is the generator of rotations with respect to the angular coordinate $\phi$ defined above, one finds that $\rho(\psi_+^1,\psi_+^2)=<\psi_+^1|e^{-2\pi H_E}|\psi^2_+>$, i.e. the density matrix is thermal with respect to the Hamiltonian $H_E$ with  inverse temperature $2\pi$. This formal proof in Minkowski space  was outlined in \cite{Kabat:1994vj}. The appropriate Euclidean Hamiltonian is then the Rindler Hamiltonian which generates the Lorentz boosts in the direction orthogonal to the surface $\Sigma$.
In \cite{Jacobson:1994fp} the proof was generalized to the case of  generic static spacetimes with bifurcate Killing horizons admitting a regular Euclidean section.

\subsection{Useful mathematical tools}
\label{section:tools}

\subsubsection{Curvature of  space with a conical singularity}
\label{section: conical curvature}

Consider a space $E_\alpha$ which is an $\alpha$-fold covering of
a smooth manifold $E$ along the Killing vector $\partial_\varphi$
generating an
Abelian isometry. Let  surface $\Sigma$ be a stationary point of
this isometry so that  near $\Sigma$  the space $E_\alpha$ looks like a direct
product,  $\Sigma \times {\cal C}_\alpha$, of the surface $\Sigma$
and a
two-dimensional cone ${\cal C}_\alpha$ with angle deficit $\delta=
2\pi(1-\alpha)$.  Outside the singular surface $\Sigma$ the space
$E_\alpha$ has the same geometry as a  smooth manifold $E$.
In particular, their curvature tensors  coincide.  However, the conical singularity at the
surface $\Sigma$ produces a
 singular (delta-function like) contribution to the curvatures. This was first demonstrated  by Sokolov and Starobinsky \cite{Solokov-Starobinsky} in  two-dimensions  by using  topological arguments. These arguments were generalized to higher dimensions  in \cite{Banados:1993qp}. One way to
extract
the singular contribution  is to use some regularization
procedure,
replacing the singular space $E_\alpha$ by  a sequence  of
regular
manifolds $\tilde{E_\alpha}$.  This procedure was developed by Fursaev and Solodukhin in \cite{Fursaev:1995ef}. In the limit  $\tilde{E_\alpha}
\rightarrow
E_\alpha$ one obtains the following results \cite{Fursaev:1995ef}:
\begin{eqnarray}
&&R^{\mu\nu}_{\ \ \alpha\beta} = \bar{R}^{\mu\nu}_{\ \
\alpha\beta}+
2\pi (1-\alpha) \left( (n^\mu n_\alpha)(n^\nu n_\beta)- (n^\mu
n_\beta)
(n^\nu n_\alpha) \right) \delta_\Sigma \, ,\nonumber \\
&&R^{\mu}_{ \ \nu} = \bar{R}^{\mu}_{ \ \nu}+2\pi(1-\alpha)(n^\mu
n_\nu)
\delta_\Sigma \, , \nonumber                            \\
&&R = \bar{R}+4\pi(1-\alpha) \delta_\Sigma\, ,
\label{singular curvature}
\end{eqnarray}
where $\delta_\Sigma$ is the delta-function, $\int_{\cal M}^{}f
\delta_\Sigma=
\int_{\Sigma}^{}f$; $n^k=n^\mu_k\partial_\mu\, , \, k=1,\, 2$ are two orthonormal
vectors
orthogonal to the surface $\Sigma$, $(n_\mu n_\nu)=\sum_{k=1}^{2}n^k_\mu
n^k_\nu$
and the quantities $\bar{R}^{\mu\nu}_{\ \ \alpha\beta}$,
$\bar{R}^{\mu}_{ \
\nu}$ and $\bar{R}$
are computed in the regular points $E_{\alpha}/\Sigma$ by the
standard method.

These formulas can be used to define the integral expressions\epubtkFootnote{It should be noted that formulas (\ref{singular curvature}) and     (\ref{curvature integral 1}), (\ref{curvature integral 2}), (\ref{curvature integral 3}), (\ref{curvature integral 4})   are valid even if subleading   terms  (as in (\ref{g-and-r}))  in the expansion of the metric near singular surface $\Sigma$ are functions of $\theta$  \cite{Fursaev:1995ef}. Such more general metrics describe what might be called a ``local Killing horizon''.}  \cite{Fursaev:1995ef}
\begin{equation}
\int_{E_{\alpha}}R = \alpha\int_{E} \bar{R}
+4\pi(1-\alpha)\int_{\Sigma}1~~~,
\label{curvature integral 1}
\end{equation}
\begin{equation}
\int_{E_{\alpha}}R^2 = \alpha\int_{E} \bar{R}^2
+8\pi(1-\alpha)\int_{\Sigma}\bar{R}+O((1-\alpha)^2)~~~,
\label{curvature integral 2}
\end{equation}
\begin{equation}
\int_{E_{\alpha}}R^{\mu\nu}R_{\mu\nu} = \alpha\int_{ E}
\bar{R}^{\mu\nu}\bar{R}_{\mu\nu}
+4\pi(1-\alpha)\int_{\Sigma}\bar{R}_{ii}+O((1-\alpha)^2)~~~,
\label{curvature integral 3}
\end{equation}
\begin{equation}
\int_{E_{\alpha}}R^{\mu\nu\lambda\rho}R_{\mu\nu\lambda\rho}
=\alpha\int_{E}
\bar{R}^{\mu\nu\lambda\rho}\bar{R}_{\mu\nu\lambda\rho}
+8\pi(1-\alpha)\int_{\Sigma} \bar{R}_{ijij}
+O((1-\alpha)^2)~~~,
\label{curvature integral 4}
\end{equation}
where $\bar{R}_{ii}=\bar{R}_{\mu\nu}n_i^{\mu}n_i^{\nu}$ and
$\bar{R}_{ijij}=\bar{R}_{\mu\nu\lambda\rho}n^{\mu}_in^{\lambda}_
i
n^{\nu}_j n^{\rho}_j $. We use a shorthand notation for the  surface integral $\int_\Sigma\equiv\int_\Sigma \sqrt{\gamma}d^{d-2}\theta$.

 The terms proportional to $\alpha$ in
(\ref{curvature integral 1})-(\ref{curvature integral 4})
are defined on the regular space $E$.
The terms $O((1-\alpha)^2)$ in (\ref{curvature integral 2})-({\ref{curvature integral 4}) are something like a square of the $\delta$-function. They are not well-defined
and depend on
the way the singular limit
 ${\tilde E}_{\beta} \rightarrow{\cal E}_{\beta}$ is taken. Those
terms however are not important in the calculation of the
entropy since they are of higher order  in $(1-\alpha)$. There are however certain invariants, polynomial in the Riemann tensor, in which the terms $O((1-\alpha)^2)$ do not appear at all.
These invariants are thus well defined on the manifolds with conical singularity. Below we consider two examples of such invariants \cite{Fursaev:1995ef}.

\paragraph*{Topological Euler number.} 
The topological  Euler number of a
$2p$-dimensional smooth manifold ${\cal E}$ is given by the integral\epubtkFootnote{Note that in ref.\cite{Fursaev:1995ef} there is a typo in eq.(3.9) defining $c_p$. This does not affect the conclusions of ref.\cite{Fursaev:1995ef} since they are based on the relation $c_{p-1}=8\pi p c_p$ rather than on the explicit form of $c_p$.
}
\be
&&\chi= \int_{{ E}}^{}{\cal E}_{2p} \sqrt{g}d^{2p}x\, , \nonumber \\
&&{\cal E}_{2p}= c_p\epsilon_{\mu_1 \mu_2...\mu_{2p-1}\mu_{2p}}\epsilon^{\nu_1 \nu_2
... \nu_{2p-1}
\nu_{2p}} R^{\mu_1 \mu_2}_{\ \ \nu_1 \nu_2} ...R^{\mu_{2p-1} \mu_{2p}}_{\ \
\nu_{2p-1} \nu_{2p}}\, , \
c_p={1 \over 2^{3p}\pi^p p!} ~~~.
\label{Topological Euler number}
\ee
Suppose that ${E}_\alpha$ has several singular surfaces (of dimension $2(p-1)$) $\Sigma_i$,  each with  conical deficit $2\pi (1-\alpha_i)$. The Euler characteristic of this manifold is \cite{Fursaev:1995ef}
\begin{equation}
\chi [{E}_{\alpha}]=
\int_{{ E}_{\alpha}/\Sigma}{\cal E}_{2p} +
\sum_{i}(1-\alpha_i) \chi [\Sigma_i]\, .
\label{Euler number conical space}
\end{equation}
A special case is when ${ E}_\alpha$ possesses a continuous Abelian isometry. The singular surfaces $\Sigma_i$ are the fixed point sets  of this isometry so that  all surfaces have the same angle deficit $\alpha_i=\alpha$.  The Euler number in this case is \cite{Fursaev:1995ef}
\be
\chi[{ E}_\alpha]=\alpha \chi[{E}_{\alpha=1}]+ (1-\alpha)\sum_i \chi[\Sigma_i]\, .
\lb{Euler number conical manifold}
\ee
An interesting consequence of this formula is worth mentioning. Since the introduction of a conical singularity can be considered as the limit of certain  smooth deformation, under which the topological number does not change, one has 
$\chi[{E}_\alpha]=\chi[{ E}_{\alpha=1}]$. Then one obtains an interesting  formula reducing the number $\chi$
of a manifold ${ E}_{}$ to that of the fixed points set of
its abelian isometry \cite{Fursaev:1995ef}
\begin{equation}
\chi[{ E}_{\alpha=1}]=\sum_{i}\chi[\Sigma_i]\, .
\label{sumrule}
\end{equation}
A simple check shows that (\ref{sumrule}) gives the correct result for 
 the Euler number of the sphere $S^d_\alpha$. Indeed, the fixed points 
of $2$-sphere $S^2_{\alpha}$ are its "north" and "south" poles.
Each of these points has $\chi=1$ and
one gets from (\ref{sumrule}): $\chi[S^2]=1+1=2$. On the other hand,
the singular surface of $S^d_{\alpha}$ ($d\geq3$) is $S^{d-2}$
and from (\ref{sumrule}) the known identity
$\chi[S^d]=\chi[S^{d-2}]$ follows.  Note that equation (\ref{sumrule}) is
valid for spaces with continuous abelian isometry  and it may be violated
for an  orbifold with conical singularities.

\paragraph*{Lovelock gravitational action.}
The general Lovelock gravitational action is introduced on a d-dimensional Riemannian manifold as
the following polynomial \cite{Lovelock:1971yv}
\begin{equation}
W_L =
\sum_{p=1}^{k_d} \lambda _p \int {1 \over 2^{2p} p!}
\delta_{[\mu_1\mu_2...\mu_{2p-1}\mu_{2p}]}^{[\nu_1 \nu_2 ... \nu_{2p-1}
\nu_{2p}]} R^{\mu_1 \mu_2}_{\ \ \nu_1 \nu_2} ...R^{\mu_{2p-1} \mu_{2p}}
_{\ \ \nu_{2p-1} \nu_{2p}}\equiv
\sum_{p=1}^{k_d} \lambda _p W_p\, ,
\label{Lovelock}
\end{equation}
where $\delta_{[...]}^{[...]}$ is the totally antisymmetrized product
of the Kronecker symbols
and $k_d$ is $(d-2)/2$ (or $(d-1)/2$) for even (odd) dimension $d$. If the dimension of spacetime is $2p$, the action $W_p$ reduces to the Euler number 
(\ref{Topological Euler number}}) and is thus topological. In other dimensions the action (\ref{Lovelock}) is not topological, although it has some nice properties
which make it interesting. In particular, the field equations which follow from (\ref{Lovelock}) are quadratic in derivatives even though the action itself is 
polynomial in curvature.

On a conical manifold ${\cal M}_{\alpha}$, the Lovelock action is
 the sum of  volume and surface parts \cite{Fursaev:1995ef}
\begin{equation}
W_L[{\cal M}_{\alpha}]= W_L[{\cal M}_{\alpha}/\Sigma] + 2\pi (1-\alpha)
\sum_{p=0}^{k_d-1} \lambda _{p+1} W_{p}[\Sigma]\, ,
\label{LovelockC}
\end{equation}
where the first term is the action computed at the regular points. As in the case of the topological Euler number, all terms quadratic in $(1-\alpha)$ mutually cancel in
(\ref{LovelockC}). The surface term in (\ref{LovelockC}) takes the form of
the Lovelock action  on the singular surface $\Sigma$.
It should be stressed that integrals $W_{p}[\Sigma]$ are defined completely
in terms of the intrinsic Riemann curvature $R^{ij}_{\ \ \ kn}$ of $\Sigma$
\begin{equation}
W_p[\Sigma]={1 \over 2^{2p} p!}\int_{\Sigma}\delta^{[i_1...i_{2p}]}
_{[j_1...j_{2p}]}R^{i_1i_2}_{\ \ \ j_1j_2}...R^{i_{2p-1}i_{2p}}_{\ \ \ j_{2p-1}
j_{2p}}~~~
\label{W_p}
\end{equation}
and $W_0\equiv \int_{\Sigma}$.   Eq.(\ref{LovelockC}) allows one to compute the entropy in the Lovelock gravity by applying the replica formula. In ref.\cite{Jacobson:1993xs} this entropy was derived in the Hamiltonian approach, whereas arguments based on the
dimensional continuation of the Euler characteristics have been used for its derivation in \cite{Banados:1993qp}.

\subsubsection{The heat kernel expansion on a space with a conical singularity} 
The useful tool to compute the effective action on a space with a conical singularity is  the heat kernel method already discussed in  section
\ref{Heat kernel}. In section \ref{Explicit-calculation} we have shown how, in flat space, using the Sommerfeld formula (\ref{Sommerfeld}), to  compute the contribution  to the heat kernel due to the singular surface $\Sigma$. This calculation can be generalized to an arbitrary curved space $E_\alpha$ that  possesses, at least locally,  an Abelian isometry with a fixed point.
To be more specific we consider a  scalar field operator
${\cal D} =-(\nabla^2+ X)$, 
where $X$ is some scalar function.  Then  the trace of the heat kernel $K=e^{-s{\cal D}}$ has the following small $s$
expansion
\begin{eqnarray}
\tr K_{E_\alpha} (s)={1 \over (4\pi s)^{d \over
2}}
\sum_{n=0}^{}{} {a}_n s^n\, ,
\label{K expansion}
\end{eqnarray}
where the   coefficients in the expansion decompose into bulk (regular) and  surface (singular) parts
\begin{equation}
{a}_n={a}^{reg}_n + a^\Sigma_{n}\, .
\label{coefficient a}
\end{equation}
The regular coefficients are  the same as  for a smooth space. The first few coefficients are
\begin{eqnarray}
&&a_0^{reg}=\int_{E_\alpha} 1 \ \ , \ \ a^{reg}_1=\int_{E_\alpha}({1 \over 6}\bar{R}+X)\, ,
\nonumber \\
&&a^{reg}_2 =\int_{E_\alpha}\left({1 \over 180} \bar{R}^2_{\mu\nu\alpha\beta} -
{1 \over 180} \bar{R}^2_{\mu\nu} +{1 \over 6} \nabla^2(X+{1\o 5}\bar{R})+{1\o 2}(X+{1\o 6}\bar{R})^2\right)
\label{a-regular}
\end{eqnarray}
The coefficients  due to the singular surface $\Sigma$ (the stationary
point of the
isometry) are
\begin{eqnarray}
&&a_{0}^\Sigma =0; \ \ \ a^\Sigma_{1}={\pi \over
3}{(1-\alpha)(1+\alpha)
\over
\alpha}
\int_{\Sigma}^{}1 \, , \label{a-singular} \\
&&a^\Sigma_{2}={\pi \over 3} {(1-\alpha)(1+\alpha) \over \alpha}
\int_{\Sigma}^{}({1 \over 6}\bar{R}+X) -{\pi \over 180}
{(1-\alpha)(1+\alpha)(1+\alpha^2) \over \alpha^3}
\int_{\Sigma}^{}(\bar{R}_{ii}
-2\bar{R}_{ijij})\, .
\nonumber
\end{eqnarray}
The form of the regular coefficients (\ref{a-regular}) in the heat kernel expansion has been well studied in  physics and mathematics literature (for a review see  \cite{Vassilevich:2003xt}). The  surface coefficient $a_1^\Sigma$ in (\ref{a-singular}) was calculated by the mathematicians McKean and Singer \cite{McKean-Singer} (see also \cite{Cheeger}). 
In  physics literature this term has appeared in the work of Dowker  \cite{Dowker:1977zj}. (In the context of cosmic strings  one has focused more on the Green's function  rather on the heat kernel  \cite{Allen:1990mm}, \cite{Frolov:1987dz}.)
The coefficient $a_2^\Sigma$
has been first obtained by Fursaev \cite{Fursaev:1994in} although  in some special cases it was known before  in works of Donnelly \cite{Donnelly-1}, \cite{Donnelly-2}.

It should be noted, that due to the fact that the surface $\Sigma$ is a fixed point of the Abelian isometry, all components of the extrinsic curvature of the surface $\Sigma$ vanish. This explains why the extrinsic curvature does not appear in the surface terms (\ref{a-singular}) in the heat kernel expansion.

\subsection{General formula for  entropy in the replica method, relation to the Wald entropy}

As a consequence of the expressions (\ref{singular curvature}) for the curvature of space with a conical singularity 
that were presented in section \ref{section: conical curvature} one obtains a general expression for the entropy. Consider a Euclidean general covariant action  
\be
W[g_{\mu\nu}, \varphi_A]=-\int d^d x\sqrt{g}\, {\cal L}(g_{\mu\nu}, R^{\alpha\beta}_{\ \ \mu\nu}, \nabla_\sigma R^{\alpha\beta}_{\ \ \mu\nu}, ..., \varphi_A)\, ,
\lb{general covariant action}
\ee
which describes the gravitational field coupled to some matter fields $\varphi_A$. In the replica trick we first introduce a conical singularity at the horizon surface $\Sigma$ with a small angle deficit
$\delta=2\pi(1-\alpha)$ so that the Riemann curvature obtains a delta-like surface contribution (\ref{singular curvature}) and the gravitational action (\ref{general covariant action})
becomes a function of $\alpha$. Then applying the replica formula 
$$S=(\alpha\partial_\alpha-1)W(\alpha)|_{\alpha=1}$$
 we get 
\be
S=2\pi\int_\Sigma Q_{\alpha\beta\mu\nu}\left((n^\mu n^\alpha)(n^\nu n^\beta)-(n^\mu n^\beta)(n^\nu n^\alpha)\right)\, 
\lb{general formula for entropy}
\ee
for  the entropy associated to  $\Sigma$, where tensor $Q_{\alpha\beta\mu\nu}$ is defined as  a  variation of the action (\ref{general covariant action}) with respect to the Riemann tensor,
\be
Q^{\mu\nu}_{\ \ \alpha\beta}={1\o \sqrt{g}}{\delta W[g_{\mu\nu}, \varphi_A]\o \delta R^{\alpha\beta}_{\ \ \mu\nu}}\, .
\lb{tensor Q}
\ee
If the action (\ref{general covariant action}) is local and it does not contain covariant derivatives of the Riemann tensor then  the tensor $Q^{\mu\nu}_{\ \ \alpha\beta}$
is a partial derivative of the Lagrangian, 
\be
Q^{\mu\nu}_{\ \ \alpha\beta}={\partial {\cal L}\o \partial R^{\alpha\beta}_{\ \ \mu\nu}}\, .
\lb{tensor Q-2}
\ee
Now, as was observed by Myers and Sinha \cite{Myers:2010tj}  (see also \cite{Azeyanagi:2007bj}), one can re-express 
\be
\sum_{i,j=1}^2(n_i^\mu n_i^\alpha)(n_j^\nu n_j^\beta)-(n_i^\mu n_i^\beta)(n_j^\nu n_j^\alpha)=\epsilon^{\mu\nu}\epsilon^{\alpha\beta}\, ,
\lb{re-expression}
\ee
where $\epsilon^{\alpha\beta}=n_1^\alpha n_2^\beta-n_2^\alpha n_1^\beta$ is the two-dimensional volume form in the space transverse to the horizon surface $\Sigma$.
Then for a local action (\ref{general covariant action}) polynomial in the Riemann curvature the entropy (\ref{general formula for entropy}) takes the form
\be
S=2\pi \int_\Sigma {\partial {\cal L}\o \partial R^{\alpha\beta}_{\ \ \mu\nu}} \epsilon_{\mu\nu}\epsilon^{\alpha\beta}\, ,
\lb{Wald's entropy-0}
\ee
which is exactly the Wald entropy  \cite{Wald:1993nt}, \cite{Jacobson:1993vj}. It should be noted that Wald's Noether charge method is an on-shell method so that the metric in 
the expression for the Wald entropy  is supposed to satisfy the field equations. On the other hand,  the conical singularity method is an off-shell method  valid for any metric that describes a black hole horizon. The relation between the on-shell and the off-shell descriptions will be discussed in section \ref{section: Euclidean approach}.

\subsection{UV divergences of entanglement entropy for a  scalar field}
\label{section: UV-divergences}

For a bosonic field described by a field operator $\cal D$ the partition function is $Z(\alpha)=\det^{-1/2} {\cal D}$. The corresponding effective action $W(\alpha)=-\ln Z(\alpha)$ on a space with a conical singularity, $E_\alpha$, is expressed in terms of the heat kernel $K_{E_\alpha}(s)$ in a standard way
\be
W(\alpha)=-{1\over 2}\int_{\epsilon^2}^\infty {ds\o s}\, \tr K_{E_\alpha}(s)\, ,
\lb{effective action}
\ee
The entanglement entropy is computed using the replica trick as
\be
S=(\alpha\partial_\alpha-1)W(\alpha)|_{\alpha=1}\, .
\label{entropy-BH}
\ee
Using the small $s$ expansion one can, in principle, compute all UV divergent terms in the entropy. The surface terms are however known only for the  first few  terms in the expansion (\ref{K expansion}). This allows us to derive an explicit form for the  UV divergent terms  in the entropy.

\paragraph*{In two dimensions} the horizon is just a point and the entanglement entropy diverges logarithmically  \cite{Callan:1994py}, \cite{Kabat:1994vj}, \cite{Dowker:1994fi}, \cite{Fiola:1994ir}, \cite{Solodukhin:1994yz}
\be
S_{d=2}={1\over 6}\ln{1\over \epsilon}\, .
\lb{2d-entropy}
\ee
\paragraph*{In three dimensions} the horizon is  a circle and the  entropy 
\be
S_{d=3}={A(\Sigma)\over 12\sqrt{\pi}\epsilon}\, 
\lb{3d-entropy}
\ee
is linearly divergent.

\paragraph*{The leading UV divergence in $d$ dimensions} can be computed directly by using the form of the coefficient $a_1^\Sigma$ (\ref{a-singular}) in  the  heat kernel expansion
\cite{Callan:1994py}
\be
S_{d}={1\over 6(d-2)(4\pi)^{d-2\o 2}}{A(\Sigma)\over \epsilon^{d-2}}\, .
\lb{entropy dimension d}
\ee
It is identical to expression (\ref{Sent}) for the entanglement entropy in flat Minkowski spacetime. This has a simple explanation.
To leading order the spacetime near the black hole horizon is approximated by the flat Rindler  metric. The leading UV divergent 
term in the entropy is thus the entanglement entropy of the Rindler horizon. The curvature corrections then show up in the subleading UV divergent terms and in the UV finite terms.

\paragraph*{The four-dimensional case} is the most interesting since in this dimension there appears a logarithmic subleading term in the entropy. For a scalar field described by a field operator $-(\nabla^2+X)$ the UV divergent
terms in the entanglement entropy of a generic $4$-dimensional black hole read \cite{Solodukhin:1994st}
\be
S_{d=4}={A(\Sigma)\o 48\pi\epsilon^2}-{1\o 144\pi}\int_\Sigma \left({R}+6X-{1\o 5}({R}_{ii}-2{R}_{ijij})\right)\ln\epsilon\, .
\lb{UV divergence}
\ee
We note that for a massive scalar field $X=-m^2$. 

Of special interest is the case of the 4d {\it conformal} scalar field. In this case $X=-{1\o 6}{R}$ and the entropy (\ref{UV divergence}) takes the form 
\be
S_{conf}={A(\Sigma)\o 48\pi\epsilon^2}+{1\o 720\pi}\int_\Sigma ({R}_{ii}-2{R}_{ijij})\ln\epsilon\, .
\lb{UV conformal}
\ee
The logarithmic term in (\ref{UV conformal}) is invariant under the simultaneous  conformal transformations of bulk metric $g_{\mu\nu}\rightarrow e^{2\sigma}g_{\mu\nu}$ and the metric on the surface $\Sigma$, $\gamma_{ij}\rightarrow e^{2\sigma}\gamma_{ij}$. This is a general feature of the logarithmic term in the entanglement entropy of a conformally invariant field.

\medskip

Let us consider some particular examples.

\subsubsection{The Reissner-Nordstr\"{o}m black hole.}
A black hole of particular interest is the charged black hole described by the Reissner-Nordstr\"{o}m metric,
\be
&&ds^2_{RN}=g(r)d\tau^2+g^{-1}(r)dr^2+r^2(d\theta^2+\sin^2\theta d\phi^2)\, , \nonumber \\
&&g(r)=1-{(r-r_+)(r-r_-)\o r^2}\, .
\lb{metric-RN}
\ee
This metric has a vanishing Ricci scalar, $\bar{R}=0$. It has inner and out horizons, $r_-$ and $r_+$ respectively, defined by
\be
r_\pm=m\pm \sqrt{m^2-q^2}\, ,
\lb{inner-outer-horizon}
\ee
where $m$ is the mass of the black hole and $q$ is the electric charge of the black hole. The two vectors normal to the horizon are characterized by the non-vanishing components
$n_1^\tau=g^{-1/2}(r)$, $n_2^r=\sqrt{g(r)}$. The projections of the Ricci and Riemann tensors on the subspace orthogonal to $\Sigma$ 
are
\be
{R}_{ii}=-{2r_-\over r^3_+}\, , \,\, \, {R}_{ijij}={2r_+-4r_-\over r_+^3}\, .
\lb{Ricci-Riemann}
\ee
Since ${R}=0$ for the Reissner-Nordstr\"{o}m metric, the entanglement entropy of a massless, minimally coupled,  scalar field $(X=0)$ and of a conformally coupled scalar field $X=-{1\o 6}{R}$ coincide \cite{Solodukhin:1994st},
\be
S_{RN}={A(\Sigma)\o 48\pi\epsilon^2}+{1\o 90}({2r_+-3r_-\o r_+})\ln{r_+\o\epsilon}+s({r_-\o r_+})\, ,
\lb{entropy-RN}
\ee
where $A(\Sigma)=4\pi r_+^2$ and $s({r_-\o r_+})$ represents the UV finite term. 
Since $s$ is dimensionless it may depend only on the ratio ${r_-\o r_+}$ of the parameters which characterize the geometry of the black hole.

\medskip
If the black hole geometry is characterized by just one dimensionful parameter, the UV finite term in (\ref{entropy-RN}) becomes an irrelevant constant.
Let us consider two cases when this happens.  
\medskip

\noindent{\bf The Schwarzschild black hole. } In this case $r_-=0$ ($q=0$) and $r_+=2m $ so that the entropy, found by Solodukhin  \cite{Solodukhin:1994yz}, is
\be
S_{Sch}={A(\Sigma)\o 48\pi\epsilon^2}+{1\o 45}\ln {r_+\o \epsilon}\, .
\lb{entropy-Sch}
\ee
Historically this was the first time when the subleading logarithmic term in entanglement  entropy was computed. The leading term in this entropy is the same as in the Rindler space, when the actual black hole spacetime is approximated by flat Rindler spacetime. This approximation is sometimes argued to be valid in the limit of infinite mass $M$. We see, however, that, even in this limit,  there always exists the logarithmic subleading term in the entropy of the black hole that was absent in the case of the Rindler horizon. The reason for this difference is purely topological. The Euler number of the black hole spacetime is non-zero while it vanishes for the Rindler spacetime, respectively the Euler number of the black hole horizon (a sphere) is 2 while it is zero for the Rindler horizon (a plane).

\medskip
\noindent{\bf The extreme charged black hole.}  The extreme geometry is obtained in the limit $r_-\rightarrow r_+$ ($q=m$). The entropy of the extreme black hole is found to  take the form \cite{Solodukhin:1994st}
\be
S_{ext}={A(\Sigma)\o 48\pi\epsilon^2}-{1\o 90}\ln {r_+\o \epsilon}\, .
\lb{entropy-ext}
\ee
Notice that we have omitted the irrelevant constants  $s(0)$ and $s(1)$ in (\ref{entropy-Sch}) and (\ref{entropy-ext}) respectively.

\subsubsection{The dilatonic charged black hole}
The metric of a dilatonic black hole which has mass $m$, electric charge $q$ and magnetic charge
$P$ takes the form \cite{Gibbons:1987ps}:
\begin{equation}
ds^2=g(r)d\tau^2+g^{-1}(r)dr^2+R^2(r)d(d\theta^2+\sin^2\theta d\phi^2)
\lb{dilatonic-BH}
\end{equation}
with the metric functions
\begin{equation}
g(r)={(r-r_+)(r-r_-) \over R^2(r)}, \ \ \ R^2(r)=r^2-D^2\, ,
\lb{dilatonic metric}
\end{equation}
where $D$ is the dilaton charge, $D={P^2-q^2 \over 2m}$. The outer and 
the inner
horizons  are defined by
\begin{equation}
r_{\pm}=m\pm \sqrt{m^2+D^2-P^2-q^2}\, .
\end{equation}
The entanglement entropy is defined for the outer horizon at $r=r_+$.
The Ricci scalar of metric (\ref{dilatonic-BH}) 
$$
{R}=-2D^2{(r-r_+)(r-r_-)\o (r^2-D^2)^3}\,.
$$
vanishesat  the outer horizon, $r=r_+$.  Therefore, the entanglement entropy associated with the outer horizon is the same for a minimal scalar field ($X=0$) and for a conformally coupled scalar field ($X=-{1\o 6}r{R})$,
\begin{equation}
S_{dilat}={A_\Sigma \over 48\pi \epsilon^2}+{1 \over 90}
 ({3r_+(r_+-r_-) \over (r^2_+-D^2)}-1)
 \log {r_+\over \epsilon}+s({r_-\over r_+}, {D\o r_+})  \, ,
 \lb{S-dilaton}
\end{equation}
where $A_\Sigma=4\pi(r^2_+-D^2)$ is the area of the outer horizon.

It is instructive to consider the black hole with only electric charge
(the magnetic charge $P=0$ in this case). This geometry is characterized by two parameters: $m$ and $q$. In this case one finds
$$
r_+=2m-{q^2\o 2m}\, , \,\, r_-={q^2\o 2m}\, , \, \, r^2_+-D^2=4m(m-{q^2\o 2m})\, 
$$
so that expression (\ref{S-dilaton}) takes the form
\begin{equation}
S_{dilaton}={A_\Sigma \over 48\pi \epsilon^2}+{1\o 180}(1+3(1-{q^2\o 2m^2}))\ln{r_+\o \epsilon}+s({q\o m})\, .
\lb{S-dilat-P=0}
\end{equation}
In the {\it extremal} limit, $2m^2=q^2$, the area of the outer horizon 
vanishes,
$A_\Sigma=0$, and the whole black hole entropy is determined only by the
logarithmically divergent term\epubtkFootnote{Equations (\ref{S-dilaton}), (\ref{S-dilat-P=0}) and (\ref{S-dilat-extr}) correct some errors in   equations (27)-(29) of \cite{Solodukhin:1994st}.}   (using a different  ``brick wall'' method a similar conclusion was reached in \cite{Ghosh:1994wb})
\begin{equation}
S_{ext-dil}={1 \over 180} \log {r_+\over \epsilon}\, .
\lb{S-dilat-extr}
\end{equation}
In this respect the extreme dilatonic black hole is similar to a two-dimensional black hole.
Notice that (\ref{S-dilat-extr}) is positive as it should be since the entanglement entropy is, by definition, a positive quantity.

\medskip

The calculation of the entanglement entropy of a static black hole  is discussed in the following papers 
\cite{Fursaev:1994pq}, \cite{Frolov:1995xe}, \cite{Fursaev:1996uz}, \cite{DeNardo:1996kp}, \cite{Fursaev:1997th}, \cite{Emparan:1994qa}, \cite{Bytsenko:1997ru}, \cite{Zerbini:1996zw},
\cite{Cognola:1995yb}, \cite{Cognola:1995km}, \cite{Cognola:1993qg}, \cite{Iellici:1998np}, \cite{Iellici:1996jx}, \cite{Moretti:1997wi}, \cite{Solodukhin:1994yz}, \cite{Solodukhin:1994st}, \cite{Ghosh:1998bi}, \cite{Ghosh:2002fb}, \cite{Ghosh:1994wb}, \cite{Ghosh:1994mm}, \cite{Ghosh:1997hz}.

\subsection{Entanglement Entropy of the Kerr-Newman black hole}

The  geometry of the rotating black hole is more subtle than that of a static black hole: near the horizon the rotating spacetime is no longer a product of a horizon sphere $S_2$ and a two-dimensional disk.
The other difficulty with applying the technique of the heat kernel to this case is that the Euclidean version of the geometry requires  the rotation parameter to be complex.
Nevertheless with some care these difficulties can be overcome and the entanglement entropy of a rotating black hole can be computed along the same lines as for a static black hole
\cite{Mann:1996bi}. In this section we briefly review the results of Mann and Solodukhin \cite{Mann:1996bi}.

\subsubsection{Euclidean geometry of Kerr-Newman  black hole}
First we describe the Euclidean  geometry in the near-horizon limit of  the Kerr-Newmann black hole.
The Euclidean Kerr-Newman metric can be written in the form
\begin{equation}
ds^2_E={\hat{\rho}^2\over \hat{\Delta}}dr^2+{\hat{\Delta}\hat{\rho}^2\over
(r^2-\hat{a}^2)^2}\omega^2+\hat{\rho}^2(d\theta^2+\sin^2\theta \tilde{\omega}^2)\, ,
\label{5}
\end{equation}
where the Euclidean time is $t=\imath \tau$ and
the rotation and charge parameters have also been transformed
$a=\imath \hat{a},~q=\imath \hat{q}$, so
that the metric (\ref{5}) is purely real. Here
$\hat{\Delta}(r)=(r-\hat{r}_+)(r-\hat{r}_-)$,
where $\hat{r}_{\pm}=m\pm \sqrt{m^2+\hat{a}^2+\hat{q}^2}$,
the quantities $\omega$ and $\tilde{\omega}$ take the form 
\begin{equation}
\omega={(r^2-\hat{a}^2)\over \hat{\rho}^2}
(d\tau-\hat{a}\sin^2\theta d \phi ) \, , \,\,\,
\tilde{\omega}={(r^2-\hat{a}^2)\over 
\hat{\rho}^2}(d\phi+{\hat{a}\over (r^2-\hat{a}^2)}d\tau) 
\label{4-4}
\end{equation}
with $\hat{\rho}^2=r^2-\hat{a}^2 \cos^2 \theta$.
This space-time has a pair of orthogonal Killing vectors
\begin{equation}
K=\partial_\tau-{\hat{a}\over r^2-\hat{a}^2}\partial_\phi~,
~~\tilde{K}=\hat{a}\sin^2\theta \partial_\tau+\partial_\phi \label{4a}
\end{equation}
which are the respective analogs of the
vectors $\partial_\tau$ and $\partial_\phi$ in the (Euclidean)
Schwarzschild case.
The horizon surface $\Sigma$ defined by $r=\hat{r}_+$ is the stationary 
surface of the Killing vector $K$. Near this surface
the metric (\ref{5}) is approximately
\be
ds^2_E=ds^2_\Sigma+\hat{\rho}^2_+ ds^2_{C_2}\, ,
\lb{near-horizon limit}
\ee
where $\hat{\rho}^2_+=\hat{r}^2_+ -\hat{a}^2\cos^2\theta$ and
\begin{equation}
ds^2_\Sigma
=\hat{\rho}_+^2d\theta^2+{(\hat{r}^2_+-\hat{a}^2)^2\over\hat{\rho}_+^2}   
\sin^2\theta d\psi^2
\label{8}
\end{equation}
is the metric on the horizon surface $\Sigma$
up to $O(x^2)$, where variable $x$ is defined by  the relation $(r-\hat{r}_+)={\gamma x^2 \over 4}$, and
$\gamma=2\sqrt{m^2+\hat{a}^2+\hat{q}^2}$.
 The angle co-ordinate 
$\psi=\phi+{\hat{a} \over (\hat{r}^2_+-\hat{a}^2)} \tau$
and is well-defined on $\Sigma$. 
The metric $ds^2_{C_2}$ is that of a two-dimensional disk $C_2$ 
\begin{equation}
ds^2_{C_2}=dx^2+{\gamma^2 x^2\over 4 \hat{\rho}_+^4}d\chi^2~~
\label{11}
\end{equation}
attached to $\Sigma$ at a point ($\theta,~\psi$),
where $\chi=\tau-\hat{a}\sin^2\theta\ \phi$ is an angle
co-ordinate on $C_2$.

Regularity of the metric near the horizon implies the identifications
$\psi \leftrightarrow \psi +2\pi$ 
and  $\chi  \leftrightarrow \chi+4\pi\gamma^{-1}\hat{\rho}^2_+$.
For this latter condition to 
hold, independently of $\theta$ on the horizon, 
it is also necessary to
identify $(\tau,~\phi )$
with $(\tau+2\pi \beta_H,~\phi-2\pi \Omega \beta_H)$, 
where $\Omega={\hat{a} \over(\hat{r}^2_+- \hat{a}^2)}$ is the
(complex) angular velocity and $\beta_H={(\hat{r}_+^2-\hat{a}^2)/ 
\sqrt{m^2+\hat{a}^2+\hat{q}^2}}$. 
The identified points have the same coordinate $\psi$.

Near $\Sigma$ we therefore have the following description
of the Euclidean  Kerr-Newman geometry:
attached to every point $(\theta, \psi$) 
of the horizon is a two-dimensional disk $C_2$ with coordinates
($x, \chi$). The periodic identification of points on $C_2$ 
holds independently for different points on the horizon $\Sigma$,
even though $\chi$ is not a global coordinate.
As in the static case, there is an Abelian 
isometry generated by the Killing vector $K$, whose fixed set is
$\Sigma$. Locally we have $K=\partial_\chi$.
The periodicity is in the direction of the vector $K$ and
the resulting Euclidean space $E$ is a regular manifold.

Now consider closing the trajectory of $K$ with an arbitrary period
$\beta\neq \beta_H$. This implies the identification
$(\tau+2\pi \beta,~\phi-2\pi \Omega \beta)$, and the metric on
$C_2$ becomes
\begin{equation}
ds^2_{C_{2,\alpha}}=dx^2+\alpha^2x^2d\bar{\chi}^2\, ,
\label{12}
\end{equation}
where $\chi=\beta \hat{\rho}^2_+ (\hat{r}_+^2-\hat{a}^2)^{-1}\bar{\chi}$ 
is a new angular coordinate, with period $2\pi$. This is
the metric of a two dimensional cone with angular deficit 
$\delta=2\pi (1-\alpha)$, $\alpha\equiv {\beta\over \beta_H}$.
With this new identification the metric (\ref{5}) now describes the 
Euclidean conical space $E_\alpha$ with singular surface $\Sigma$.

The difference of the Kerr-Newman metric from the static case considered
above is that the Euclidean space near the bifurcation surface  is {\it not} a direct product of the surface $\Sigma$ and two-dimensional  cone $C_{2,\alpha}$. Instead, it is a nontrivial foliation of $C_{2,\alpha}$ over $\Sigma$. This foliation however shares certain common features with
the static case. Namely, the invariants constructed from quadratic combinations of extrinsic curvature of $\Sigma$ vanish $\Sigma$ identically.

\subsubsection{Extrinsic curvature of the horizon}
In the case of a static black hole we have argued that the presence of an Abelian isometry with horizon being the stationary point of the isometry guarantees that the extrinsic curvature
identically vanishes on the horizon.  In fact this is also true in the case of rotating black hole. The role of the Abelian isometry generated by the Killing vector $K$ is less evident
in this case. That is why in this subsection, following the analysis of ref.\cite{Mann:1996bi}, we explicitly evaluate the extrinsic curvature for the Kerr-Newman black hole and demonstrate that quadratic invariants, that can be constructed with the help of the extrinsic curvature, vanish on the horizon. 

With respect to the  Euclidean metric (\ref{5}) we may define a pair of 
orthonormal vectors $\{n_a=n_a^\mu \partial_\mu~,~~a=1,2 \}$:
\begin{equation}
n_1^r=\sqrt{\hat{\Delta} \over \hat{\rho}^2}\, ;~~
n_2^\tau={(r^2-\hat{a}^2)\over \sqrt{\hat{\Delta} \hat{\rho}^2}}~,~~n_2^\phi={-
\hat{a}\over  \sqrt{\hat{\Delta} \hat{\rho}^2}}\, .
\label{normal vectors Kerr-Newman}
\end{equation}
Covariantly these are
\begin{equation}
 n^1_r=\sqrt{\hat{\rho}^2 \over \hat{\Delta}}\, ; ~~n^2_\tau= \sqrt{\hat{\Delta} \over \hat{\rho}^2}~,~~n^2_\phi=-\sqrt{\hat{\Delta} \over \hat{\rho}^2} \hat{a}\sin^2\theta\, .
\label{normal co-vectors Kerr-Newman}
\end{equation} 
The vectors $n^1$ and $n^2$ are normal to the horizon surface $\Sigma$ 
(defined as $r=r_+,~\Delta (r=r_+)=0)$, which is a
two-dimensional surface with induced metric $\gamma_{\mu\nu}=g_{\mu\nu}-
n^1_\mu n^1_\nu -n^2_\mu n^2_\nu$. 
With respect to the normal vectors $n^a, ~a=1,2$ 
one defines  the extrinsic 
curvatures of the surface $\Sigma$: 
$\kappa^a_{\mu\nu}=-\gamma_\mu^\alpha \gamma_\nu^\beta \nabla_\alpha n^a_\beta$.
The exact expression for the components of extrinsic curvature is given in \cite{Mann:1996bi}.
The trace of the extrinsic curvature, $\kappa^a=\kappa^a_{\mu\nu}g^{\mu\nu}$,
\begin{equation}
\kappa^1=-{2r\over \hat{\rho}^2}\sqrt{\hat{\Delta}\over \hat{\rho}^2}~,~~\kappa^2=0
\label{extrinsic-KN}
\end{equation}
vanishes when restricted to the horizon surface 
$\Sigma$ defined by condition $\hat{\Delta}(r=\hat{r}_+)=0$. 
Moreover, the quadratic combinations 
\begin{eqnarray}
\kappa^1_{\mu\nu} \kappa_{1}^{\mu\nu}= {2r^2\hat{\Delta}\over \hat{\rho}^6}\, , \,\,\, \kappa^2_{\mu\nu} \kappa_{2}^{\mu\nu}= {2\hat{a}^2\cos^2\theta \hat{\Delta}
 \over \hat{\rho}^6} 
\label{trace extrinsic-KN }
\end{eqnarray}
vanish on the horizon $\Sigma$. Consequently, we have
$\kappa^a_{\mu\nu} \kappa^{a\mu\nu}=0$ on the horizon.

\subsubsection{Entropy}

Applying the conical singularity method to calculate the entanglement entropy of a rotating black hole we have to verify that i) the curvature singularity at the  horizon of a stationary black hole behaves in the same way as in the static case and ii) there are no extra surface terms  in the heat kernel expansion for the rotating black hole. 
The first point was explicitly checked in \cite{Mann:1996bi}: the curvature formulas (\ref{curvature integral 1})-(\ref{curvature integral 4}) are still valid in the stationary case.
Regarding the second point, it was shown by Dowker \cite{Dowker:1994bj} that  for a generic metric with conical singularity  at  some surface $\Sigma$ 
the only modification of the surface terms in the heat kernel expansion (\ref{a-singular})  are due to the extrinsic curvature of $\Sigma$.
 For example, the surface  coefficient $a_2^\Sigma$ may be modified by integrals over $\Sigma$ of terms $\kappa^a\kappa_a$ and $\kappa^a_{\mu\nu} \kappa^{a\mu\nu}$. Since, as was shown in the previous section, these terms identically vanish for the Kerr-Newman metric there is no modification of the surface terms in this case. The expression for the entropy (\ref{UV divergence}) thus
remains unchanged in the case of rotating black hole. The Ricci scalar for the Kerr-Newmann metric is zero, ${R}=0$.
The integrals of the projections of Ricci and Riemann tensors over horizon surface are
\begin{eqnarray}
&&\int_\Sigma {R}_{ijij}=8\pi{(\hat{r}^2_++\hat{q}^2)\over \hat{r}^2_+}+
4\pi{\hat{q}^2\over \hat{r}^2_+}{(\hat{r}^2_+-\hat{a}^2)\over \hat{a}\hat{r}_+}
\ln ({\hat{r}_++\hat{a}\over \hat{r}_+-\hat{a}}) \nonumber \\
&&\int_\Sigma {R}_{ii}=4\pi{\hat{q}^2\over\hat{r}^2_+}\left( 1 +
{(\hat{r}^2_+-\hat{a}^2)\over 2\hat{a}\hat{r}_+}
\ln ({\hat{r}_++\hat{a}\over \hat{r}_+-\hat{a}})\right)\, .
\label{Ricci-Riemann-KN}
\end{eqnarray}
The analytic continuation of these expressions back to real values
of the parameters $a$ and $q$ requires  the substitution
\begin{eqnarray}
&&\hat{q}^2=-q^2,~~\hat{a}^2=-a^2,~~\hat{r}_+=r_+ \nonumber \\
&&{1\over \hat{a}}\ln ({\hat{r}_++\hat{a}\over \hat{r}_+-\hat{a}})=
{2\over a} \tan^{-1}({a\over r_+})\, .
\label{analytic KN} 
\end{eqnarray}
With these identities the quantum entropy of the Kerr-Newman black hole reads \cite{Mann:1996bi}
\be
S_{KN}={A(\Sigma) \over 48\pi \epsilon^2} 
+{1\over 45}
\left( 1 -{3q^2\over 4r^2_+}  (1+{(r^2_++a^2)\over a r_+} \tan^{-1}({a\over r_+}))\right)\ln {1\o \epsilon}\, ,
\label{Kerr-Newman}
\ee
where $A(\Sigma)=4\pi (r^2_++a^2)$ is area of the horizon $\Sigma$.
In the limit $a\rightarrow 0$ this expression reduces to that of the
Reissner-Nordstr\"{o}m black hole (\ref{entropy-RN}). An interesting and a still  somewhat puzzling feature of this result is that, 
in the case of the Kerr black hole,  described by the Kerr-Newman metric with vanishing electric charge ($q=0$), the logarithmic term in the entropy does not depend on the rotation parameter $a$ and is the same as in the case of the Schwarzschild black hole. In particular for the extreme Kerr black hole ($q=0,\, m=a$) one has
\be
S_{Kerr}={A(\Sigma)\o 48\pi\epsilon^2}+{1\o 45}\ln{r_+\over \epsilon}\, .
\lb{Kerr}
\ee
The entropy of the Kerr black hole in the ``brick wall'' model was calculated in \cite{Cognola:1997dv} and a result different from (\ref{Kerr-Newman})   was found. 
The subsequent study in \cite{Frolov:1999gy} has, however, confirmed (\ref{Kerr-Newman}). 

\subsection{Entanglement entropy as one-loop quantum correction}
A natural point of view on the entanglement entropy of black hole is that this entropy, as was  suggested by Callan and Wilczek \cite{Callan:1994py}, is the first quantum correction to the Bekenstein-Hawking entropy\epubtkFootnote{This statement should be taken with some care. Entanglement entropy is a small correction compared to the Bekenstein-Hawking entropy if the UV cutoff $1/\epsilon$ is, for example, of order of few GeV (energy scale of the Standard Model). However, the two entropies are of the same order if the cutoff is at the Planck scale.
I thank G. 't Hooft for his comments on this point.}
   .
Indeed, the Bekenstein-Hawking entropy $S_{BH}$ can be considered as classical, or tree-level,  entropy. If we restore the presence of the Planck constant $\hbar$ the Bekenstein-Hawking entropy $S_{BH}$ is proportional to $1/\hbar$ while the entanglement entropy $S_{ent}$ is an $\hbar^0$ quantity. The total entropy of black hole is then a sum 
\be
S=S_{BH}+S_{ent}\, ,
\lb{total entropy}
\ee
where all particles that exist in Nature contribute to the entanglement entropy $S_{ent}$.

\subsection{The statement on the renormalization of the entropy}

As defined in the  previous sections the   entanglement entropy is a UV divergent quantity.
The other well-known  quantity which possesses  UV divergences is the effective action.
The standard way to handle the UV divergences in the action is to absorb them into redefinition of the
couplings which appear in the gravitational action. In four dimensions the gravitational action should also include the terms
 quadratic in the Riemann curvature.  The renormalization procedure then is well studied and is described in the textbooks (see for instance \cite{Birrell-Davies}). 
The idea now is that  exactly the same procedure renormalizes the UV divergences in the entropy. In order to demonstrate this statement
consider a minimally coupled scalar field. For simplicity suppose that the mass of the field vanishes. 
The bare (tree-level) gravitational action in four dimensions is the sum of the Einstein-Hilbert term and all possible combinations quadratic in the Riemann curvature,
\begin{equation}
W_{gr}=\int{}\sqrt{g}d^4x \left( -{1 \over 16\pi G_B} (R+2\Lambda_B)
+c_{1,B}R^2+c_{2,B}
R^2_{\mu \nu} +c_{3,B} R^2_{\mu\nu\alpha\beta} \right)\, ,
\label{bare action}
\end{equation}
where $G_B,\,, \Lambda_B,\, c_{1,B},\, c_{2,B},\, c_{3,B}$ are the bare coupling constants in the gravitational action.

The UV divergences of the gravitational action are computed by the method of the heat kernel using the small $s$ expansion
(\ref{a-regular}). For a minimal massless field ($X=0$ in the scalar field equation) one finds
\be
W_{div}(\epsilon)=-{1\o 64\pi^2\epsilon^4}\int_E 1-{1\o 192\pi^2\epsilon^2}\int_E R+{1\o 16\pi^2}\int_E\left({1\o 180}R^2_{\alpha\beta\mu\nu}-{1\o 180}R^2_{\alpha\beta}+{1\o 72}R^2\right)\, \ln\epsilon\, ,
\lb{action UV divergences}
\ee
These divergences are removed by  standard renormalization of the gravitational couplings in the bare gravitational action
\begin{equation}
W_{gr}(G_B, c_{i,  B}, \Lambda_B)+W_{div}(\epsilon)=W_{gr} (G_{ren}, c_{i,
ren},\Lambda_{ren})\, ,
\label{renormalization of action}
\end{equation}
where $G_{ren}$ and $c_{i,ren}$ are the renormalized couplings expressed in terms of the bare ones and the UV parameter $\epsilon$
\be
&&{1 \over G_{ren}}={1 \over G_{B}}+ {1 \over 12\pi \epsilon^2}\,
, \ \ c_{1,ren}=c_{1,B}+{1 \over 32\pi^2}{1\o 36}\ln
{\epsilon} \nonumber \\
&&c_{2,ren}=c_{2,B} -{1 \over 32 \pi^2} {1 \over 90} \ln {
\epsilon}
\, , \ \ c_{3,ren}=c_{3,B} +{1 \over 32 \pi^2} {1 \over 90} \ln {
\epsilon}\, .
\lb{renormalized constants}
\ee

The tree-level entropy can be obtained by means of  the same replica trick, considered in the previous sections,  upon 
introduction of the conical singularity with a small angle deficit $2\pi(1-\alpha)$, $S(G_B,c_{i,B})=(\alpha\partial_\alpha-1)W_{gr}(\alpha)$. The conical singularity at
the horizon
$\Sigma$ manifests itself in that a part of the Riemann tensor for
such a
manifold $E_\alpha$ behaves as a distribution having support on
the surface
$\Sigma$.   Using formulas (\ref{curvature integral 1})-(\ref{curvature integral 4}) one finds
for the tree-level entropy
\begin{equation}
S(G_B, c_{i,B})= {1 \over 4G_B} A(\Sigma)-\int_{\Sigma} \left( 8\pi
c_{1,B} R
+4\pi c_{2,B}R_{ii}+8\pi  c_{3,B}R_{ijij} \right)\, .
\label{tree level entropy}
\end{equation}
The Bekenstein-Hawking entropy  $S={1 \over  4G}A(\Sigma)$ is thus
modified due to the  presence of 
$R^2$-terms in the action  (\ref{bare action}). 
It should be noted that (\ref{tree level entropy}) exactly coincides with the entropy
computed by
the Noether charge method of Wald \cite{Wald:1993nt}, \cite{Jacobson:1993vj} (the relation between Wald's method and the method of conical singularity is discussed in \cite{Iyer:1995kg}).

The UV divergent part of the entanglement entropy of black hole has been already calculated, see (\ref{UV divergence}).
For a minimal massless scalar one has 
\be
S_{div}={A(\Sigma)\o 48\pi\epsilon^2}-{1\o 144\pi}\int_\Sigma \left({R}-{1\o 5}({R}_{ii}-2{R}_{ijij})\right)\ln\epsilon\, .
\lb{S-div}
\ee
The main point now is that the sum of the UV divergent part (\ref{S-div}) of the entanglement
entropy
 and  the tree-level entropy
(\ref{tree level entropy})
\begin{equation}
S(G_B, c_{i,B})+S_{div}(\epsilon)=S(G_{ren}, c_{i, ren})
\label{entropy renormalized}
\end{equation}
takes again the tree-level form (\ref{tree level entropy}) if expressed
in terms of 
the renormalized coupling constants $G_{ren},  \  c_{i,  ren}$ defined in (\ref{renormalized constants}).
Thus, the UV divergences in  entanglement entropy can be handled by
the standard renormalization of the  gravitational couplings. So that
no separate
renormalization procedure for  the entropy is required.

It should be noted that  the proof of the renormalization statement is based
on a nice property of  the heat kernel coefficients $a_n$ (\ref{coefficient a}) on space with conical singularity.  Namely,  up to
$(1-\alpha)^2$ terms the exact coefficient
$a_n=a_n^{reg}+a_n^\Sigma$   on the conical space  $E_\alpha$ is
equal to the regular volume coefficient
${a}_n^{reg}$ expressed in terms of the complete curvature, regular part plus a delta-like contribution,
using relations (\ref{singular curvature})
\begin{equation}
a_n (E_\alpha )=   a^{reg}_n(\bar{R})+a^\Sigma_N={a}^{reg}_n (\bar{R}+R^{sing}) +
O((1-\alpha)^2)\, .
\label{nice relation}
\end{equation}
The terms quadratic in $R^{sing}$ are not well defined. However these terms are proportional to $(1-\alpha)^2$ and do not
affect the entropy calculation. 
Thus,  neglecting   terms of order $(1-\alpha)^2$ in the calculation of entropy the renormalization of entropy
(\ref{entropy renormalized})
directly follows from the renormalization of the effective action
(\ref{renormalization of action}).

That the leading $1/\epsilon^2$ divergence in the entropy can be handled by the standard renormalization of  Newton's constant $G$ has  been suggested by Susskind and Uglum \cite{Susskind:1994sm} and Jacobson
\cite{Jacobson:1994iw}.
That one has to renormalize also the  higher curvature couplings in the gravitational action in order to remove all divergences in the entropy of the Schwarzschild black hole was suggested by Solodukhin 
\cite{Solodukhin:1994yz}.  For a generic static black hole the renormalization statement 
was proved  by Fursaev and Solodukhin in  \cite{Fursaev:1994ea}. In a different approach based on t'Hooft's ``brick wall model'' the renormalization was verified
for the Reissner-Nordstr\"{o}m black hole by Demers,  Lafrance and Myers \cite{Demers:1995dq}. For the rotating black hole described by the Kerr-Newman metric the renormalization of the entropy was demonstrated by Mann and Solodukhin \cite{Mann:1996bi}. The non-equilibrium aspect (as defining the rate in a semiclassical decay of hot flat space by black hole
nucleation) of the black hole entropy and the renormalization was discussed by Barbon and Emparan \cite{Barbon:1995im}. 

\subsection{Renormalization in theories with a modified propagator}
\label{section: theories with a modified propagator, covariant}

Let us comment briefly on the behavior of the entropy in theories described by a wave operator  ${\cal D}=F(-\nabla^2)$ which is a function of the standard Laplace operator $\nabla^2$.
In flat space this was analyzed in section \ref{section: theories with a modified propagator}.
As is shown in \cite{Nesterov:2010yi} there is a precise relation between the small $s$ expansion of the heat kernel of operator $F(-\nabla^2)$ and that of the Laplace operator
$-\nabla^2$. The latter heat kernel has the standard decomposition
\be
\Tr e^{s\nabla^2}={1\o (4\pi )^{d/2}}\sum_{n=0} a_n s^{n-d/2}\, .
\lb{standard decomposition}
\ee
The heat kernel of operator $F(-\nabla^2)$ then  has the decomposition  \cite{Nesterov:2010yi}
\be
\Tr e^{-sF(-\nabla^2)}={1\o (4\pi )^{d/2}}\sum_{n=0} a_n {\cal T}_n(s)\, ,
\lb{modified decomposition}
\ee
where 
\begin{equation} \lb{exact coefficient}
  {\cal T}_n(s) = \left\{
              \begin{array}{ll}
                 P_{d-2n}(s)
                  &  n<d/2 \\
                (-\partial_q)^{n-d/2} e^{-s F(q)}\Big|_{q=0}
                  &  n\geq d/2
              \end{array}
    \right.
 \end{equation}
 In even dimension $d$ the term ${\cal T}_{d/2}(s)=1$.
  This decomposition is valid both for regular manifolds and manifolds with a conical singularity. If a conical singularity is present, the coefficients $a_n$ have the standard decomposition into regular $a_n^{reg}$ and surface $a_n^\Sigma$ parts as in (\ref{coefficient a}). The surface term for $n=1$ is just the area of the surface $\Sigma$ while the surface terms with $n\geq 2$ contain surface integrals of $(n-1)$-th power of the Riemann curvature. Thus (\ref{modified decomposition}) is a decomposition in powers of the curvature of the spacetime. 
  
The functions $P_n$ are defined in (\ref{functions-Pn}). In particular, if $F(q)=q^k$ ($k>0$) one finds that
\be
P_n(s)=s^{-{n\o 2k}}{\Gamma({n\o 2k})\o \Gamma({n\o 2})k}\, .
\lb{monom}
\ee
 The terms with $n\leq d/2$ in decomposition (\ref{modified decomposition}) produce the UV divergent terms in the effective action and entropy. The term $n=d/2$ gives rise to the logarithmic UV divergence. 
In $d$ dimensions the area term in the entropy is the same as in flat spacetime (see eq.(\ref{Sent-general-F})).
In four dimensions ($d=4$) the UV divergent terms in the entropy are
\be
S={A(\Sigma)\o 48\pi}\int_{\epsilon^2}^\infty {ds\o s}\, P_2(s)-{1\o 144\pi}\int_\Sigma \left({R}-{1\o 5}({R}_{ii}-2{R}_{ijij})\right)\ln\epsilon\, .
\lb{d=4 modified divergence}
\ee
We note that an additional contribution to the logarithmic term may come from the first term in (\ref{d=4 modified divergence}) (for instance, this is so for the Laplace operator modified  by the mass term,
$F(q)=q+m^2$).

In the theory with operator $F(-\nabla^2)$  Newton's constant is renormalized as \cite{Nesterov:2010yi}
\be
{1\o G_{ren}}={1\o G_B}+{1\o 12\pi}\int_{\epsilon^2}^\infty {ds\o s}\, P_2(s)
\lb{renormalization modified}
\ee
while the higher curvature couplings $c_i$, $i=1,2,3$ in the effective action are renormalized in the same way as in (\ref{renormalized constants}).
The renormalization of $G$ and $\{c_i\}$  then makes finite both  the effective action and the entropy exactly in the same way as in the case of the Laplace operator $-\nabla^2$.
Thus the renormalization statement generalizes to the theories with modified wave operator $F(-\nabla^2)$.

\subsection{Area law: generalization to  higher spin fields }
In this section we will focus only on the leading UV divergent term, proportional to the area of the horizon. The proportionality of the entanglement entropy to the area is known as  the  ``area law''.
As we have discussed already for the case of a  scalar field,  this term in the entanglement entropy of a black hole is the same as in flat spacetime.
In flat Minkowski spacetime, for a field of spin $s$, massive or massless, including the gauge fields,  
the calculation of entanglement entropy effectively reduces to the scalar field calculation, provided the number of  scalar fields is equal to the number of physical degrees of freedom of the spin-$s$ field in question. The contribution of fermions comes with the weight $1/2$. 
Thus we can immediately write down  the general expression for the entanglement entropy of a quantum field of spin $s$ in $d$ dimensions,
\be
S_{(s,d)}={{\cal D}_s(d)\over 6(d-2)(4\pi)^{d-2\o 2}}{A(\Sigma)\over \epsilon^{d-2}}\, ,
\lb{entropy-arbitrary spin}
\ee
where ${\cal D}_s(d)$ is (with weight $1/2$ for fermionic fields) the number of physical (on-shell) degrees of freedom of a particle of 
spin $s$ in $d$ dimensions. 
For gauge fields this assumes gauge fixing.
In particular one has
\be
{\cal D}_{1/2}(d)={2^{[d/2]}\o 2}
\lb{spinors}
\ee
for Dirac fermions,
\be
{\cal D}_1(d)=(d-2)\cdot {\rm dim}\,  G\, ,
\lb{vectors}
\ee
for the  gauge vector fields (including the contribution of ghosts), where ${\rm dim} \, G$ is the dimension of the gauge group,
\be
{\cal D}_{3/2}(d)=(d-2)\, {2^{[d/2]}\o 2}
\lb{gravitino}
\ee
for the Rarita-Schwinger particles of spin $3/2$ with gauge symmetry (gravitinos), and 
\be
{\cal D}_2(d)={d(d-3)\over 2}\, 
\lb{graviton}
\ee
for massless spin-2 particles (gravitons).

The result for  Dirac fermions was first obtained by Larsen and Wilczek \cite{Larsen:1994yt}, \cite{Larsen:1995ax} and later in a paper of Kabat \cite{Kabat:1995eq}.
The contribution of the gauge fields to the entropy was derived  by Kabat \cite{Kabat:1995eq}. The entropy of the Rarita-Schwinger spin-3/2 particle
and of a  massless graviton was analyzed by Fursaev and Miele\epubtkFootnote{Among other things the authors of \cite{Fursaev:1996uz} observe certain, surprising,  non-smooth behavior of the
heat kernel coefficients for the spin-3/2 and spin-2  fields in the limit of vanishing angle deficit.}  \cite{Fursaev:1996uz}.

\subsection{Renormalization of entropy due to fields of different spin}
The effective action of a field of spin $s$ can be written as  
\be
W_{(s)}={(-)^{2s}\o 2}\int^\infty_{\epsilon^2}{ds\o s}\tr e^{-s\Delta^{(s)}}\, .
\lb{effective action spin s}
\ee
The second order covariant operators acting on  the spin-$s$ field can be represented in the following general form
\be
\Delta^{(s)}=-\nabla^2 +X^{(s)}\, ,
\lb{diff-operator}
\ee
where  the matrices $X_{(s)}$ depend on the chosen representation of the quantum field and are  linear in the Riemann tensor.
Here are some examples \cite{Christensen:1978md}, \cite{Christensen:1979iy}
\be
&&X^{(0)}=\xi\, R~,~~X^{(1/2)}_{AB}={1\o 4}R\delta_{AB}~,~~X^{(1)}_{\mu\nu}=\pm R_{\mu\nu}\, ,\nonumber \\
&&X^{(3/2)}_{AB,\mu\nu}={1\o 4}R\,\delta_{AB}g_{\mu\nu}-{1\o 2}R_{\mu\nu\alpha\beta}(\gamma^\alpha\gamma^\beta)_{AB}\, ,\nonumber \\
&&X^{(2)}_{\mu\nu ,\alpha\beta}={1\o 2} R(g_{\mu\alpha}g_{\nu\beta}+g_{\mu\beta}g_{\nu\alpha})-R_{\alpha\mu}g_{\beta\nu}-R_{\beta\nu}g_{\alpha\mu}-R_{\mu\alpha\nu \beta}-R_{\nu\alpha\mu\beta}\, ,
\lb{matrices X}
\ee
where $\gamma^\alpha_{AB}$ are gamma-matrices. 
The coefficient $a^{reg}_1$ in the small $s$ expansion (\ref{K expansion})-(\ref{a-regular}) of the heat kernel of operator (\ref{diff-operator}) has the general form
\be
a_1^{(s)}=\int_E({{D}_s(d)\o 6}R-\tr X^{(s)})\, ,
\lb{a2-general spin}
\ee
where $D_s(d)$ is the dimension of the representation of spin $s$,
\be
&&D_{s=0}=1\, , \,\, D_{s=1/2}=2^{[d/2]}\, ,\,\, D_{s=1}=d\, , \nonumber\\
&&D_{s=3/2}= d\, 2^{[d/2]}\, , \,\, D_{s=2}={(d-1)(d+2)\over 2}\, .
\lb{dimensions representation}
\ee
$D_s(d)$ can be interpreted as the number of off-shell degrees of freedom.

\medskip

\noindent Let us consider some particular cases.
\medskip

\paragraph*{ Dirac fermions ($s=1/2$).} The partition function for Dirac fermions is $Z_{1/2}=\det^{1/2}\Delta^{(1/2)}$. In this case $\tr X_{(1/2)}={1\o 4}\, 2^{[d/2]}\,R$ and hence
\be
a_1^{(s=1/2)}=-{{\cal D}_{1/2}\o 6}\int_E R\, ,
\lb{a1}
\ee
where ${\cal D}_{1/2}$ was introduced in (\ref{spinors}). We note that the negative sign in (\ref{a1}) in combination with negative sign for fermions in the effective action 
(\ref{effective action spin s}) gives the total positive contribution to Newton's constant. The renormalization of  Newton's constant due to Dirac fermions is
\be
{1\o 4G_{ren}}={1\o 4G}+{1\o (4\pi)^{d-2 \o 2}(d-2)}  {{\cal D}_{1/2}\o 6}{1\o \epsilon^{d-2}}\, .
\lb{Gren-spin-1/2}
\ee
Comparison of this equation with the UV divergence of entropy (\ref{entropy-arbitrary spin}) for spin-$1/2$ shows that
the leading UV divergence in the entropy of spin-$1/2$ field is handled  by the renormalization of Newton's constant in the same manner as it was for a scalar field.

\paragraph*{The  Rarita-Schwinger field ($s=3/2$).}  The partition function, including gauge fixing and the Faddeev-Popov ghost contribution, in this case is
\be
Z_{3/2}={\det}^{1/2} \Delta^{(3/2)}{\det}^{-1} \Delta^{(1/2)}\, ,
\lb{gravitino partfunction}
\ee
so that  the appropriate heat kernel coefficient is 
\be
a_1=a_1^{(3/2)}-2 a^{(1/2)}=-{{\cal D}_{3/2}\o 6}\int_E R\, ,
\lb{spin-3/2}
\ee
where ${\cal D}_{3/2}$ is introduced in (\ref{gravitino}). The renormalization of Newton's constant
\be
{1\o 4G_{ren}}={1\o 4 G}+{1\o (4\pi)^{d-2 \o 2}(d-2)}  {{\cal D}_{3/2}\o 6}{1\o \epsilon^{d-2}}\, 
\lb{Gren-spin-gravitino}
\ee
then, similarly to the case of Dirac fermions,  automatically renormalizes the entanglement entropy (\ref{entropy-arbitrary spin}).

\medskip

This property, however, does not  hold for all fields. The main role in the  mismatch between the UV divergences in the entanglement entropy and in Newton's constant
is played by the non-minimal coupling terms $X^{(s)}$  which appear in the  field operators (\ref{diff-operator}).

\subsection{The puzzle of non-minimal coupling}
\label{section: non-minimal coupling}

The simplest case to consider is 
\paragraph*{A non-minimally coupled scalar field.}  In this case one has  $\tr X^{(0)}=\xi R$, where $\xi$ is the parameter of non-minimal coupling. The renormalization of Newton's constant 
\be
{1\o 4G_{ren}}={1\o 4G}+{1\o (4\pi)^{d-2 \o 2}(d-2)}({1\o 6}-\xi){1\o \epsilon^{d-2}}\, 
\lb{non-minimal scalar}
\ee
is modified due to  the presence of the  non-minimal coupling $\xi$ in the scalar field operator.
At the same time, the entropy calculation on a Ricci flat background (${R}=0$) is not affected by the non-minimal coupling since the field operator for this background is identical to the minimal one. This simple reasoning shows that the area law in the case of a non-minimally coupled scalar field is the same as in the case of the minimal  scalar field (\ref{entropy dimension d}). Clearly, there is a mismatch between the renormalization of 
Newton's constant and the renormalization of  the entanglement entropy. One concludes that, in the presence of non-minimal coupling, when the Riemann tensor appears explicitly in the action of quantum field, the UV divergence of the entanglement entropy {\it can not} be handled by the 
standard renormalization of the Newton'c constant. The mismatch in the entropy is 
\be
S^{non-min}_\xi={(-\xi)\over (d-2)(4\pi)^{d-2\o 2}}{A(\Sigma)\over \epsilon^{d-2}}\, .
\lb{S-non-minimal}
\ee
It is an important fact that there is no known  way to give a statistical meaning to this entropy. Moreover, (\ref{S-non-minimal}) does not  have a definite sign  and may become negative if
$\xi$ is positive. In some respect this term is similar to the classical Bekenstein-Hawking entropy: both entropies, at least in the framework of the conventional field theory,  do not have a well-defined statistical meaning. There is a hope that in string theory the terms similar to  (\ref{S-non-minimal})
may acquire  a better meaning. This question is however still open.

We should note that on a space with a conical singularity one can  consider the Ricci scalar in the non-minimal scalar operator as the complete curvature including the $\delta$-like singular term as in  (\ref{singular curvature}). Then the differential operator $-(\nabla^2+\xi R)$ contains a  delta-like potential concentrated on the horizon surface $\Sigma$. 
The presence of this potential modifies the surface terms in the heat kernel in such a way that \cite{Solodukhin:1995ak}
\be
a_1^\Sigma(\xi)=a_1^\Sigma(\xi=0)-4\pi \xi (1-\alpha)\int_\Sigma 1+O(1-\alpha)^2\, ,
\lb{heat kernel with delta term}
\ee
where $a_1^\Sigma (\xi=0)$ is the surface term $a_1^\Sigma$  (\ref{a-singular}) without the non-minimal coupling, the term $O(1-\alpha)^2$ is ill-defined (something like $\delta^2(0)$),
however it does not affect the entropy calculation. If we now apply the replica trick and calculate the entropy corresponding to the theory with the heat kernel with the surface term (\ref{heat kernel with delta term}) we get that \cite{Solodukhin:1995ak}, \cite{Larsen:1995ax}, \cite{Barvinsky:1995dp}
\be
S_{div}={1\o (d-2)(4\pi)^{d-2\o 2}\epsilon^{d-2}}({1\o 6}-\xi)A(\Sigma)\, .
\lb{entropy with non-min coupling}
\ee
This divergence takes the form  consistent with the UV divergence   of Newton'constant (\ref{non-minimal scalar}). However, we can not interpret this entropy as a contribution to the entanglement entropy since the presence of the delta-like potential in the Euclidean field operator is not motivated from the point of view of the original Lorentzian theory, for which the entanglement entropy is calculated. Moreover, (\ref{entropy with non-min coupling}) is not positive  if $\xi>1/6$ while the entanglement entropy is supposed to be a  positive quantity.

\medskip

\noindent Similar features are shared by other non-minimally coupled fields.

\medskip

\paragraph*{Abelian vector field.} After  gauge fixing the partition function of an Abelian gauge field is 
\be
Z={\det}^{-1/2}\, \Delta_+^{(1)}\cdot \det \Delta^{(0)}\, .
\lb{Z-vector field}
\ee
where  $\Delta_+^{(1)}$ is operator defined in (\ref{diff-operator}) with sign $+$ in the matrix $X^{(1)}_{\mu\nu}$ (\ref{matrices X}). 
For the effective action $W_{eff}=-\ln Z$ we find,
\be
W_{eff}=-{1\o 2}\int_{\epsilon^2}^\infty {ds\o s}{1\o (4\pi s)^{d/2}}(\int_E 1\ +a_1 \, s+..)\, ,
\lb{W-vector}
\ee
where 
\be
a_1=a_1^{(s=1)}-2a_1^{(s=0)}=({{\cal D}_1(d)\o 6}  -1)\int_E R  \, ,
\lb{a1-vector}
\ee
and ${\cal D}_1(d)=d-2$ is the number of on-shell degrees of freedom of the Abelian vector field. The renormalization of Newton's constant is
\be
{1\o 4G_{ren}}={1\o 4G}+{1\o (4\pi)^{d-2\o 2}(d-2)} ( {{\cal D}_{1}(d)\o 6}-1) {1\o \epsilon^{d-2}}\, .
\lb{G-ren-vector-1}
\ee
Comparison with (\ref{entropy-arbitrary spin})  shows that there is again a mismatch between the UV divergences in the entropy and in Newton's
constant.  This mismatch originates  from the non-minimal term $X^{(1)}_{\mu\nu}$ in the Laplace type field operator for the vector field.

\paragraph*{Massless graviton.} The partition function of a massless graviton in $d$ dimensions, after  gauge fixing and adding the Faddeev-Popov ghost contribution, is
\be
Z={\det}^{-1/2}\Delta^{(2)}\cdot\det\Delta_{-}^{(1)} \cdot {\det}^{-1/2}\Delta^{(0)}\, ,
\lb{Z-graviton}
\ee
where  $\Delta_-^{(1)}$ is operator defined in (\ref{diff-operator}) with sign $-$ in the matrix $X^{(1)}_{\mu\nu}$ (\ref{matrices X}). The operator $\Delta^{(2)}$ governs the dynamics of the tensor perturbations which satisfy the condition $\nabla^\mu(h_{\mu\nu}-{1\o 2}g_{\mu\nu}h)=0$, the operator $\Delta^{(0)}$ is due to the contribution of the conformal mode while
the determinant of operator $\Delta_{-}^{(1)}$ is due to the Faddeeev-Popov ghosts. 
Hence one has in this case that
\be
a_1=a_1^{s=2}-2a_1^{(s=1)}+a_1^{(s=0)}=({{\cal D}_2(d)\o 6}-c(d))\int_E R\, ,\ \ c(d)={d^2-d+4\o 2}\, ,
\lb{a1-graviton}
\ee
where ${\cal D}_2(d)={d(d-3)\o 2}$ is the number of on-shell degrees of freedom of a massless spin-$2$ particle.  The  renormalization of Newton's constant is
\be
{1\o 4G_{ren}}={1\o 4G}+{1\o (4\pi)^{d-2\o 2}(d-2)} ( {{\cal D}_{2}(d)\o 6}-c(d)) {1\o \epsilon^{d-2}}\, .
\lb{G-ren-vector}
\ee
Again, we observe the mismatch between the UV divergent terms in the entropy (\ref{entropy-arbitrary spin}) and in Newton's constant, this time due to the  graviton.

\medskip

To summarize, the UV divergences in the entanglement entropy of minimally coupled scalars and fermions are properly renormalized by the redefinition of Newton's constant. It happens that each minimally coupled field (no matter bosonic or fermionic)  contributes positively to Newton's constant and positively to the entropy of black hole. The contributions to both quantities come proportionally  that allows the simultaneous 
renormalization of both quantities.
The mismatch between the UV divergences in the entropy and in Newton's constant appears for gauge bosons: the Abelian (and non-Abelian)  vector fields and gravitons.
The source of the mismatch are those non-minimal terms $X^{(s)}$ in the field operator which contribute {\it negatively} to Newton's constant and do not make any contribution to the entanglement entropy of the black hole. At the time this review is being written, 
the appropriate treatment of the entropy of non-minimally coupled fields is not yet available.

\subsection{Comments on the entropy of interacting fields}

So far we have considered
   free fields in a fixed gravitational (black hole) background.  The interaction can be included by adding a potential
term $\int d^dx V(\psi)$ to the classical action, here
$\psi=\{\psi^i~, ~i=1,..,N\}$ is  set of fields in
question.
In the  one-loop approximation one splits
$\psi=\psi_c+\psi_q$, where $\psi_c$ is the classical
background field and $\psi_q$ is the quantum field. The
integration over $\psi_q$ then  reduces to calculation of the
functional determinant of operator ${\cal D}+M^2(\psi_c)$,
where $M^2_{ij}=\partial^2_{ij} V(\psi_c)$.   The fields $\psi_c$ representing the classical background  are in general  functions on the curved space-time. In some cases these fields are constants   that minimize the potential
$V(\psi)$.  The matrix $M_{ij}(x)$ plays the role of  an $x$-dependent mass matrix.
In the approximation when one can neglect the derivatives of the matrix $M$ the heat kernel
of  operator ${\cal D}+M^2$ is presented as the product  $\tr  e^{-s{\cal D}}\cdot \tr e^{-sM^2}$.
Using the already calculated trace of the heat kernel $\tr e^{-s{\cal D}}$ on space with a conical singularity  one obtains  at one-loop the entanglement entropy of the interacting fields. In $d$ dimensions one obtains \cite{Solodukhin:2009sk}
\be
S_{(d)}={1\over 12(4\pi)^{d-2\over
2}}\int_\Sigma\Tr[M^{d-2}\Gamma(1-{d\over 2},M^2\epsilon^2)]~~,
\label{Eint}\ee where we used that $\int_{\epsilon^2}^\infty
s^{-d/2}e^{-M^2s}=\Gamma(1-{d\over 2},M^2\epsilon^2)$. Using the
asymptotic behavior $\Gamma
(-\alpha,x)=\alpha^{-1}x^{-\alpha}+..$, we find that the leading
UV divergence of the entropy (\ref{Eint}) is again (multiplied by
$N$) (\ref{entropy dimension d}) and is thus not affected by the presence of the
interaction in the action. The interaction however shows up in the
sub-leading UV divergent and the UV finite terms. For instance, in
four dimensions on a flat background we find \be S={A(\Sigma)\over 48\pi} {N\over
\epsilon^2}+{1\o 48\pi}\int_\Sigma \left((\gamma-1)\Tr M^2+\tr M^2\ln(\epsilon^2 M^2)\right)\, .\lb{ent4}
\ee 
We see that the leading UV divergent term proportional to the area is not modified by the presence of the self-interaction.
The mass matrix $M^2$ is function of the background field $\psi_c$. The result (\ref{ent4}) thus indicates that
at  tree-level the entropy should contain terms additional to those of the standard area law, which depend on the value of
field $\psi$ at the horizon. In order to illustrate this point consider a  $\psi^4$ model of a single field
(a two-dimensional   model of this type was considered in \cite{Kabat:1995jq}, in four dimensions the role of self-interaction was discussed in \cite{Cognola:1993qg})
\be
W[\psi]={1\o 2}\int d^4x\sqrt{g}\left((\nabla\psi)^2+\xi R \psi^2+{\lambda\o 6}\psi^4\right)\, ,
\lb{psi-4}
\ee
where we included the term with the non-minimal coupling. In fact if we had not  included this term, it would have been generated by the quantum corrections
due to the self-interaction of the field $\psi^4$. This is a well-known fact, established in \cite{Nelson:1982kt}. The renormalized  non-minimal  coupling in the model (\ref{psi-4}) is 
\be
\xi_{ren}=\xi-{\lambda\o 8\pi^2}({1\o 6}-\xi)\ln\epsilon\, ,
\lb{xi-renormalization}
\ee
where we omit the terms of higher order in $\lambda$.
 Splitting in (\ref{psi-4})
the field $\psi$ into classical and quantum parts we find  $M^2=\xi R+\lambda\psi_c^2$. Suppose for simplicity that the background metric is
flat. Then to leading order in $\lambda$ the entanglement entropy (omitting the UV finite terms) is
\be
S_{div}={A\o 48\pi\epsilon^2}+{\lambda\o 24\pi}\int_\Sigma\psi_c^2\, \ln\epsilon\, .
\lb{entropy-lambda}
\ee
This entropy should be considered as a quantum correction to the tree-level entropy
\be
S_{tree}={A\o 4G_N}-2\pi\xi\int_\Sigma\psi^2_c\, ,
\lb{entropy-xi}
\ee
which follows from the action $W_{gr}+W[\psi]$.
We see that the logarithmic divergences in (\ref{entropy-lambda}) and (\ref{xi-renormalization})  agree if $\xi=0$. On the other hand, the renormalization of $\xi$ (\ref{xi-renormalization}) does not make the total entropy
 $S_{tree}+S_{div}$ completely UV finite. This is yet another manifestation of the puzzling behavior of the non-minimal coupling.

\section{ Other related methods}
 \subsection{ Euclidean path integral and thermodynamic entropy }
 \label{section: Euclidean approach}
In 1977 Gibbons and Hawking \cite{Gibbons:1976ue} developed a method based on the Euclidean path integral for studying the thermodynamics of black holes. In this method one obtains what may be called a thermodynamical entropy. One deals with metrics which satisfy the gravitational field equations 
and thus avoids the appearance of metrics with conical singularities. The entanglement entropy, on the other hand,  has a well-defined statistical meaning.
In ordinary systems the thermodynamical entropy and the statistical (microscopical) entropies coincide. For black holes the exact relation between the two entropies can be  seen from the following reasoning\epubtkFootnote{As I have learned recently (private communication  of R. Myers), Jacobson and Myers (unpublished) had similar ideas back in the 90-s.}   \cite{Solodukhin:1995ak}.

 Consider a gravitationally coupled  system (gravity plus quantum matter fields)
at some arbitrary temperature $T=(\beta)^{-1}$. A standard way to describe a thermal state of a field system is to use an Euclidean
path integral over  all fields in question
defined on manifold with  periodicity $2\pi\beta$ along the time-like Killing vector.
Suppose that it is a priori known that the system includes a black hole. Thus there exists a 
 surface $\Sigma$
(horizon) which is a fixed point of the isometry generated by the killing vector.  This  imposes an extra condition on the possible class of metrics in the
path integral. The other condition to be imposed on  metrics in the path integral 
is the asymptotic behavior  at infinity: provided   the mass $M$ and the electric charge $Q$ of the gravitational configuration are fixed, one has to specify the fall-off of the metrics for large values of $r$. Thus, the Euclidean path integral is
\be
Z(\beta, M,Q)=\int{\cal D}g_{\mu\nu}\int {\cal D\psi} e^{-W_{gr}[g]+W_{mat}[\psi, g]}\, ,
\lb{functional integral}
\ee
where the integral is taken over $\beta$-periodic fields $\psi(\tau,x^i)=\psi(\tau+\beta,x)$ and over  metrics which satisfy the  following conditions:

\medskip

\noindent i) $g_{\mu\nu}$ possesses an Abelian isometry  with respect to the Killing vector $\partial_\tau$;

\medskip

\noindent  ii) there exists a surface $\Sigma$ (horizon) where the Killing vector $\partial_\tau$ becomes null; 

\medskip

\noindent iii) asymptotic fall-off of metric $g_{\mu\nu}$ at large values of radial coordinate $r$ is fixed by the mass $M$ and electric charge $Q$ of the configuration. 

\medskip

Since the inverse temperature $\beta$ and mass $M$ in the path integral are two independent 
parameters, the path integral (\ref{functional integral}) is mostly over metrics which have a conical singularity at the surface $\Sigma$.
The integration in (\ref{functional integral}) can be done in two steps. First, one computes the integral over matter fields $\psi$ on the background of a metric which satisfies conditions i), ii) and iii).
The result of this integration is  the quantity (\ref{W3}) used in the computation of the entanglement entropy,
\be
\int {\cal D\psi} e^{-W_{mat}[\psi, g]}=e^{-W[\beta,g]}\, .
\lb{psi-integration}
\ee
Semiclassically,  the functional integration over metrics in (\ref{functional integral}) can be  performed in a saddle-point approximation,
\be
Z(\beta,Q,M)=e^{-W_{tot}[\beta,g(\beta)]}\,  ,
\lb{Z-saddle}
\ee
where metric $g_{\mu\nu}(\beta)$ is a solution to the saddle-point equation
\be
{\delta W_{tot}[\beta,g]\o \delta g}=0\, ,\, W_{tot}=W_{gr}[\beta,g]+W[\beta,g]\, ,
\lb{equilibrium}
\ee
whith the inverse temperature $\beta$ kept fixed. The solution of this equation is a regular (without conical singularities) metric $g_{\mu\nu}(\beta)$. This is an on-shell metric which incorporates the quantum corrections due to the vacuum  polarization by the matter fields.
It can be also called an equilibrium configuration which corresponds to the fixed temperature $\beta^{-1}$.
In the saddle point approximation there is a constraint relating the charges at infinity $M$ and $Q$ and the inverse temperature $\beta$: $\beta=\beta(M,Q)$.  

The thermodynamic entropy is defined  by the total response of the free energy $F=-\beta^{-1}\ln Z(\beta)$ to a small change of the temperature,
\be
S_{TD}=\beta^2d_\beta F=(\beta d_\beta
-1)W_{tot}(\beta,g_{\beta}) \lb{thermodynamical entropy}
\ee
and involves, in particular, the derivative of the equilibrium configuration $g_{\mu\nu}(\beta)$ with respect to $\beta$
\be
d_\beta W_{tot}=\partial_\beta W_{tot}[\beta, g]+{\delta W_{tot}[\beta, g]\o \delta g_{\mu\nu}}{\delta g_{\mu\nu}\o \delta\beta}\, .
\lb{variation}
\ee
For an equilibrium configuration, satisfying (\ref{equilibrium}), the second term in (\ref{variation}) vanishes and thus the total derivative with respect to $\beta$ coincides with a partial derivative.

Thus, in order to compute the thermodynamical entropy one may proceed in two steps. First, for a generic metric which satisfies the conditions i), ii) and iii) compute the off-shell entropy using the replica method, i.e. by introducing a small conical singularity at horizon. This computation is done by taking a partial derivative with respect to $\beta$. Second, consider this off-shell entropy for an equilibrium configuration which solves equation (\ref{equilibrium}).
Since for the classical gravitational action (\ref{bare action}) one finds $(\beta\partial_\beta-1)W_{gr}[\beta, g]=S(G_B, c_{iB})$  (\ref{tree level entropy}) and for the quantum effective action one obtains the entanglement entropy  $(\beta\partial_\beta-1)W[\beta, g]=S_{ent}$ the relation between the entanglement entropy and thermodynamical entropy is given by  
\be
S_{TD}=S(G_B, c_{iB})+S_{ent}\, .
\lb{TD versus entanglement}
\ee
Therefore, the entanglement entropy constitutes only a (quantum) part of the thermodynamical entropy of the black hole. 
The thermodynamical entropy is defined for equilibrium configurations satisfying the quantum corrected Einstein equations (\ref{equilibrium}).
Thus these configurations are not classical solutions to the Einstein equations but incorporate the quantum (one-loop) corrections.
These configurations are regular metrics without conical singularities.  The UV divergences in the free energy for these configurations  are renormalized in a standard way and thus for the thermodynamical entropy the renormalization statement discussed above holds automatically\epubtkFootnote{This is true for minimally coupled matter fields. In the presence of non-minimal couplings there appear extra terms in the thermodynamical entropy which are absent in the entanglement entropy, as we discussed earlier.}.

In flat spacetime the quantum (one-loop) thermodynamical and statistical entropies coincide as was shown by Allen \cite{Allen:1986qi} due to the fact that the corresponding partition functions  differ by  terms proportional to $\beta$. 
In the presence of black holes the exact relation between the two entropies
has been a subject of some debate (see for example \cite{Frolov:1994zi}, \cite{Solodukhin:1996vx}). The analysis made  in  \cite{Fursaev:1997th} however  shows 
that in the presence of black hole the Euclidean and statistical free energies coincide provided an appropriate method of regularization is used to regularize  both quantities.

\subsection{ 't Hooft's brick wall model}
                             
In 1985 t' Hooft \cite{'tHooft:1984re} proposed a model which was one of the first successful demonstrations that an entropy that scales as an area
can be associated, in a rather natural way, with a black hole horizon.     The idea of t' Hooft's calculation was to consider a thermal gas of Hawking particles propagating just outside the black hole horizon.   The entropy in the canonical description   of the system is calculated by means of the WKB approximation.
 Provided the temperature of the gas is equal to the Hawking one the result of this calculation  is unambiguous.  There is however an important subtlety:
the density of states of a Hawking particle becomes infinite as one gets closer to  horizon. The reason for this is simple. 
Close to the horizon all particles effectively propagate in the so-called optical metric. The later is conformally related to the black hole metric
\be
ds^2_{BH}=-g(r)dt^2+g^{-1}(r)dr^2+r^2d\omega_{d-2}
\lb{BH metric}
\ee
as follows
\be
ds^2_{opt}=g^{-1}(r)ds^2_{BH}=-dt^2+g^{-2}(r)dr^2+r^2g^{-1}(r)d\omega^2_{d-2}\, ,
\lb{optical}
\ee
where $d\omega^2_{d-2}$ is the metric of the $(d-2)$-unit sphere.
 In the optical  metric the near-horizon region, where the metric function in (\ref{BH metric}) can be approximated as
 $$
 g(r)={4\pi\o \beta_H}(r-r_+)+O(r-r_+)^2\, ,
 $$ 
 occupies an infinite volume. 
Clearly, the infinite volume contains  an infinite number of states.
In order to regularize this infinity t' Hooft   introduced a {\it brick wall}, an imaginary boundary  at some small  distance  $\epsilon$ 
from the actual horizon.  The regularized optical volume then is divergent when $\epsilon$ is taken to zero
\be
V_{opt}=\Omega_{d-2}\int_{r_\epsilon} dr r^{d-2}g^{-d/2}\sim A(\Sigma){\beta_H^{d-1}\o \epsilon^{d-2}}\, ,
\lb{regularized volume}
\ee
where $A(\Sigma)=r^{d-2}_+\Omega_{d-2}$ is the area of the  horizon and $\epsilon\sim\sqrt{\beta_H(r_\epsilon-r_+)}$ is the invariant distance between the brick wall ($r=r_\epsilon$) and the actual horizon ($r=r_+$).
The entropy of a gas of massless particles at temperature $T=\beta^{-1}_H$ confined in the volume $V_{opt}$ in $d$ spacetime dimensions  
\be
S_{BW}\sim V_{opt}\beta_H^{1-d}\sim {A(\Sigma)\o \epsilon^{d-2}}\, 
\lb{BW1 entropy}
\ee
in the optical metric, is  proportional to the horizon area. We should note that the universal behavior of the regularized optical volume (\ref{regularized volume}) in the limit of small $\epsilon$
and its proportionality to the horizon area in this limit was important in establishing the result (\ref{BW1 entropy}).

\subsubsection{ WKB approximation, Pauli-Villars fields}
In the original  calculation by 't Hooft  one considers a minimally coupled scalar field which satisfies the Klein-Gordon equation
\be
(\nabla^2-m^2)\varphi=0\, 
\lb{Klein-Gordon}
\ee
on the background of a black hole metric (\ref{BH metric}) and imposes a brick wall  boundary condition 
\be
\varphi(x)=0\ {\rm at} \ r=r_\epsilon\, .
\lb{BW condition}
\ee
Consider for simplicity the four dimensional case. 
Expanding the scalar field in spherical coordinates $\varphi=e^{-i\omega t}Y_{l,m}(\theta,\phi) f(r)$ one finds that equation (\ref{Klein-Gordon}) becomes
\be
\omega^2 g^{-1}(r)f(r)+r^{-2}\partial_r(r^2g(r)\partial_rf(r))-({l(l+1)\o r^2}+m^2)f(r)=0\, .
\lb{field equation}
\ee
One uses the WKB approximation in order to find a solution of this equation. In this approximation one represents $f(r)=\rho(r)e^{iS(r)}$, where
$\rho(r)$ is a slowly varying function of $r$ while $S(r)$ is a rapidly varying phase. One neglects derivatives of $\rho(r)$ and the second derivative
of $S(r)$ on obtains the radial function in the form
\be
f(r)=\rho(r)e^{\pm i\int^r{dr\o g(r)}k(r,l,E)}\, , \ \ k(r,l,E)=\sqrt{E^2-(m^2+{l(l+1)\o r^2})g(r)}\, ,
\lb{WKB}\ee
valid in the region where $k^2(r)\geq  0$. The latter condition defines a maximal radius $r_{\omega,l}$ which is a solution to the equation $k^2(r_{\omega,l})=0$. For a fixed value of the energy  $E$, 
by increasing the mass $m$ of the particle or the angular momentum $l$, the radius $r_{\omega,l}$ approaches $r_+$ so that the characteristic region where the solution (\ref{WKB}) is valid 
is in fact the near horizon region. One imposes an extra Dirichlet condition $\varphi=0$ at $r=r_{\omega,l}$ so that the one-particle spectrum becomes discrete
\be
2\int_{r_\epsilon}^{r_{\omega,l}}{dr\o g(r)}k(r,\omega,l)=2\pi n\, ,
\lb{quantization condition}
\ee
where $n$ is an integer. This relation is used to count the number of one-particle states that correspond to fixed values of energy $\omega$ and angular momentum $l$,
\be
n(\omega,l)={1\o \pi}\int_{r_\epsilon}^{r_{\omega,l}}{dr\o g(r)}k(r,\omega,l)\, .
\lb{number of states}
\ee
Calculating the total number of  states which have the same energy $E$, one has to sum over $l$. This sum can be approximated by an integral
\be
n(\omega)=\int dl(2l+1)n(\omega,l)={2\o 3}\int_{r_\epsilon}^{r_\omega} {r^2\o g^2(r)}k^3(r,\omega)\, , \ \ k(r,\omega)=\sqrt{\omega^2-m^2g(r)}\, ,
\lb{modes}
\ee
where $r_\omega$ is determined by condition that $k(r,\omega)=0$.

In the near horizon region one approximates the metric function $g(r)$ in (\ref{optical}) by the first two terms in the expansion in powers of $(r-r_+)$,
\be
g(r)={4\pi\o \beta_H}(r-r_+)+C(r-r_+)^2\, ,
\lb{metric function-1}
\ee
where $\beta_H$ is the inverse Hawking temperature and $r_+$ is the horizon radius. Constant $C$ is related to the curvature of spacetime near the horizon. The radial position of the brick wall is $r_\epsilon=r_++{\pi\epsilon^2\o \beta_H}$,
where $\epsilon$ is the geodesic distance between the brick wall and the horizon.
Focusing only on the brick wall divergent terms,  one obtains for the number of states (\ref{modes})
\be
n(\omega)={r_+^2\beta_H^3 \omega^3\o 24\pi^4\epsilon^2}+\left({r_+^2\beta_H^2 \omega^3\o 24\pi^4}(\beta_H C-{4\pi \o r_+})+{r_+^2\beta_H m^2 \omega\o 2\pi^2}\right)\ln\epsilon\, .
\lb{BW divergences}
\ee

In a thermal ensemble of scalar particles at fixed temperature $T=\beta^{-1}$, each state in the one-particle spectrum can be occupied by any integer number of quanta.  
One gets for the free energy
\be
\beta F=\int_{0}^\infty d\omega{dn(\omega)\o d\omega} \ln (1-e^{-\beta \omega})~\, \lb{free energy}
\ee
or, integrating by parts,
\be
\beta F=-\beta \int_{0}^\infty {n(\omega)\o e^{\beta \omega}-1} d\omega\, .
\lb{free energy-2}
\ee
Substituting here (\ref{BW divergences}) and using the integrals
\be
\int_0^\infty {d\omega\omega\o e^{\beta \omega}-1}={\pi^2\o 6\beta^2}\, , \ \ \int_0^\infty {d\omega\omega^3\o e^{\beta \omega}-1}={\pi^4\o 15\beta^4}\, .
\lb{integrals}
\ee
one calculates the divergent terms in the free energy
\be
F=-{r_+^2\o 360 \epsilon^2}{\beta^3_H\o \beta^4}-\left({r_+^2\o 360}(\beta_H C-{4\pi\o r_+}){\beta_H^2\o \beta^4}+{r_+^2 m^2\o 12}{\beta_H\o \beta^2}\right)\ln\epsilon
\lb{BW free energy}
\ee
and,  using equation $S=\beta^2\partial_\beta F|_{\beta=\beta_H}$, the entropy
\be
S={r_+^2\o 90\epsilon^2}+\left( {r_+^2\o 90\beta_H}(\beta_H C-{4\pi \o r_+})+{1\o 6}r_+^2 m^2\right) \ln\epsilon\, .
\lb{BW entropy}
\ee
Due to the relations 
\be
Cr^2_+-{4\pi r_+\o \beta_H}={1\o 8\pi}\int_\Sigma (R_{ii}-2R_{ijij})\ \ {\rm and} \ \ 4\pi r_+^2=\int_\Sigma 1
\lb{horizon relations}
\ee
this expression for the entropy can be rewritten in a completely geometric form
\be
S={A(\Sigma)\o 360\pi \epsilon^2}+{1\o 720\pi}\int_\Sigma (R_{ii}-2R_{ijij}+30 m^2)\ln\epsilon\, .
\lb{BW entropy 2}
\ee
The leading term proportional to the area was first calculated in the seminal paper of  't Hooft \cite{'tHooft:1984re}. The area law in the brick wall model was also  studied in
\cite{Mann:1990fk},  \cite{Susskind:1994sm},  \cite{Barbon:1994ej}, \cite{Barbon:1994wa}, \cite{Mukohyama:1998rf}.

It is an important observation made by Demers, Lafrance and Myers in    \cite{Demers:1995dq}  that the brick wall divergences are in fact the UV divergences. This can be seen in the Pauli-Villars regularization
as was first done in \cite{Demers:1995dq}.  
Applying the Pauli-Villars regularization scheme for 
the   four-dimensional scalar field theory studied here, one introduces
five regulator fields $\{\varphi_i,~i=1,...,5\}$ of different statistics
and masses $\{m_i,~i=1,...,5\}$ dependent on the UV cut-off
$\mu$ \cite{Demers:1995dq}. Together with
the original scalar $\varphi_0=\varphi$ ($m_0=m$) 
these fields satisfy two constraints
\be\sum_{i=0}^5 \Delta_i=0\ m\ {\rm and} \ \ \sum_{i=0}^5 \Delta_im_i^2=0\, ,
\lb{PV constraints}
\ee
where $\Delta_i=+1$ for the commuting fields, and $\Delta_i=-1$
for the anticomuting fields. 
Not deriving the exact expressions for $m_i$, we just quote here
the following  asymptotic behavior
\begin{eqnarray}
&&\sum_{i=0}^5 \Delta_im^2_i\ln m^2_i =\mu^2 b_1+m^2\ln{m^2\over \mu^2}
+m^2 b_2~~, \nonumber \\
&&\sum_{i=0}^5 \Delta_i\ln m^2_i =\ln{m^2\over \mu^2}~~,
\label{PV asymptotes}
\end{eqnarray}
(where $b_1$ and $b_2$ are some constants), valid in the 
limit $\mu \rightarrow \infty$.
The total  free energy is the sum of all contributions, from the original scalar field and the regulators
\begin{equation}
\beta{F}_{}=\beta\sum_{i=0}^5 \Delta_i  F^i\, .
\label{total free energy}
\end{equation}
It is clear that due to the constraints (\ref{PV constraints}) all brick wall divergences (with respect to the parameter $\epsilon$) in the free energy (\ref{total free energy}) and in the entropy  cancel. On the other hand, both the free energy and the entropy become divergent if the Pauli-Villars regulator $\mu$ is taken to infinity, thus confirming their identification as UV divergences. For the free energy one finds
\begin{equation}
F= -{1\over 24} {\beta_H\over \beta^2}r^2_+ 
\sum_{i=0}^5 \Delta_iM^2_i\ln M^2_i 
-{1\over 1440}{\beta_H^3\over \beta^4} r^2_+ C 
\sum_{i=0}^5 \Delta_i\ln M^2_i~~.
\label{free energy PV}
\end{equation}
and for the entropy one has
\begin{eqnarray}
S={1\over 48\pi} A_{\Sigma}
\sum_{i=0}^5 \Delta_im^2_i\ln m^2_i+
{1\over 1440\pi}\int_\Sigma (R_{ii}-
2R_{ijij} )
\sum_{i=0}^5 \Delta_i\ln m^2_i ~~.
\label{entropy PV}
\end{eqnarray}

Several remarks are in order.

1) Comparing the entropy calculated in the brick wall model, (\ref{BW entropy 2}) or (\ref{entropy PV}), with the entanglement entropy (\ref{UV divergence}) we see that the structure of the UV divergent terms in two entropies is similar. The logarithmic terms in (\ref{UV divergence}) and (\ref{entropy PV}) (or (\ref{BW entropy 2})) are identical if the black hole metric 
has vanishing Ricci scalar, $R=0$. This is the case, for example, for the Reissner-Nordstr\"{o}m black hole considered in \cite{Demers:1995dq}. The logarithmic terms in the two calculations  are however  different  if the Ricci scalar is non-zero.  This discrepancy appears to arise due to certain limitations of the WKB approximation. In the exact solutions, known explicitly, for  example, for a scalar field in a constant curvature spacetime, the mass $m$  always appears  in  combination $m^2-{1\o 6}R$. This, however, is not seen in the WKB approximation (\ref{WKB}).
In fact, if one makes this substitution everywhere in the above brick wall calculation, the Ricci scalar would appear in the brick wall entropy in a manner which agrees with the 
entanglement calculation  (\ref{UV divergence}). Moreover, an alternative calculation \cite{Fursaev:1997th} of the density of states which does not make use of the WKB approximation results in an expression for the entropy which  agrees with (\ref{UV divergence}).

2) The similarity between the two entropies suggests that the UV divergences in the brick wall entropy (\ref{entropy PV}) can be renormalized by the renormalization of the couplings in the gravitational action in the same way as for the entanglement entropy. That this indeed works was  demonstrated in \cite{Demers:1995dq}.

3) For a non-minimally coupled field we have the same problem as in the case of  the entanglement entropy. For metrics with $R=0$ the non-minimal coupling does not show up in
the scalar field equation and does not change the density of states, the free energy and entropy. On the other hand, the non-minimal coupling affects the renormalization of Newton's constant, even if the background metric is Ricci flat.  An attempt was made in \cite{Solodukhin:1996jt} to modify the Dirichlet boundary condition at the brick wall and replace it by a more sophisticated condition, which would depend on the value  of the non-minimal coupling $\xi$, so that the resulting entropy would have the UV divergences consistent with the renormalization of  Newton's constant. This attempt however can not be considered as   successful since it does not reproduce the expected behavior of the entropy for large positive values of $\xi$. 

The calculation of the brick wall entropy for a rotating black hole is more complicated due to the presence of the superradiance modes in the spectrum.
This issue is considered in papers \cite{Cognola:1997dv}, \cite{Frolov:1999gy}, \cite{Kim:2005pb}, \cite{ChangYoung:2008pw}, \cite{Jing:2001qh}, \cite{Wu:2003qc}, \cite{Ho:1998du}, \cite{Jing:1999vy}, \cite{Kenmoku:2005zh}.

\subsubsection{ Euclidean path integral approach in terms of optical metric}
 
 \paragraph*{Field equation in optical metric.}
 Consider a slightly more general equation than (\ref{Klein-Gordon}), by including a non-minimal coupling,
 \be
 (-\nabla^2+\xi R +m^2)\phi=0\, 
 \lb{KG nonminimal}
 \ee
 in the background of the black hole metric $g_{\mu\nu}$, which takes the form (\ref{BH metric}). The optical metric is conformally related to the black hole metric,
 $\bar{g}_{\mu\nu}=e^{2\sigma}g_{\mu\nu}$, where $e^{2\sigma}=1/|g_{tt}|$ (in the metric (\ref{BH metric}) we have that $g_{tt}=g(r)$).
 The equation (\ref{KG nonminimal}) can be rewritten entirely in terms of   the optical metric  $\bar{g}_{\mu\nu}$ (\ref{optical})       as follows
 \be
 &&(\partial_t^2+\hat{H}^2)\varphi_{opt}=0 \, \, , 
 \hat{H}^2=-\Delta_{opt}+{\cal  V}\, , \nonumber \\
 && {\cal V}=  e^{-2\sigma}\left((\xi R+m^2)-{(d-2)\o 2}\nabla^2\sigma-{(d-2)^2\o 4}(\nabla\sigma)^2\right)\, ,
 \lb{optical equation}
 \ee  
 where $\varphi_{opt}=e^{{(d-2)\o 4}\sigma} \varphi$, $\Delta_{opt}$ is the Laplace operator for spatial part $\gamma^{opt}_{ij}=\bar{g}_{ij}$ of the optical metric while the scalar curvature $R$ and the covariant derivative $\nabla$ are defined with respect to original metric $g_{\mu\nu}=e^{-2\sigma}\bar{g}_{\mu\nu}$. We notice that, since $e^{-2\sigma}=g(r)$, the effective potential $\cal V$ in (\ref{optical equation}) vanishes at the horizon. This is a general feature of wave equations in the black hole background: the fields become, effectively, massless in the near horizon region.
  The frequency $\omega$ which appears in equation  (\ref{field equation}) in the brick wall calculation  is thus 
 an eigenvalue of the operator $\hat{H}^2$,
 \be
 \hat{H}^2\varphi_\omega=\omega^2\varphi_\omega\, .
 \lb{eigen function}
 \ee             

\paragraph*{The canonical free energy and Euclidean path integral.}      The canonical free energy (\ref{free energy})
\be
F=\beta^{-1}\sum_{\omega} n(\omega)\ln(1-e^{-\beta \omega})\, ,
\lb{free energy 3}
\ee
where $\omega$ are eigenvalues of the spatial operator $\hat{H}^2$, $n(\omega)$ is the degeneracy of the energy level $\omega$,  can be represented 
in terms of the Euclidean path integral for a field theory with wave operator $(\partial_\tau^2+\hat{H}^2)$, provided that the Euclidean time $\tau$ is a circle with period $\beta$.
(This property was first clearly formulated by Allen  \cite{Allen:1986qi}.) It order to see  this   in a rather elementary way, we first notice that
\be
\ln(1-e^{-\beta \omega})=-{\beta\omega\o 2}+\ln\beta\omega+\sum_{k=1}^\infty \ln(1+{\omega^2\beta^2\o 4\pi^2 k^2})\, .
\lb{sum representation}
\ee      
The sum in this expression can be rewritten as a difference of two sums
\be
\sum_{k=1}^\infty \ln(1+{\omega^2\beta^2\o 4\pi^2 k^2})=\sum_{k=-\infty}^\infty {1\o 2}\ln (\omega^2+{4\pi^2 k^2\o \beta^2})-\sum_{k=1}^\infty \ln ({4\pi^2 k^2\o \beta^2})-\ln\omega 
\lb{sum 2}
\ee
Each of these sums should be understood in terms of the  zeta-function regularization. In particular, using the properties of the Riemann $\zeta$-function, we find
\be
\sum_{k=1}^\infty \ln ({4\pi^2 k^2\o \beta^2})=\lim_{z\rightarrow 0}{d\o dz}\sum_{k=1}^\infty \ln ({4\pi^2 k^2\o \beta^2})^{-z}=-\ln\beta
\lb{sum1}
\ee
Collecting together (\ref{sum representation}), (\ref{sum 2}) and (\ref{sum1}) one obtains that
\be
F=-{1\o 2}\sum_\omega n(\omega)\omega +{1\o 2\beta} \sum_{\omega} n(\omega) \sum_{k=-\infty}^\infty \ln (\omega^2+{4\pi^2 k^2\o \beta^2})
\lb{free energy 2}
\ee
The second term in (\ref{free energy 2}) can be expressed in terms of the Euclidean path integral. This can be seen as follows.  In the  Euclidean formulation one first makes a Wick rotation of time $t\rightarrow -i\tau$. The  effective action $W_{opt}$ then is defined by means of the Euclidean path integral
\be
e^{-W_{opt}}=\int {\cal D}\varphi_{opt}\, e^{-\int \varphi_{opt}(-\partial_\tau^2+\hat{H}^2)\varphi_{opt}}\, .
\lb{effective action optical}
\ee      
At finite temperature one closes the Euclidean time by identifying $\tau$ and $\tau+\beta$. The eigen values of operator $\partial_\tau$ then are $i{2\pi\o \beta}k$, where $k=0,\pm 1,\pm 2, ..$. The effective action can be expressed in terms of the logarithm of the determinant
\be
W_{opt}=-{1\o 2}\ln\det (-\partial_\tau^2+\hat{H}^2)=-{1\o 2}\sum_\omega n(\omega)\sum_{k=-\infty}^{+\infty}\ln(\omega^2+{4\pi^2 k^2\o \beta^2})\, .
\lb{effective action-opt}
\ee
Comparing this with (\ref{free energy 2}) and  defining the vacuum energy as
\be
E_0={1\o 2}\sum_\omega n(\omega)\omega     
\lb{vacuum energy}
\ee
one arrives at an  expression for the free energy (\ref{free energy 2}) 
\be
F=\beta^{-1}W_{opt}-E_0\, .
\lb{Free energy-effective action}
\ee                 
  
\paragraph*{Evaluation of the effective action in the optical metric.}

The effective action $W_{opt}$ (\ref{effective action optical}) can be calculated using the heat kernel method. One notes that the heat kernel of the operator
$-\partial_\tau^2+\hat{H}^2$ takes the form of a product of two heat kernels, for commuting operators $\partial_\tau^2$ and $\hat{H}^2$. The heat kernel for operator
$\partial_\tau^2$ is computed explicitly, provided the periodicity condition, $\tau\rightarrow \tau+\beta$, is imposed. One finds for the trace
\be
\tr e^{s\partial_\tau^2}={\beta\o \sqrt{4\pi s}}\sum_{k=-\infty}^\infty e^{-{k^2\beta^2\o 4s}}\, .
\lb{heat kernel circle}
\ee
So that the effective action takes the form
\be
W_{opt}=-{1\o 2}\int_{\epsilon^2}^\infty {ds\o s}{\beta\o \sqrt{4\pi s}}\sum_{k=-\infty}^\infty e^{-{k^2\beta^2\o 4s}}\, \Tr K_{\hat{H}^2}\, .
\lb{action heat kernel}
\ee
The operator $\hat{H}^2=-\Delta_{opt}+{\cal V}$  (\ref{optical equation}) is defined  for the $(d-1)$-dimensional metric $\gamma_{ij}^{opt}$, the spatial part of the optical metric
(\ref{optical}). The trace of the heat kernel of operator $\hat{H}^2=-\Delta_{opt}+{\cal V}$  (\ref{optical equation}) can be represented as  a series  expansion in powers of $s$,
\be
\tr K_{\hat{H}^2}={1\o (4\pi s)^{d-1\o 2}}\left(\int \sqrt{\gamma_{opt}}+s\int \sqrt{\gamma_{opt}}({1\o 6}R_{opt}-{\cal V})+O(s^2)\right)\, ,
\lb{heat kernel of H}
\ee
where the integration is taken over the spatial part of the optical metric (\ref{optical}), $R_{opt}$ is the Ricci scalar of $(d-1)$-metric $\gamma_{ij}^{opt}$.                   
                       
For $n\neq 0$ the integration over the proper time $s$ in (\ref{action heat kernel}) is regularized for small $s$ due to the thermal exponential factor,    so that the UV regulator
$\epsilon$ can be removed. Interchanging the sum and the integral one obtains
\be
\sum_{n=1}^\infty \int_0^\infty {ds\o s}s^{m-{d\o 2}} e^{-{n^2\beta^2\o 4s}}=({\beta\o 2})^{2m-d}\zeta(d-2m)\Gamma({d\o 2}-m)\, , \ \ \  m=0,\, 1,\, 2,\, ..
\lb{integral over s}
\ee
Only the term with $n=0$ in   (\ref{action heat kernel}) contains the UV divergences. This term in the effective action is proportional to the inverse  temperature $\beta$ and thus
it does not make any contribution to the  entropy. On the other hand, the $n=0$ term gives the free energy at zero temperature. Thus, only the zero temperature contribution to the free energy is UV divergent as was shown by Dowker and Kennedy \cite{Dowker:1978md}. In fact the usual way    to renormalize the free energy is to subtract the $n=0$ term in (\ref{action heat kernel}) or, equivalently, to subtract the zero temperature free energy, $F^R=F(T)-F(T=0)$.      
  With this regularization and using (\ref{integral over s}) one obtains a sort of high temperature expansion of the effective action. In $d$ dimensions one finds for the regularized action \cite{Dowker:1988jw}, \cite{Dowker:1989gp}
  \be
  W^R_{opt}=-{\beta^{1-d}\o \pi^{d/2}}\zeta(d)\Gamma ({d\o 2})V_{opt}-{\beta^{3-d}\o 4\pi^{d/2}}\zeta (d-2)\Gamma({d\o 2}-1)\int\sqrt{\gamma_{opt}}({1\o 6}R_{opt}-{\cal V})+O(\beta^{4-d})\, ,
  \lb{higher temperature expansion}
  \ee                          
 where for $d=4$ the term $O(\beta^{4-d})$ contains also logarithmic term $\ln\beta$.  In four dimensions the first two terms in (\ref{higher temperature expansion}) are the only terms in the effective action which are divergent when the integration in the optical metric is taken up to the horizon. The first term in (\ref{higher temperature expansion}) and the respective term in the free energy and entropy is the contribution of a thermal gas in $(d-1)$ spatial volume $V_{opt}$ at temperature $T=\beta^{-1}$ in flat spacetime. The other terms in (\ref{higher temperature expansion}) are curvature corrections to the flat spacetime result as discussed by Dowker and Schofield \cite{Dowker:1988jw}, \cite{Dowker:1989gp}.
 
 Let us focus on the four-dimensional case. Defining the regularized free energy  $F^R=\beta^{-1}W^R_{opt}$ and entropy $S_{opt}=\beta^2\partial_\beta F^R$
 one finds (provided one  imposes the condition  $\beta=\beta_H$, after taking the derivative with respect to $\beta$)
 \be
 S^{opt}_{d=4}={2\pi^2\o 45}T^3_HV_{opt}+{1\o 12}T_H\int\sqrt{\gamma_{opt}}({1\o 6}R_{opt}-{\cal V})\, ,
 \lb{optical entropy}
 \ee
where we omit terms which are finite when the volume integration is extended to the horizon. The important observation now is that
\be
\int\sqrt{\gamma_{opt}}({1\o 6}R_{opt}-{\cal V})=\int\sqrt{\gamma_{opt}}e^{-2\sigma}(({1\o 6}-\xi)R-m^2)\, ,
\lb{observation}
\ee
where $e^{-2\sigma}=g(r)$ and $R$ is the scalar curvature of the original black hole metric. We recall that the latter is conformally related to the optical metric,
$g^{opt}_{\mu\nu}=e^{2\sigma}g_{\mu\nu}$,  so the relation between the scalar curvature in two spacetimes is
\be
R_{opt}=e^{-2\sigma}(R-6\nabla^2\sigma-6(\nabla\sigma)^2)\, .
\lb{conformal transfo curvature}
\ee
Using this relation and the form of the potential term $\cal V$ one arrives at (\ref{observation}).

Introducing the cut-off $\epsilon$ as before, $r_\epsilon=r_++{\pi\epsilon^2\o \beta_H}$,
one finds for the volume in the optical metric
\be
V_{opt}={\beta^3_H A(\Sigma)\o 16\pi^3\epsilon^2}+{\beta_H\o 32\pi^3}\int_\Sigma (R_{ii}-2R_{ijij})\ln\epsilon +O(\epsilon)
\lb{optical volume}
\ee
 and
 \be
  \int\sqrt{\gamma_{opt}}({1\o 6}R_{opt}-{\cal V}=(({1\o 6}-\xi)R(r_+)-m^2){\beta_H\o 2\pi}A_+\ln\epsilon^{-1}+O(\epsilon)\, ,
  \lb{a1 term}
  \ee
 where $A_+=4\pi r^2_+$ is the horizon area and $R(r_+)$ is the value of the scalar curvature at the horizon.
 
Putting everything together, one finds that the entropy in the optical metric 
 is
 \be
 S^{opt}_{d=4}={A_+\o  360\pi\epsilon^2}+\int_\Sigma\left({1\o 720\pi}(R_{ii}-2R_{ijij})-{1\o 24\pi}(({1\o 6}-\xi) R-m^2)\right)\ln\epsilon\, .
 \lb{optical entropy 2}
 \ee
 Comparing this result with the entanglement entropy (\ref{UV divergence}), computed earlier, we find  complete agreement.
Notice that $\epsilon$ here is  in fact the  IR regulator, the brick wall cut-off, which  regularizes the integration in the radial direction in the 
optical metric. As in t' Hooft's original calculation, this divergence can be transformed into a UV divergence by introducing the Pauli-Villars regulator fields of characteristic mass $\mu$.
The UV divergences of the entropy when $\mu$ is taken to infinity then are of the same type as in  (\ref{optical entropy 2}). This was analyzed in \cite{Fursaev:1997th}. 

The Euclidean path integral approach in the optical metric was considered in \cite{Barvinsky:1994jca}, \cite{deAlwis:1995}, \cite{deAlwis:1994ej}, \cite{Barbon:1994ej}, \cite{Barbon:1994wa}, \cite{Moretti:1996wd}, \cite{Iellici:1996gv}.
 
We remind the reader that the entanglement entropy (\ref{UV divergence}) is obtained by using the Euclidean path integral in the original black hole metric with a conical singularity
at the horizon (for $\beta\neq \beta_H$). The black hole metric and the optical metric are related by a conformal transformation.  This transformation is singular at the horizon and in fact produces a topology change: there appears a new boundary at $r=r_+$ in the optical metric which was a tip of the cone in the original black hole metric.  Because of this singular behavior of the conformal transformation, the exact relation between the two Euclidean path integrals is more subtle than for a regular conformal transformation. That the UV divergences in the entropy calculated  in these two approaches coincide suggests that the equivalence between the two approaches might  extend to the UV finite terms. Although in arbitrary dimension 
this equivalence may be difficult to prove, the analysis in two dimensions \cite{Solodukhin:1996vx} shows that the entanglement entropy and the brick wall entropy are indeed equivalent. 
 This is, of course, consistent with the formal proof outlined in Section \ref{section:formal proof}.

\section{Some particular cases}
In this section we shall consider some particular examples in which the entanglement entropy, including the UV finite terms, can be calculated explicitly.

\subsection{ Entropy of 2d black hole  }
In two dimensions the conformal symmetry plays a special role. This has many manifestations. In particular, the conformal symmetry can be used in order to
completely reproduce, for a conformal field theory (CFT), the UV finite part of the corresponding  gravitational effective action.  This is done by integration of the conformal anomaly. For regular two-dimensional spacetimes the  result is the well-known non-local Polyakov action. In the presence of a conical singularity the derivation is essentially the same 
although one has to take into account the contribution of the singularity. Consider a two-dimensional CFT characterized by a central charge $c$.  For a regular two-dimensional manifold the Polyakov action can be written in the form
\be
W_{PL}[M]={c\o 48\pi}\int_M({1\o 2}(\nabla\psi)^2+\psi R)\, ,
\lb{Polyakov action}
\ee
where the field equation for the field $\psi$ is $\nabla^2\psi=R$. On a manifold $M^\alpha$ with a conical singularity with angle deficit $\delta=2\pi(1-\alpha)$
the Polyakov action is modified by the contribution from the singularity at the horizon $\Sigma$  (which is just a point in two dimensions) so that \cite{Solodukhin:1994yz}, \cite{Fursaev:1994ea}
\be
W_{PL}[M^\alpha]=W_{PL}[M^\alpha / \Sigma]+{c\o 12}(1-\alpha)\psi_h +O(1-\alpha)^2\, ,
\lb{Polyakov action on conical space}
\ee
where $\psi_h$ is the value of the field $\psi$ on the horizon. Applying the replica method to the Polyakov action (\ref{Polyakov action on conical space}) one obtains that the corresponding contribution to the entanglement entropy from the UV finite term in the effective action is 
\be
S_{fin}={c\o 12}\psi_h\, .
\lb{finite entropy}
\ee
This result agrees with a derivation of Myers \cite{Myers:1994sg} who used the  Noether charge method of Wald \cite{Wald:1993nt} in order to calculate the entropy. The easiest way to compute the function $\psi$ is to use the conformal gauge 
$g_{\mu\nu}=e^{2\sigma}\delta_{\mu\nu}$ in which $\psi=2\sigma$. Together with the UV divergent part, the complete entanglement entropy in two dimensions is
\be
S={c\o 6}\sigma_h+{c\o 6}\ln{\Lambda\o \epsilon}\, ,
\lb{complete entropy}
\ee
where $\Lambda$ is an IR cut-off.

Let the black hole geometry be described by a 2d metric
\begin{equation}
ds^2_{bh}=f(x)d\tau^2+{1\over f(x)}dx^2, \label{B}
\end{equation}
where the metric function $f(x)$ has a simple zero at $x=x_+$. Assume that this black hole is placed inside a box of finite size $L$ so that
$x_+\leq x \leq L$. In order to get a regular space one closes the Euclidean time $\tau$ with period $\beta_H$, 
 $\beta_H={4\pi \over f'(x_+)}$. It is easy to see
that (\ref{B}) is conformal to the flat disk of radius $z_0$ ($\ln
z={{2\pi \over \beta_H}\int^x_L {dx \over f(x)}}$):
\begin{eqnarray}
&&ds^2_{bh}=e^{2\sigma }z_0^2(dz^2+z^2d\tilde{\tau}^2 )~~, \\
&&\sigma={1\over 2} \ln f(x)-{2\pi \over \beta_H}\int^x_L {dx\over
f(x)}
 +\ln {\beta_H  \over 2\pi z_0}, \nonumber
\label{4}
\end{eqnarray}
where
 $\tilde{\tau}={2\pi \tau \over \beta_H}$ ($0\leq \tilde{\tau} \leq2\pi$),
 $0 \leq z\leq 1$.
So that the entanglement entropy of the 2d black hole takes the form
 \cite{Solodukhin:1996vx}, \cite{Frolov:1996hd}
\begin{equation}
S={c\over 12} \int^L_{x_+}{dx \over f(x)}({4\pi \over
\beta_H}-f')+ {c\over 6}\ln ({\beta_H  f^{1/2}(L)\over \epsilon}),
\label{S}
\end{equation}
where we omit the irrelevant term that is a function of $(\Lambda
, z_0)$ but not of the parameters of the black hole and have
retained dependence on the UV regulator $\epsilon$. 

 As was shown in
\cite{Solodukhin:1996vx}, the entanglement entropy (\ref{S}) is
identical to the entropy of the thermal atmosphere of quantum
excitations outside  horizon in the "brick wall" approach of  't Hooft
\cite{'tHooft:1984re}.

The black hole resides inside a finite size box  and $L$ is the
coordinate of the boundary of the box. The coordinate invariant
size of the subsystem complimentary to the black hole is $L_{\tt
inv}=\int_{x_+}^Ldx/\sqrt{f(x)}$. Two limiting cases are of
interest.  In the first,  the size of the system $L_{\tt inv}$ is
taken to infinity. Then, assuming that the black hole space-time
is asymptotically flat, we obtain that the entanglement entropy
(\ref{S}) approaches the entropy of the thermal gas,  
\be
S={c\pi\over 3}L_{\tt inv}T_H\, .
\lb{thermal entropy-entanglement}
\ee
This calculation illustrates an important feature of the entanglement entropy of a black hole placed in a box of volume $V$. 
Namely,  the entanglement entropy contains a contribution of the thermal gas that,  in the limit of large volume in dimension $d$, 
takes the form (\ref{Sthermal}). This is consistent with the thermal nature of the reduced density matrix obtained from the Hartle-Hawking
state by tracing over modes inside the horizon.

The other interesting  case is
when  $L_{\tt inv}$ is small. In this case we find the universal
behavior \be S={c\over 6}\left(\ln{L_{\tt inv}\over
\epsilon}+{R(x_+)\over 24}L^2_{\tt
    inv}+O(L^3_{\tt inv})\right) .
\lb{Lsm} \ee The universality of this formula lies in the fact that
it does not depend on any characteristics of the  black hole (mass,
temperature) other than the value of the curvature $R(x_+)$ at the
horizon.

\bigskip

Consider two particular examples.

\medskip

\noindent {\bf 2d de Sitter spacetime} is characterized by the metric function $f(x)=1-{x^2\o l^2}$, the Hawking temperature $T_H=1/2\pi l$. In this spacetime the size of the box is bounded from above, $L_{inv}\leq \pi l$.  The corresponding entanglement entropy 
\be
S_{dS_2}={c\o 6}\ln \left({1\o \pi T_H\epsilon}\tan (T_H\pi L_{inv})\right)
\lb{de Sitter entropy}
\ee
is a periodic function of $L_{inv}$.

\medskip

\noindent {\bf The string inspired black hole} \cite{Witten:1991yr}, \cite{Mandal:1991tz} is described by the metric function $f(x)=1-e^{-\lambda x}$. It described an asymptotically flat spacetime. The Hawking temperature is 
$T_H={\lambda \o 4\pi}$. The entanglement entropy in this case is
\be
S_{str}={c\o 6}\ln \left({1\o 2\pi T_H\epsilon}\sinh (2\pi T_H L_{inv})\right)\, .
\lb{string inspired black hole}
\ee

The entropy in these two examples resembles the entanglement entropy in flat spacetime at zero temperature (\ref{entropy-T}) and at a finite temperature (\ref{entropy-Tl})  respectively.

\subsection{ Entropy of 3d BTZ black hole}

\subsubsection{BTZ black hole geometry}  
The black hole solution in three-dimensional gravity with negative cosmological constant was first obtained in \cite{Banados:1992wn} (see also  \cite{Banados:1992gq} for the global analysis of the solution).                           
We start with the black hole metric written in a form 
that makes it similar to the four-dimensional Kerr metric. 
Since we are interested in its thermodynamic behaviour,
we write the metric in the Euclidean form:
\begin{equation}
ds^2=f(r)d\tau^2+f^{-1}(r) dr^2+r^2(d\phi+N(r)d\tau)^2~~,
\label{BTZ metric}
\end{equation}
where the metric functions $f(r)$ and $N(r)$ read
\begin{equation}
f(r)={r^2\over l^2}-{j^2\over r^2}-m={(r^2-r_+^2)(r^2+|r_-|^2)\over l^2 r^2}\, \, , \ \ N(r)=-{j\over r^2}
\label{BTZ metric function}
\end{equation}
and we use the notation
\begin{equation}
r^2_+={ml^2\over 2}(1+\sqrt{1+({2j\over m l})^2})~,
~~|r_-|^2={ml^2\over 2}(\sqrt{1+({2j\over m l})^2}-1)~~
\label{horizons}
\end{equation}
Obviously one has that $r_+|r_-|=jl$. The coordinate $\phi$ in (\ref{BTZ metric}) is assumed to be periodic with period $2\pi$.

In order to transform the metric (\ref{BTZ metric}) to Lorentzian  singnature we
need to make the analytic transformation
$\tau\rightarrow i t$, $j\rightarrow -ij$ so that
\begin{eqnarray}
&&r_+\rightarrow r^L_+=\sqrt{ml^2\over 2}~\left(1+\sqrt{1-({2j\over m l})^2}~\right)^{1/2}~~, \nonumber \\
&&|r_-|\rightarrow \imath ~r_-^L=\sqrt{ml^2\over 2}~\left(1-\sqrt{1-({2j\over m l})^2}~\right)^{1/2}~~,
\label{analytic}
\end{eqnarray}
where $r^L_+$ and $r^L_-$ are the values in the Lorentzian space-time. These 
are the  respective radii of the outer and inner horizons  of the 
Lorentzian  black hole in $(2+1)$ dimensions. Therefore
we must always apply the transformation (\ref{analytic}) after carrying out
all calculations in the Euclidean geometry  in order to
obtain the result for the Lorentzian  black hole.
The Lorentzian version of the metric (\ref{BTZ metric}) describes a black hole with mass $m$ and angular 
momentum $J=2j$.
The outer horizon is located at $r=r_+$, the respective  inverse Hawking temperature is 
\begin{equation}
\beta_H={4\pi\over f'(r_+)}={2\pi r_+ l^2 \over r^2_++|r_-|^2}\, .
\label{temperature}
\end{equation}
In the $(\tau , r)$ sector of the metric (\ref{BTZ metric}) there is 
no conical singularity at the horizon if the Euclidean time $\tau$ 
is periodic with period $\beta_H$. 
The horizon $\Sigma$ is a one-dimensional space with metric
$
ds^2_{\Sigma}=l^2d\psi^2~~,
$
where $\psi={r_+\over l} \phi-{|r_-|\over l^2}\tau$ is a natural coordinate on the horizon.

The BTZ space is obtained from the three-dimensional maximally symmetric hyperbolic space $H_3$ (sometimes called the global Euclidean anti-de Sitter space) by making certain identifications.
In order to see this one may use the coordinate transformation
\begin{eqnarray}
&&\psi={r_+\over l} \phi-{|r_-|\over l^2}\tau~, ~~\theta={r_+\over l} \tau
+{|r_-|\over l^2}\phi \nonumber \\
&&\cosh^{-1}\rho  =({r^2_++|r_-|^2 \over r^2+|r_-|^2})^{1/2}\, .
\label{transformation to H3}
\end{eqnarray}                           
In new coordinates $(\rho, \theta,\psi)$ the BTZ metric  takes the form
\begin{equation}
ds^2=l^2 \left( d\rho^2+\cosh^2\rho d\psi^2+\sinh^2\rho d\theta^2 \right)\, ,
\label{global AdS}
\end{equation}                          
which is the metric on the hyperbolic space $H_3$. In this metric the BTZ geometry     is defined by identifications

$i).~~\theta\rightarrow \theta+2\pi 
$

$
ii). ~~\theta\rightarrow \theta+2\pi {|r_-|\over l}~,~~\psi\rightarrow\psi+
2\pi{r_+\over l}\, .
$

\noindent The outer horizon $r=r_+$ in the coordinate system $(\rho,\theta,\psi)$ is located at $\rho=0$ and $\psi$ is the angular coordinate on the horizon.
 Notice that the geodesic distance $\sigma$ between two points with coordinates $(\rho,\psi,\theta)$ and $(\rho,\psi',\theta')$ is
\begin{equation}
\sinh^2{\sigma \over 2l}=\cosh^2\rho\, \sinh^2{\psi-\psi'\over 2}+\sinh^2\rho \, \sin^2{\theta-\theta' \over 2}\, .
\label{geodesic distance}
\end{equation}
                        
\subsubsection{Heat kernel on regular BTZ geometry}
Consider a scalar field with the operator ${\cal D}=-(\nabla^2+\xi/l^2)$. The maximally symmetric constant curvature space is a nice example of a curved space in which  the heat equation 
$(\partial_s+{\cal D})K(x,x',s)=0$ has a simple, exact, solution. The heat kernel in this case is a function of the geodesic distance $\sigma$ between two points $x$ and $x'$.
On the global space $H_3$ one finds                       
\begin{equation}
K_{H_3}(\sigma , s)={1\over (4\pi s)^{3/2}}{\sigma /l \over \sinh (\sigma / l)}
e^{-{\sigma^2\over 4s}-\mu{s\over l^2}}\, ,
\label{heat kernel on H3}
\end{equation}
where $\mu=1-\xi$. The regular BTZ  geometry is defined by identifications $i)$ and $ii)$ defined above. As is seen from (\ref{geodesic distance}) the geodesic distance and  the heat kernel (\ref{heat kernel on H3}), expressed   in coordinates $(\rho,\psi,\theta)$, are automatically invariant under indentification $i)$.   It remains thus to maintain the identification $ii)$. This is done
by summing over images 
\begin{equation}
K_{BTZ}(x,x',s)=\sum_{n=-\infty}^{+\infty} K_{H_3}(\rho ,~ \rho ',~
\psi-\psi '+2\pi{r_+\over l}n,~\theta-\theta '+2\pi{|r_-|\over l}n)~~.
\label{heat kernel BTZ}
\end{equation}
 Using the path integral representation of the heat kernel we would say that
the $n=0$ term in (\ref{heat kernel BTZ}) is due to the direct way of connecting
points $x$ and $x'$ in the path integral. On the other hand, 
the $n\neq 0$ terms are due to uncontractible winding paths that 
go $n$ times around the circle.

\subsubsection{Heat kernel on conical BTZ geometry }

The conical BTZ geometry which is relevant to the entanglement entropy calculation   is obtained from global hyperbolic space $H_3$    by the replacing the identification
$i)$ as   follows
$$
i'). \ \theta\rightarrow \theta +2\pi\alpha
$$
and not changing the identification $ii).$  For $\alpha \neq 1$ this Euclidean space   has a conical singularity at the horizon ($\rho=0$).
The heat kernel on the conical BTZ geometry is constructed via the heat kernel  (\ref{heat kernel BTZ})
on the regular BTZ space by means of the Sommerfeld formula (\ref{Sommerfeld})
\begin{equation}
K_{BTZ_{\alpha}}(x,x',s)=K_{BTZ}(x,x',s)+{1\over 4\pi\alpha}
\int_{\Gamma}\cot {w\over 2\alpha}~~K_{BTZ}(\theta-\theta '+w,s)~~dw\, ,
\label{heat kernel conical BTZ}
\end{equation}
where $K_{BTZ}$ is the heat kernel (\ref{heat kernel BTZ}). 
The contour $\Gamma$ is defined in (\ref{Sommerfeld}).

For the trace of  the heat kernel (\ref{heat kernel conical BTZ}) one finds \cite{Mann:1996ze} after computing by residues the contour integral 
\begin{eqnarray}
&&Tr K_{BTZ_{\alpha}}=\left( {V_{BTZ_\alpha}\over l^3}+{A_+\over l}(2\pi\alpha )
c_2(\alpha )~\bar{s}~ \right){e^{-\mu\bar{s}}\over (4\pi \bar{s})^{3/2}}
\nonumber \\
&&+2\pi {e^{-\mu\bar{s}}\over (4\pi \bar{s})^{3/2}}{A_+\over l}~\bar{s}~
\sum_{n=1}^{\infty}{\sinh {\Delta\psi_n\over \alpha}\over \sinh {\Delta\psi_n}}
~~{e^{-{\Delta\psi^2_n\over 4\bar{s}}}\over (\sinh^2{\Delta\psi_n\over 2\alpha}+\sinh^2{\gamma_n\over 2\alpha})}~~,
\label{3.13}
\end{eqnarray}
where $\bar{s}=s/l^2$, $\gamma_n=A_-n/l$ and $\Delta \psi_n=A_+n/l$ ($A_+=2\pi r_+$ and $A_-=2\pi r_-$). Notice that we have already made the analytical continuation to the values of $r_+$ and $r_-$ in the Lorentzian geometry.

\subsubsection{The entropy} 
When the trace of the heat kernel    on the conical geometry is known one may compute the entanglement entropy by using the replica trick. The entropy then
is the sum of UV divergent and UV finite   parts \cite{Mann:1996ze}
\be
S_{ent}=S_{div}+S_{fin}\, ,
\lb{BTZ entropy}
\ee
where the UV divergent part is
\be
S_{div}={A_+\o 24\sqrt{\pi}}\int_{\epsilon^2}^{\infty}{ds\over s^{3/2}}e^{-\mu s/ l^2}={A_+\o 12\sqrt{\pi}}(\epsilon^{-1}-{\sqrt{\mu\pi }\o l})\, .
\lb{UV div entropy}
\ee
This divergence is renormalized by the standard renormalization of Newton's constant
\begin{equation}
{1\over 16\pi G_{ren}}={1\over 16\pi G_B}+{1\over 12}~
{1\over (4\pi)^{3/2}}~\int_{\epsilon^2}^{\infty}{ds\over s^{3/2}}e^{-\mu s/ l^2~}\, 
\label{renormalized GN}
\end{equation}
in the three dimensional gravitational action.

The  UV finite part in the entropy is                            
\begin{eqnarray}
&&S_{fin}=\sum_{n=1}^\infty
s_n \, , \nonumber \\
&&s_n={1\over 2n}{e^{-\sqrt{\mu}\bar{A}_+ n}\over (\cosh \bar{A}_+n-\cosh \bar{A}_-n)}
(1+\bar{A}_+ n \coth \bar{A}_+ n \nonumber \\
&&- {(\bar{A}_+n \sinh \bar{A}_+ n -
\bar{A}_-n \sinh \bar{A}_-n )\over (\cosh \bar{A}_+n-\cosh \bar{A}_-n)}
)\, ,
\label{UV finite entropy}
\end{eqnarray}                           
where $\bar{A}_\pm=2\pi r_\pm/l$.                             

After the renormalization of Newton's constant the complete entropy of the BTZ black hole, $S_{BTZ}=S_{BH}+S_{ent}$, is a rather complicated function of
the area of inner and outer horizons. Approximating in (\ref{UV finite entropy}) the infinite sum by an integral one finds \cite{Mann:1996ze}
\begin{eqnarray}
&&S={A_+\over 4G_{ren}}+\int_{\bar{A}_+}^\infty
~s(x)~dx \, ,\nonumber \\
&&s(x)={1\over 2x}{e^{-\sqrt{\mu}x}\over (\cosh x-\cosh kx)}
\left(1+x \coth x - {(x\sinh x -
kx \sinh kx )\over (\cosh x-\cosh kx)}
\right)\, ,
\label{complete entropy BTZ}
\end{eqnarray}
where  $k=A_-/A_+$.
The second term in the right hand side of (\ref{complete entropy BTZ}) can be 
considered  to be the one-loop
quantum (UV-finite) correction to the classical entropy of black hole.

For large enough $\bar{A}_+\equiv {A_+\over l}>>1$ the integral in (\ref{complete entropy BTZ}) goes to zero exponentially
 and we have the classical Bekenstein-Hawking formula for entropy.
On the other hand, for  small $\bar{A}_+$, the integral in (\ref{complete entropy BTZ})
 behaves logarithmically 
so that one has \cite{Mann:1996ze}
\begin{equation}
S_{BTZ}={A_+\over 4G}+{\sqrt{\mu}\over 6}{A_+\over l}-{1\over 6}\ln {A_+\over l}+
O(({A_+\over l})^2)~~.
\label{log term in BTZ}
\end{equation} 
 This 
logarithmic behavior for small values of $A_+$ (provided the ratio  $k=A_-/A_+$ is fixed) is universal and
independent of the constant $\xi$ 
(or $\mu$) in the field operator 
and the area of the inner horizon  ($A_-$) of the black hole. Hence
the rotation parameter $J$ enters (\ref{log term in BTZ}) only via the area $A_+$ 
of the outer horizon.

The other interesting feature of the entropy (\ref{complete entropy BTZ}) is that it always develops a minimum which is a solution to the equation
\be
{l\o 4 G_{ren}}=s({A_+\o l})\, .
\lb{minimal entropy}
\ee
This  black hole of minimal entropy may be interesting in the context of the final stage of the Hawking evaporation          
in three dimensions.   
As follows from the analysis of Mann and Solodukhin  \cite{Mann:1996ze}, the minimum of the entropy occurs for a hole whose horizon area  is of the Planck length, $A_+\sim l_{PL}$ (in threee dimensions
$l_{PL}\sim G_{ren}$).

\subsection{ Entropy of d-dimensional extreme black holes}
The extremal black holes  play a special role in gravitational theory.  These black holes are characterized by vanishing Hawking temperature $T_H$ which means that in the metric
(\ref{BH metric}) the near-horizon expansion in the metric function $g(r)$ starts with the quadratic term $(r-r_+)^2$. Topologically, the true extremal geometry is different from the non-extremal one.  Near the horizon the non-extremal static geometry looks like a product of a two-dimensional disk (in the plane $(r,\tau)$) and a $(d-2)$-dimensional sphere.
The horizon then is the center in the polar coordinate system on the disk.  Contrary to this, an extremal geometry in the near-horizon limit is a product of a two-dimensional cylinder
and a $(d-2)$-dimensional sphere. Thus, the horizon in the extremal case is just another boundary rather than a regular inner point as in the non-extremal geometry.
However, one may consider a certain limiting procedure in which one approaches the extremal case staying all the time in the class of non-extremal geometries. This limiting procedure is what we shall call the ``extremal limit''. A concrete procedure of this type was suggested by Zaslavsky  \cite{Zaslavsky:1997ha}.
One considers a sequence of non-extreme black holes in a cavity at $r=r_B$ and finds that there  exists a set of data $(r_+,r_B,r_-)$ such that the limit
${r_+\over r_-}\rightarrow 1,~{r_B\over r_+}\rightarrow 1$ is well-defined.
 Even if one may have started with a rather general non-extremal metric the limiting geometry is characterized by very few parameters. In this sense one may talk about ``universality'' of the extremal limit. In fact, in the most interesting (and tractable) case the limiting geometry is the product of two-dimensional hyperbolic space $H_2$ with the $(d-2)$-dimensional sphere.  Since the limiting geometry belongs to the non-extreme class its classical  entropy is proportional to the horizon area in accord with the Bekenstein-Hawking formula. The entanglement entropy of the limiting geometry then is a one-loop quantum correction to the classical result. The universality we have just mentioned suggests that this correction possesses a universal behavior in the extreme limit and, since the limiting geometry is rather simple, the limiting entropy can be found explicitly. The latter was indeed shown by Mann and Solodukhin in \cite{Mann:1997hm}.

\subsubsection{Universal extremal limit}

Consider a static spherically-symmetric  metric in the following form
\begin{eqnarray}
ds^2=g(r)d\tau^2+{1\over g(r)}dr^2+r^2d\omega^2_{d-2}\, ,
\label{non-extremal metric}
\end{eqnarray}
where $d\omega^2_{d-2}$ is the metric on the $(d-2)$-dimensional unit sphere,
describing a non-extreme hole with an outer horizon located at $r=r_+$. The analysis can be made for a more general metric, in which $g_{\tau\tau}\neq g^{-1}_{rr}$, the limiting geometry however is the simplest  in the case we consider in (\ref{non-extremal metric}). 
The function $g(r)$ in (\ref{non-extremal metric}) can be expanded as follows
\begin{eqnarray}
g(r)=a(r-r_+)+b(r-r_+)^2+O((r-r_+)^2)\, .
\label{metric function}
\end{eqnarray}
It is convenient to consider the geodesic distance $l=\int g^{-1/2}dr$ as a
radial coordinate. Retaining the first two terms in (\ref{metric function}), we find, for $r>r_+$, that
\be
&&(r-r_+)={a\over b} \sinh^2 ({lb^{1/2}\over 2})\, , \nonumber \\
&&g(lb^{1/2})=({a^2\over 4b})\sinh^2({lb^{1/2}})\, .
\label{transform}
\ee
In order to avoid the appearance of a conical singularity at $r=r_+$,
the Euclidean time $\tau$ in (\ref{non-extremal metric}) must be compactified with
period $4\pi/a$
which goes to infinity in the extreme limit
$a\rightarrow 0$. However, rescaling $\tau \rightarrow \phi=\tau{a/ 2}$
yields a new variable $\phi$ having period $2\pi$. Then, taking into
account (\ref{transform})  one finds for the metric (\ref{non-extremal metric})
\begin{eqnarray}
ds^2=\frac{1}{b}\left(\sin^2xd\phi^2+dx^2\right)
+(r_++{a\over b}\sinh^2{x\over 2})^2d\omega^2_{d-2}\, ,
\label{approximate metric}
\end{eqnarray}
where we have introduced the variable $x=lb^{1/2}$.
To obtain the extremal limit one just takes $a\rightarrow 0$. 
The limiting geometry 
\begin{equation}
ds^2=\frac{1}{b}\left(\sin^2 xd\phi^2+dx^2\right)+ r^2_+d\omega^2_{d-2}
\label{limiting metric}
\end{equation}
is that of the direct product of a 2-dimensional space and a $(d-2)$-sphere and is
characterized by a pair of dimensional parameters
$b^{-1/2}$ and $r_+$.
The parameter $r_+$ sets the radius of the $(d-2)$-dimensional sphere while
the parameter $b^{-1/2}$ is the curvature radius
 for the $(x,\phi)$ 2-space. Clearly, this two-dimensional space is the negative constant curvature space $H_2$.
This is the universality we mentioned above: although the non-extreme geometry
is in general described by an infinite number of parameters associated with the
determining function  $g(r)$ the geometry in the extreme limit 
depends only on two parameters ${b}$ and $r_+$. 
Note that the coordinate $r$ is inadequate for describing the
extremal limit (\ref{limiting metric}) since the coordinate transformation (\ref{transform})
is singular when $a\rightarrow 0$. The limiting metric (\ref{limiting metric}) is characterized by a finite temperature, determined by the $2\pi$ 
periodicity in angular coordinate $\phi$.

The limiting geometry (\ref{limiting metric})
is that of a direct product $H_2\times S_2$ of 2d hyperbolic space $H_2$ with
radius $l=b^{-1/2}$ and a 2d sphere $S_2$ with radius $l_1=r_+$.
It is worth noting that the limiting geometry
(\ref{limiting metric})  precisely merges near the horizon with the
geometry of the original metric (\ref{non-extremal metric}) in the sense that all the 
curvature tensors for both metrics coincide.
This is in contrast with, say, the situation in which the Rindler metric is considered
to approximate the geometry of a non-extreme black hole:
the curvatures of both spaces do not merge in general.

For a special type of extremal black hole, when  $l=l_1$,  the limiting geometry is characterized by just one dimensionful parameter. This is the case for   the Reissner-Nordstr\"{o}m black hole in four dimensions. The limiting 
extreme geometry in this case  is the well-known 
Bertotti-Robinson space characterized by just one parameter $r_+$.
This space has remarkable properties in the context of supergravity theory that are not the subject of
the present review.

\subsubsection{Entanglement entropy in the extremal limit}
\label{section: entropy in the extremal limit}

Consider now a scalar field propagating on the background of the limiting geometry (\ref{limiting metric}) and described by the operator
\be
{\cal D}=-(\nabla^2+X)\, \, , \ X=-\xi R_{(d)} \, ,
\lb{field operator}
\ee
where $R_{(d)}$ is the Ricci scalar. For the metric (\ref{limiting metric}) characterized by two dimensionful parameters $l$ and $l_1$, one has that $R_{(d)}=-2/l^2+(d-2)(d-3)/l_1^2$.  For a $d$-dimensional conformally coupled scalar field we have $\xi={d-2\o 4(d-1)}$. In this case
\be
X_{conf}=-{(d-2)(d-4)\o 4l^2}+{(d-2)\o 2(d-1)}(1/l^2-1/l_1^2)\, .
\lb{E conf}
\ee

 The calculation
of  the respective entanglement entropy goes along the same lines as before. First, one allows the coordinate $\phi$, which plays the role of the Euclidean time, to
have period $2\pi\alpha$. For $\alpha\neq 1$ the metric (\ref{limiting metric}) then describes the
space $E^\alpha = H^\alpha_2\times S_{d-2}$, where $H^\alpha_2$
is the hyperbolic space coinciding with $H_2$ everywhere except the 
point $x=0$ where it has a conical singularity with an
angular deficit $\delta=2\pi (1-\alpha )$.      The heat kernel of the Laplace operator $\nabla^2$ on $E^\alpha$ is given by the product 
$$
K_{E^\alpha}(z,z',s)=K_{H_2^\alpha}(x,x',\phi , \phi ',s)~K_{S_{d-2}}(\theta ,
\theta ', \varphi, \varphi ',s)~~
$$
where $K_{H_2^\alpha}$ and $K_{S^2}$ are the heat kernels, of the Laplace operator on respectively $H_2^\alpha$ and $S_{d-2}$.
The effective action reads
\begin{equation}
W_{eff}[E^\alpha ]=-{1\over 2}\int^\infty_{\epsilon^2}{ds\over s} Tr K_{H_2^\alpha} Tr K_{S_{d-2}} e^{Xs}~~,
\label{effective action on E}
\end{equation}
where $\epsilon$ is a UV cut-off.
On spaces with constant curvature the heat kernel function 
is known explicitly \cite{Camporesi:1990wm}. In particular, on a
2d space $H_2$ of  negative constant curvature the heat kernel has the following integral representation:
\begin{equation}
K_{H_2}(z,z',s)={1\over l^2}{\sqrt{2}e^{-{\bar{s}/ 4}}
\over (4\pi \bar{s})^{3/2}}\int_\sigma^\infty {dyye^{-{y^2/ 4\bar{s}}}\over
\sqrt{\cosh y-\cosh \sigma }}~~,
\label{heat kernel H2}
\end{equation}
where $\bar{s}=sl^{-2}$. In equation (\ref{heat kernel H2}) $\sigma$ is the geodesic
distance between  
the points  on $H_2$.  Between two points $(x, \phi )$ 
and $(x, \phi+\Delta \phi )$ the geodesic distance is given by
$
\sinh^2 {\sigma \over 2}=\sinh^2 x \sin^2 {\Delta \phi \over 2}~~.
$
The heat kernel on the conical hyperbolic space $H_2^\alpha$ can be obtained from (\ref{heat kernel H2})  by applying the Sommerfeld formula (\ref{Sommerfeld}). Skipping                the technical details,   available in   \cite{Mann:1997hm},     let us just quote the result for the trace
\be
&&\Tr K_{H^\alpha_2}=\alpha \Tr K_{H_2}+(1-\alpha){e^{-\bar{s}/4}\o (4\pi\bar{s})^{1/2}}k_H(\bar{s})+O(1-\alpha)^2\, , \nonumber \\
&&k_H (\bar{s})=\int^\infty_0 dy {\cosh y\over \sinh^2y}(1-{2y\over \sinh 2y})
e^{-{y^2/ \bar{s}}}\, ,
\lb{trace K conical H2}
\ee
where $\bar{s}=s/l^2$.  

Let us denote $\Theta_{d-2}(s)=\Tr K_{S_{d-2}}(s)$  the trace of the heat kernel of Laplace operator  $-\nabla^2$ on $(d-2)$-dimensional sphere of unite radius. 
The entanglement entropy in the extremal limit then takes the form \cite{Mann:1997hm}, \cite{Solodukhin-2010}
\be
S_{ext}={1\o 4\sqrt{\pi}}\int^\infty_{\epsilon^2/l^2}{ds\o s^{3/2}}k_H(s)\Theta_{d-2}(s{l^2\o l_1^2})e^{-s/4}e^{sXl^2}\; .
\lb{entropy in extreme limit}
\ee
The function $k_H(s)$ has  the  following small-$s$ expansion
\be
k_H(s)=\sqrt{\pi s}({1\o 3}-{1\o 20}s+{17\o 1120}s^2-{29\o 4480}s^3+{1181\o 337920}s^4-{1393481\o 615014400}s^5+{763967\o 447283200}s^6+..)\; 
\lb{expansion of k}
\ee
The trace of the heat kernel on a sphere is known in some implicit form. For our purposes however a representation in a form of an expansion is more useful,
\be
\Theta_{d-2}(s)={\Omega_{d-2}\o (4\pi s)^{(d-2)/2}}\left(1+(d-2)(d-3)\sum_{n=1}^\infty a_{2n}s^n\right)\; ,  
\lb{expansion heat kernel on sphere}
\ee
where $\Omega_{d-2}={2\pi^{(d-1)/2}\o \Gamma((d-1)/2)}$ is the area of a unit radius sphere $S_{d-2}$.
The first few coefficients in this expansion can be calculated using the results collected in \cite{Vassilevich:2003xt},
\be
a_2={1\o 6}~,~~a_4={(5d^2-27d+40)\o 360}~,~~ 
a_6={(35d^4-392d^3+1699d^2-3322d+2520)\o 45360}\; .\lb{coefficients in expansion} 
\ee

We shall consider some particular cases.

\paragraph{d=4.}  The entanglement entropy in the extreme limit is
\be
S_{d=4}={l_1^2\o 12\epsilon}+s_0\ln{\epsilon\o l}+s({l_1\o l})\, , \ \ s_0={1\o 18}(6\xi-1)+{1\o 15}{l_1^2\o l^2}(1-5\xi)\, ,
\lb{d=4 extreme}
\ee
where $s({l_1\o l})$ is UV finite part of the entropy. For minimal coupling $(\xi=0)$ this result was obtained in \cite{Mann:1997hm}. The first term in (\ref{d=4 extreme})  is proportional to the horizon area $A=4\pi l^2_1$ while the second term is a logarithmic correction to the area law.
For conformal coupling $\xi=1/6$ the logarithmic term is
\be
s^{conf}_0={1\o 90}{l_1^2\o l^2}\, .
\lb{conformal s0}
\ee

\medskip

\paragraph{d=5.}  The entropy is
\be
S_{d=5}={\sqrt{\pi} l_1^3\o 72 \epsilon^3}+{\sqrt{\pi}\o 120}((2l_1^2 -5l^2)+10\xi (3l^2-l_1^2)){l_1\o l^2\epsilon}+s({l_1\o l})\, .
\lb{d=5 entropy}
\ee

\medskip

\noindent To simplify the expressions in higher dimensions  we consider only the case of the conformal coupling $\xi={d-2\o 4(d-1)}$.

\medskip

\noindent The entropy takes the form:

\medskip

\noindent {\bf d=6.} 
\be
&&S_{d=6}={l_1^4\o 144\epsilon^4}+{1\o 180}{l_1^2\o l^2}{(4l^2-5l_1^2)\o \epsilon^2}+s_0\ln{\epsilon\o l_1}+s({l\o l_1})\, ,\nonumber \\
&&s_0=-{1\o 18900}(1068 {l_1^4\o l^4}-1680{l_1^2\o l^2}+637)\, .\lb{d=6 entropy}
\ee

\medskip

\noindent{\bf d=7.}
\be
&&S_{d=7}={\sqrt{\pi}l_1^5\o 384 \epsilon^5}+{7\sqrt{\pi}\o 34560}{(25l^2-32l_1^2)l_1^3\o l^2\epsilon^3} \nonumber \\
&&+{\sqrt{\pi}\o 1935360}{(70592 l_1^4-109760l_1^2l^2+40635l^4)l_1\o l^5\epsilon}+s({l_1\o l})\, .
\lb{d=7 entropy}
\ee

\medskip

\noindent{\bf d=8.} 
\be
&&S_{d=8}={s_6\o \epsilon^6}+{s_4\o \epsilon^4}+{s_2\o \epsilon^2}+s_0\ln\epsilon +s({l_1\o l})\, , \lb{d=8 entropy}\\
&&s_6={l_1^6\o 2160}\, , \ s_4={1\o 75600}{l_1^4(-209l_1^2+160l^2)\o l^6}\, , \nonumber \\
&& s_2={1\o 352800}{l_1^2(8753l_1^4-13376 l^2l_1^2+4875l^4)\o l^6}\, ,   \nonumber\\
&&s_0={1\o 11113200}{(1102263 l_1^6-2520864l^2l_1^4+1833975l^4l_1^2-413120l^6)\o l^6}\, .
\ee

\bigskip

Two examples of the  extreme geometry  are of particular interest.

\medskip

\paragraph{Entanglement entropy of the round sphere in Minkowski spacetime.}
Consider  a sphere of radius $R$ in flat Minkowski spacetime. One can choose a spherical coordinate system $(\tau,r,\theta^i)$ so that the surface $\Sigma$ 
is defined as $\tau=0$ and $r=R$, and variables $\theta^i\; , \; i=1,..,d-2$ are the angular coordinates on $\Sigma$. The $d$-metric reads
\be
ds^2=d\tau^2+dr^2+r^2d\omega^2_{d-2}\; ,
\lb{d-metric}
\ee
where $d\omega^2_{d-2}$ is metric on $(d-2)$ sphere of unite radius.  Metric (\ref{d-metric}) is conformal to the metric
\be
ds_{ext}^2={R^2\o r^2}(d\tau^2+dr^2)+R^2d\omega^2_{d-2}\; ,
\lb{extreme limit}
\ee
which describes the product of two-dimensional hyperbolic space $H_2$ with coordinates $(\tau,r)$ and the sphere $S_{d-2}$.
Note that both spaces, $H_2$ and $S_{d-2}$, have the same radius $R$. Metric (\ref{extreme limit}) describes the spacetime which appears in the extremal
limit of a $d$-dimensional static black hole. In the hyperbolic space $H_2$ we can choose a polar coordinate system $(\rho,\phi)$  with its center at point $r=R$,
\be
r={R\o \cosh\rho-\sinh\rho\cos\phi}~,~~\tau={R\sinh\rho\sin\phi\o \cosh\rho-\sinh\rho\cos\phi}\; ,
\lb{transfo}
\ee
(for small $\rho$ one has that $r=R+\rho\cos\phi~,~~\tau=\rho\sin\phi$ as in the polar system in flat spacetime)
so that the metric takes the form
\be
ds_{ext}^2=R^2(d\rho^2+\sinh^2\rho d\phi^2)+R^2 d\omega^2_{d-2}\; .
\lb{metric extreme}
\ee
In this coordinate system the surface $\Sigma$ is defined by the condition $\rho=0$.
In the entanglement entropy of a conformally coupled scalar  field the logarithmic term $s_0$ is conformal invariant. Therefore, it is the same \cite{Solodukhin-2010} for the entropy of a round sphere of radius $R$ in Minkowski spacetime and 
in the extreme limiting geometry (\ref{metric extreme}). In various dimensions $s_0$ can be obtained from the results (\ref{conformal s0})-(\ref{d=8 entropy}) by setting
$l=l_1=R$. One finds  $s_0={1\o 90}$ in $d=4$, $s_0=-{1\o 756}$ in $d=6$ and $s_0={23\o 113400}$ in $d=8$. 
For $d>4$ the logarithmic term in the entropy of a round sphere has been calculated by  Casini and Huerta \cite{Casini:2010kt} directly in Minkowski spacetime. 
They have obtained $s_0$ in all even dimensions  up to $d=14$. Subsequently, Dowker \cite{Dowker:2010nq} has extended this result to $d=16$ and $d=18$. In arbitrary dimension
$d$ the logarithmic term $s_0$ can be expressed in terms of Bernulli  numbers as is shown in  \cite{Casini:2010kt} and  \cite{Dowker:2010nq}

\paragraph*{Entanglement entropy of the extreme Reissner-Nordstr\"{o}m black hole in $d$ dimensions.}
As was shown by Myers and Perry  \cite{Myers:1986un} a generalization of the Reissner-Nordstr\"{o}m solution to higher dimension $d>4$ is given by  (\ref{non-extremal metric}) with
\be
g(r)= {(r^{d-3}-r_+^{d-3})(r^{d-3}-r_-^{d-3})\o r^{2(d-3)}}\, .
\lb{Myers-Perry}
\ee
In the extreme limit $r_+\rightarrow r_-$.  Expanding (\ref{Myers-Perry}) near the horizon one finds, in this limit, that $b={(d-3)^2/r_+^2}$. Thus this extreme geometry is characterized by
values  of radii $l=r_+/(d-3)$ and $l_1=r_+$. In dimension $d=4$ we have $l=l_1$ as in the case considered above. In dimension $d>4$ the two radii are different, $l\neq l_1$.
For the conformal coupling  the values of logarithmic term $s_0$ in various dimensions are presented in Table 1.

\begin{table}[t]
\renewcommand{\arraystretch}{1.5}
\centering
\begin{tabular}{|c|ccc|} 
\hline
  $d$& $4$& $6$& $8$  \\
\hline
 $s_0$ &  $\frac{1}{90}$& $-\frac{2881}{756}$ & $\frac{1569275563}{1111320}$  \\[0.5ex]
 \hline  
\end{tabular} \caption{Coefficients of the logarithmic term in the entanglement entropy of extreme Reissner-Nordstr\"{o}m black hole.}
\end{table}

\section{ Logarithmic term in the entropy of generic conformal field theory}

As we have already seen,  in even dimensions, there typically appears a logarithmic term in the entanglement entropy. This term is  universal
in the sense that it  does not depend on the  scheme which is used to regularize the UV divergences. In conformal field theories the
logarithmic terms in the entropy are closely related to the conformal anomaly. In this section we discuss in detail this aspect and formulate precisely
the relation between entanglement entropy and conformal anomalies.

Consider a conformal field theory in $d$ spacetime dimensions. As we have discussed throughout this review the most efficient way to calculate the corresponding
entanglement entropy for a non-extremal black hole is to introduce a small angle deficit $\delta=2\pi
(1-\alpha)$ at the horizon surface $\Sigma$, compute the effective action $W(E^\alpha)$ on a
manifold $E^\alpha$ with a singular surface and then apply the
replica formula $S=(\alpha\partial_\alpha-1)|_{\alpha=1}W(E^\alpha)$ and obtain from it the entanglement entropy.
In $d$ dimensions the effective action 
has the general structure 
\be W_{\rm CFT}(E^\alpha)={a_d\over
\epsilon^d}-{a_{1}\over \epsilon^{d-2}}-..-{a_{n}\o \epsilon^{d-2n}}-..-a_{d/2}\ln \epsilon
+w(g^{(\alpha)})~~. \lb{effective action d dimensions} \ee 
The logarithmic term in this expansion appears only if dimension $d$ is even. Thus only even $d$ will be considered in this section.
The terms $a_d$, $a_{d-2}$,..
representing the power UV divergences, are not  universal,
while the term $a_{d/2}$ is universal and is determined by the
integrated conformal anomaly.  The term $w(g^{(\alpha)})$ is the UV finite
part of the effective action. Under a global rescaling of the 
metric on $E^\alpha$, $g^{(\alpha)}\rightarrow \lambda^2
g^{(\alpha)}$, one has \be
w(\lambda^2g^{(\alpha)})=w(g^{(\alpha)})+a_{d/2}\ln \lambda~~.
\lb{action transform} \ee 

An important property of the expansion  (\ref{effective action d dimensions}) for a quantum field theory which classically is conformally invariant  is that
the logarithmic term $a_{d/2}$ is conformal invariant  (see \cite{Vassilevich:2003xt} and references therein),
$$a_{d/2}[e^{-2\omega}g]=a_{d/2}[g]\, .
$$ 
On a manifold with a conical singularity at the surface $\Sigma$, the coefficients
$a_{d-2n}$ have  a bulk part and a surface part. To
first order in $(1-\alpha)$ one finds that \be
a_{d-2n}(E^\alpha)=\alpha a^{\rm bulk}
_{d-2n}(E)+(1-\alpha)a_{d-2n}^\Sigma +O(1-\alpha)^2\, \lb{decomposition an}
\ee
 For $n=d/2$ the oefficients $a_{d/2}^{\rm
bulk} (E)$ and $a_{d/2}(\Sigma)$ are respectively the integrated bulk
and surface conformal anomalies. The bulk and surface terms are independently invariant under conformal transformation
$g\rightarrow e^{-2\omega}g$, \be a_{d/2}^{\rm
bulk}(e^{-2\omega}g)=a_{d/2}^{\rm bulk}(g) \ {\rm and}  \
a_{d/2}^\Sigma(e^{-2\omega}g)=a_{d/2}^\Sigma (g)\, . \lb{singular conf transform} \ee 
Applying
the replica formula one obtains the entanglement entropy 
\be
&&S={s_{d-2}\over \epsilon^{d-2}}+..+{s_{d-2n}\o \epsilon^{d-2n}}+..+{s_0}^\Sigma \ln \epsilon+s(g)~~,
\nonumber \\
&&s_0=a^\Sigma_{d/2}\, , \ s_{d-2n}=a^\Sigma_{n}\, , \ n=1, .. ,d/2-1 \nonumber \\ 
&&s(\lambda^2 g)=s(g)-a_{d/2}^\Sigma\ln\lambda~~. \lb{entropy general structure} \ee
For a regular manifold each term $a_{d-2n}$ is the integral of a polynomial of degree $n$ in the Riemann curvature.  
Respectively, the surface terms $a^\Sigma_{d-2n} $ is the integral over the singular surface $\Sigma$ of polynomial of degree $n-1$ in the Riemann curvature and its
projections onto the subspace orthogonal to surface $\Sigma$.
The concrete structure of the polynomials depends on the dimension $d$.

\subsection{Logarithmic terms in 4-dimensional conformal field theory}

In four dimensions the bulk conformal anomaly is a combination of two
terms, the topological Euler term and the square of the
Weyl tensor, \be &&a^{\rm bulk}_2=A E_{(4)}+B I_{(4)}\, ,\nonumber \\
&&E_{(4)}={1\over 64}
\int_E(R_{\alpha\beta\mu\nu}R^{\alpha\beta\mu\nu}-4R_{\mu\nu}R^{\mu\nu}+R^2)\, ,
\nonumber \\ &&I_{(4)}=-{1\over
64}\int_E(R_{\alpha\beta\mu\nu}R^{\alpha\beta\mu\nu}-2R_{\mu\nu}R^{\mu\nu}+{1\over
3} R^2)\, . \lb{d=4 conformal anomaly} 
\ee 
These are, respectively, the conformal
anomalies of type A and B. 
In a theory with $n_s$ particles of spin $s$ one finds \cite{Duff:1993wm} (the contributions of fields of spin $3/2$ and $2$ can be obtained from table 2 on p.180 of the book of Birrell and Davies \cite{Birrell-Davies})
\be
&&A={1\o 90\pi^2}(n_0+11n_{1/2}+62n_1+0n_{3/2}+0n_2)\; , \nonumber \\
&&B={1\o
30\pi^2} (n_0 + 6n_{1/2} + 12n_1-{233\o 6}n_{3/2}+{424\o 3} n_2)\; .
\lb{A and B}
\ee
The surface contribution to the
conformal anomaly can be calculated directly by, for example, the
heat kernel method as in \cite{Fursaev:1994in}. The direct
computation although straightforward is technically involved. One
has however a short cut: there is a precise balance, observed in
\cite{Solodukhin:1994yz} and \cite{Fursaev:1994ea}, between the
bulk and surface anomalies, this balance is such that, to first
order in $(1-\alpha)$, one can take $a_2(E^\alpha)=a_2^{\rm
bulk}(E^\alpha)+O(1-\alpha)^2$ and use for the Riemann tensor of
$E^\alpha$ the representation as a sum of regular and singular
(proportional to a delta-function concentrated on surface
$\Sigma$) parts. The precise expressions are given in
\cite{Fursaev:1994ea}, \cite{Fursaev:1995ef}. This representation,
however, is obtained under the assumption that the surface
$\Sigma$ is a stationary point of an Abelian isometry and thus has  vanishing extrinsic curvature. Under this
assumption one finds that \cite{Fursaev:1994ea},
\cite{Fursaev:1995ef}  (see also \cite{Ryu:2006ef})
\be &&a_2(E^\alpha)=\alpha a^{\rm bulk}
_2(E)+(1-\alpha)
a_2^\Sigma +O(1-\alpha)^2~~,\nonumber \\
&&a_2^\Sigma=A a_A^\Sigma+B a_B^\Sigma ~~,\nonumber \\
 &&a_A^\Sigma={\pi\over 8}\int_\Sigma
(R_{ijij}-2R_{ii}+R) ~~,\nonumber \\
&&a_B^\Sigma= -{\pi\over 8}\int_\Sigma (R_{ijij}-R_{ii}+{1\over
3}R)~~, \lb{d=4 surface anomaly} \ee
where $R_{ijij}=R_{\alpha\beta\mu\nu}n^\alpha_i n^\beta_j n^\mu_i n^\nu_j$,
$R_{aa}=R_{\alpha\beta}n^\alpha_a n^\beta_a$.

Each surface term in (\ref{d=4 surface anomaly}) is invariant
under a sub-class of conformal transformations, $g\rightarrow
e^{-2\omega}g$, such that  the normal derivatives of $\omega$
vanish on surface $\Sigma$. The surface term due to the bulk
Euler number is, moreover, a topological invariant: using the
Gauss-Codazzi equation 
\be R=R_\Sigma +2R_{ii}-R_{ijij}-k^ik^i+\tr
k^2~~,\lb{Gauss-Codazzi} \ee where $R_\Sigma$ is the intrinsic Ricci scalar
of the surface and $k^i_{\alpha\beta}$ is the extrinsic curvature, and in the assumption of
vanishing extrinsic curvature the $a_A^\Sigma$ term, as shown in
\cite{Fursaev:1995ef}, is proportional to the Euler number of the
2d surface $\Sigma$, \be a_A^\Sigma= {\pi\over 8} \int_\Sigma
R_\Sigma~~, \lb{A conformal anomaly} \ee where $R_\Sigma$ is intrinsic curvature
of $\Sigma$.

For completeness we note that this result can be generalized to an arbitrary codimension 2 surface in $4$-dimensional spacetime.  The conformal transformation
then is generalized to any function $\omega$ with non-vanishing normal derivative at $\Sigma$. 
The terms with the normal derivatives of $\omega$ in the conformal
transformation of $a_2^\Sigma$ then can be cancelled by adding the
quadratic combinations of extrinsic curvature, $\tr k^2$ and $k_a
k_a$. The analysis presented by Solodukhin  \cite{Solodukhin:2008dh} (this analysis is based on an earlier consideration by Dowker \cite{Dowker:1994bj})  results in the following expressions
\be
&&a_2^\Sigma=A a_A^\Sigma+B a_B^\Sigma~~, \nonumber \\
 &&a_A^\Sigma={\pi\over 8}\int_\Sigma
(R_{ijij}-2R_{ii}+R-\tr k^2+k_i k_i)={\pi\over 8}\int_\Sigma R_\Sigma~~, \nonumber \\
&&a_B^\Sigma= -{\pi\over 8}\int_\Sigma (R_{ijij}-R_{ii}+{1\over
3}R -(\tr k^2-{1\over 2} k_i k_i ))~~. \lb{general surface anomaly} \ee 
This is  the most general form of the logarithmic term in the entanglement entropy in four spacetime dimensions.

Thus, as follows from  equation (\ref{d=4 conformal anomaly}), the logarithmic term in the entanglement entropy of black hole in four dimensiosn is
\be
s^{(d=4)}_0=A{\pi\over 8}\int_\Sigma
(R_{ijij}-2R_{ii}+R) -B{\pi\over 8}\int_\Sigma (R_{ijij}-R_{ii}+{1\over
3}R)\, .
\lb{d=4 log term}
\ee
For conformal fields of various spin the values of $A$ and $B$ are presented in (\ref{A and B}).

\medskip

Consider some particular examples.

\paragraph*{Extreme static geometry.} For an extreme geometry which has the structure of the product $H_2\times S_2$ and characterized by two dimensionful parameters
$l$ (radius of $H_2$) and $l_1$ (radius of $S_2$) the logarithmic term in the entropy 
\be
s_0^{ext}=A\pi^2 -{B\pi^2\o 3}(1-{l_1^2\o l^2})\, 
\lb{extreme geometry log term}
\ee
is determined by both the anomalies of type A and B. In the case of the extreme Reissner-Nordstr\"{o}m black hole one has $l=l_1$ and the logarithmic term (\ref{extreme geometry log term}) is determined only by the anomaly of type A. For a conformal scalar field one has that $A=B/3=1/90\pi^2$ and this equation reduces to (\ref{conformal s0}). 
As we already discussed, the geometry $H_2\times S_2$ for $l=l_1$ is conformal to flat 4-dimensional space. Thus the Weyl tensor vanishes in this case as does its projection to the 
subspace orthogonal to horizon $S_2$. That is why the type B anomaly does not contribute in this case to the logarithmic term.

\paragraph*{The Schwarzschild black hole.} In this case the background is Ricci flat and the logarithmic term  is determined by the difference of $A$ and $B$,
\be
s^{Sch}_0=(A-B)\pi^2\, .
\lb{schwarzschild log term}
\ee
The same is true for any Ricci flat metric.  For a conformal scalar field equation (\ref{schwarzschild log term}) reduces to (\ref{entropy-Sch}).
For a scalar field the relation of the logarithmic term in the entropy and the conformal anomaly was discussed by Fursaev \cite{Fursaev:1994te}.
 The logarithmic term vanishes if $A=B$. In this case the Riemann tensor does not appear in the conformal anomaly (\ref{d=4 conformal anomaly}) so that the anomaly vanishes if the metric is Ricci flat.  In particular, the relation $A=B$ can be found from the  ${\cal N}=4$ super-conformal gauge theory, dual to  supergravity on $AdS_5$, according to the AdS/CFT correspondence of Maldacena \cite{Maldacena:1997re}. The conformal anomaly in this theory was calculated in \cite{Henningson:1998gx}. 

\paragraph*{Non-extreme and extreme Kerr black hole.} For the Kerr black hole (characterized by mass $m$ and rotation $a$) the logarithmic term does not depend on the parameter of the rotation and it takes the same form
\be
s^{Kerr}_0=(A-B)\pi^2 \, 
\lb{kerr log term}
\ee
as in the case of the Schwarschild metric. In the extreme limit $a=m$ the logarithmic term takes same value (\ref{kerr log term}).

\subsection{Logarithmic terms in 6-dimensional conformal field theory}
In six dimensions, omitting the total derivative terms, the conformal anomaly is a combination of four different conformal invariants \cite{Bastianelli:2000hi}
\be
a^{bulk}_3=\int_{M^6}(B_1 I_1+B_2I_2+B_3 I_3+A E_6)\, ,
\lb{d=6 conformal anomaly}
\ee
where $I_1$, $I_2$ and $I_3$ are cubic in the Weyl tensor 
\be
&&I_1=C_{kijl}C^{imnj}C_{m\ \ n}^{\ kl}\, , \ I_2=C_{ij}^{\ \ kl}C_{kl}^{\ \ mn}C_{mn}^{\ \ ij}\, , \nonumber \\
&&I_3=C_{iklm}(\nabla^2 \delta^i_j+4R^i_j-{6\o 5}R\delta^i_j)C^{jklm}\, .
\lb{d=6 conformal invariants}
\ee
and $E_6$ is the Euler density (\ref{Topological Euler number})
\be
E_6={1 \over 3072\pi^3}\epsilon_{\mu_1 \mu_2...\mu_{6}}\epsilon^{\nu_1 \nu_2
... 
\nu_{6}} R^{\mu_1 \mu_2}_{\ \ \nu_1 \nu_2} ...R^{\mu_{5} \mu_{6}}_{\ \
\nu_{5} \nu_{6}}\, .
\label{d=6 Euler number}
\ee
As was shown in \cite{Bastianelli:2000hi} in a free conformal field theory with $n_0$ scalars, $n_{1/2}$ Dirac fermions and $n_B$ 2-form fields one has that\epubtkFootnote{
Note that the coefficient $b_6$ of paper \cite{Bastianelli:2000hi} is related to $a_3$ as $b_6=-a_3$.}
\be
A={8\cdot 3!\o 7!}(-{5\o 72}n_0-{191\o 72}n_{1/2}-{221\o 4}n_B)\, ,
\lb{A}
\ee
\be
B_1={1\o (4\pi )^37!}({28\o 3}n_0+{896\o 3}n_{1/2}+{8008\o 3} n_B)\, ,
\lb{B1}
\ee
\be
B_2={1\o (4\pi )^37!}(-{5\o 3}n_0+32 n_{1/2}+{2378\o 3}n_B)\, ,
\lb{B2}
\ee
\be
B_3={1\o (4\pi )^37!}(-2n_0-40 n_{1/2}-180 n_B)\, .
\lb{B3}
\ee
Applying the formulas (\ref{singular curvature}) to $I_1$, $I_2$ and $I_3$ and using the relation (\ref{Euler number conical manifold}) for the Euler number
one finds for the logarithmic term in the entanglement entropy of $4$-dimensional surface $\Sigma$ in a 6-dimensional conformal field theory
\be
s_0^{d=6}=B_1\, s_{B1}+B_2\, s_{B2}+B_3\, s_{B3}+A\, s_A\, ,
\lb{d=6 log term}
\ee
where
\be
s_A=\chi[\Sigma]\, ,
\lb{sA}
\ee
is the Euler number of the surface $\Sigma$, and
\be
s_{B1}=6\pi(C^{j\alpha\beta i}C_{\alpha \ \ \beta}^{\ ij}-C^{j\alpha\beta j}C_{\alpha \ \ \beta}^{\ ii }-{1\o 4}C^{i\alpha\beta\mu}C^{i}_{\ \alpha\beta\mu}+{1\o 20}C^{\alpha\beta\mu\nu}C_{\alpha\beta\mu\nu})\, ,
\lb{sB1}
\ee
\be
s_{B2}=6\pi(2C^{ij\alpha\beta}C_{\alpha\beta}^{\ \ ij}-C^{i\alpha\beta\mu}C^i_{\ \alpha\beta\mu}+{1\o 5}C^{\alpha\beta\mu\nu}C_{\alpha\beta\mu\nu})\, ,
\lb{sB2}
\ee
\be
s_{B3}=8\pi(\nabla^2 C^{ijij}+4R^i_{\ \alpha}C^{\alpha jij}-R_{\alpha\beta}C^{\alpha i\beta i}-{6\o 5}RC^{ijij}+C^{i}_{\ \alpha\beta\mu}C^{i\alpha\beta\mu}-{3\o 5}C^{\alpha\beta\mu\nu}C_{\alpha\beta\mu\nu})\, ,
\lb{sB3}
\ee
where  tensors with  Latin indices  are obtained by contraction with components of normal vectors $n^i_\alpha, \ i=1,2$. Note that in (\ref{sB3}) we used for brevity the notation
$\nabla^2 C^{ijij}\equiv n^i_\alpha n^j_\beta n^i_\mu n^j_\nu \nabla^2 C^{\alpha\beta\mu\nu}$. Equations (\ref{sB1}), (\ref{sB2}), (\ref{sB3}) agree with result obtained in \cite{Hung:2011xb}.

\medskip

\noindent Let's consider some examples.

\medskip

\paragraph*{6-dimensional Schwarzschild black hole.} The 6-dimensional generalization of the Schwarzschild solution is \cite{Myers:1986un}
\be
ds^2=g(r)d\tau^2+g^{-1}(r)dr^2+r^2d\omega^2_{S_4}\, , \ g(r)=1-{r_+^3\o r^3}\, ,
\lb{d=6 Schrwazschild}
\ee
where $d\omega^2_{S_4}$ is metric of unit 4-sphere. The area of horizon is $A_+={8\pi^2\o 3}r_+^4$. The Euler number of the horizon $\chi[S_4]=2$. This metric is Ricci flat so that only the Riemann tensor contributes to the Weyl tensor. The logarithmic term in this case is
\be
s_0^{Sch}=16\pi^3(-51B_1+156 B_2-192B_3)+2A\, .
\lb{log term d=6 schwarzschild}
\ee
It is interesting to note that this term vanishes in the case of the  interacting $(2,0)$ conformal theory which is dual to supergravity on $AdS_7$. Indeed in  this case one has \cite{Henningson:1998gx},  \cite{Bastianelli:2000hi}
\be
&&B_i={b_i\o (4\pi)^3 7!}\, , \ A={8\cdot 3!\o 7!}a\, , \nonumber \\
&&a=-{35\o 2}\, , \, b_1=1680\, , \, b_2=420\, , \, b_3=-140\, 
\lb{2,0 theory}
\ee
so that $s^{(2,0)}_0=0$. This is as expected.  The Riemann tensor does not appear in the conformal anomaly of the strongly interacting $(2,0)$ theory so that the anomaly identically vanishes if the spacetime is Ricci flat. This property is not valid in the case of the free $(2,0)$ tensor multiplet \cite{Bastianelli:2000hi} so that the logarithmic term of the free multiplet is non-vanishing.

\paragraph*{Conformally flat extreme geometry.} In conformally flat spacetime the Weyl tensor $C_{\alpha\beta\mu\nu}=0$ so that terms   (\ref{sB1}), (\ref{sB2}) and (\ref{sB3}) identically vanish. The logarithmic term (\ref{d=6 log term}) then is determined by the anomaly of type A only.
In particular this is the case for the extreme geometry
$H_2\times S_4$  with equal radii $l=l_1$ of two components. One has
\be
s_0^{ext}=2A\, 
\lb{log term extreme}
\ee
for this extreme geometry.
This geometry is conformal to flat  spacetime and the logarithmic term   (\ref{log term extreme})  is the same as for  the entanglement entropy                       
of a round sphere in flat 6-dimensional spacetime. This generalizes the result discussed in section \ref{section: entropy in the extremal limit} for the entropy of a round sphere due to conformal scalar field.
The result (\ref{log term extreme}),  as is shown in \cite{Casini:2011kv}, \cite{Myers:2010tj},  generalizes to a spherical entangling surface in a conformally flat spacetime of  any even dimension.

\subsection{Why might logarithmic terms in the entropy  be interesting?}

By a  logarithmic term we  mean both the logarithmically UV divergent term in the entropy  and the UV finite term which scales logarithmically with respect to the
characteristic size of the black hole. As we have seen, these terms are  identical. However, after the  UV divergences in the entropy have been renormalized  it is the UV finite term, which scales logarithmically,  that will interest us here. 

\medskip

\noindent i) First of all, the logarithmic terms are universal and do not depend on the way the entropy was calculated and on the scheme 
in which the UV divergences are  regularized.
This is in contrast with the power UV divergences in the entropy that depend both on the calculation procedure and on the  regularization scheme.

\medskip

\noindent ii)  The logarithmic terms are related to the conformal anomalies. As the conformal anomalies play an important role in the modern 
theoretical models, any new manifestation of the anomalies merits of our special attention. This may be even more important in view of ideas that 
the conformal symmetry may play a more fundamental role in Nature
than is usually thought. As is advocated by 't Hooft in a number of recent papers \cite{Hooft:2010nc}, \cite{Hooft:2010ac}, \cite{'tHooft:2009ms} 
a crucial ingredient for  understanding  Hawking radiation and entropy is to realize that gravity itself is a
spontaneously broken conformal theory.

\medskip

\noindent iii) For a large class of extremal black hole solutions, which arise in supergravity theories considered as low energy approximation of string theory, there exists
a microscopic calculation of the entropy. This calculation requires a certain amount of unbroken supersymmetry, so that the black holes in question are the BPS solutions and uses the conformal field theory tools, such as the Cardy formula. 
The Cardy formula predicts certain logarithmic corrections to the entropy (these corrections are discussed, in particular,  in \cite{Carlip:2000nv} and \cite{Solodukhin:1997yy}). One may worry whether exactly same corrections are reproduced in the macroscopic, field theoretical, 
computation of the entropy. This aspect was studied recently in  \cite{Banerjee:2010qc} for black holes in ${\cal N}=4$ supergravity and  at least some partial (for the entropy due to matter multiplet of the supergravity)  agreement with the microscopic calculation has been indeed observed.

\medskip

\noindent iv) Speaking about the already renormalized entropy of black holes and taking into account the backreaction of the quantum matter on the geometry, the black hole entropy 
can be represented as a series expansion in powers of Newton's constant, $G/r_g^2$ (the quantity $1/r_g^2$,  where $r_g=2GM$ is the size of black hole, is the scale of the curvature at the horizon; thus the ratio $G/r^2_g$ measures the strength of gravitational self-interaction at the horizon) or, equivalently, in powers of  $M^2_{PL}/M^2$. In particular for the Schwarzschild
black hole of mass $M$ in four spacetime dimensions  one finds
\be
S=4\pi {M^2\o M^2_{PL}}+\sigma\ln M +O({M^2_{PL}\o M^2})\, .
\lb{entropy M}
\ee 
The logarithmic term is the only correction to the classical Bekenstein-Hawking entropy  that is growing with the mass $M$.

\medskip

\noindent v) Although the logarithmic term is still negligibly small compared to the classical entropy for  macroscopic black holes it becomes important 
for small black holes especially at the latest stage of the black hole evaporation.  In particular, it manifests itself in a modification of the Hawking temperature as a function of mass $M$  \cite{Fursaev:1994te}. Indeed,  neglecting the terms $O({M^2_{PL}\o M^2})$ in the entropy (\ref{entropy M}) one finds
\be
1/T_H={\partial S\o \partial M}=8\pi M/M^2_{PL}+\sigma M^{-1}\, 
\lb{Hawking temperature}
\ee
so that the Hawking temperature $T_H\sim M$ for small black holes. 
Depending on the sign of the coefficient $\sigma$ in (\ref{entropy M}) there can be two different scenarios.  If $\sigma<0$ then the entropy $S(M)$ as function of mass develops a minimum at some value of $M_{min}\sim M_{PL}$. For this value of the mass the temperature (\ref{Hawking temperature})  becomes infinite. This is the final point (at least in this approximation) of the evaporation for black holes of mass $M>M_{min}$. It is reached in finite time. Not worrying about exact numerical factors one has
\be
{dM\o dt}\sim -T_H^4 A_+\sim -T_H^4 M^2
\lb{evaporation rate}
\ee
for the evaporation rate. For large black holes ($M_0\gg M_{min}$) the evaporation time is  $t_{BH}\sim M_0^3/M^4_{PL}$. This  evaporation time can be  obtained by solving 
(\ref{evaporation rate}) with the classical expression for the Hawking temperature $T_H\sim M^2_{PL}/M$, i.e. without the correction as in (\ref{Hawking temperature}). Thus, if there is no logarithmic term in (\ref{entropy M}) any black hole evaporates in finite time.   If the correction term is present, it  becomes important for $M_0\sim M_{min}$. Then, assuming that $M_{min}\sim M_{PL}$, one finds  $t_{BH}\sim (M_0-M_{PL})^5/M^4_{PL}$ for the evaporation time. On the other hand, a black hole of   mass $M_0<M_{min}$, if it exists, evaporates down to zero mass in infinite time. Similar behavior is valid for black holes of arbitrary mass $M_0$  if $\sigma>0$ in equations (\ref{entropy M}), (\ref{Hawking temperature}). The evaporation rate considerably slows down for small black holes since the Hawking temperature $T_H\sim M$ for small $M$. The black hole then evaporates to zero mass in infinite time. Asymptotically, for large time $t$, the mass of black hole decreases as $M(t)\sim t^{-1/5}$. Thus, the sub-Planckian black holes if ($\sigma<0$) and any black holes
if ($\sigma>0$) live much longer than  one would have  expected if one used the classical expression for the entropy and for the Hawking temperature.

\section{Holographic description of entanglement entropy of black hole}
\label{section: Holographic description}

A popular trend in  modern fundamental physics is to reconsider various, sometimes very well known, phenomena from the point of view of
 {\it holography}. Holography is a rather general statement  that the physics inside a spatial region can be  understood by looking at a
certain theory defined on the boundary of the region. This  so-called {\it holographic principle} was first formulated by 't Hooft \cite{'tHooft:1993gx} and later generalized by Susskind  \cite{Susskind:1994vu}.
A review of the holographic principle was given in  \cite{Bousso:2002ju}.  A concrete realization of  holography is the AdS/CFT correspondence  \cite{Maldacena:1997re}, \cite{Witten:1998qj},
\cite{Gubser:1998bc}. According to this correspondence the theory of supegravity
(more precisely string theory the low energy regime of which is described by a supegravity) in a $(d+1)$-dimensional anti-de Sitter spacetime (AdS) is equivalent to a quantum conformal field theory (CFT) defined on a $d$-dimensional boundary of the anti-de Sitter. There is a precise dictionary of how phenomena on one-side of the correspondence can be translated into
phenomena on the other side. The correspondence has proved to be extremely useful, both for better understanding the gravitational physics and the quantum field theory.
If $d=4$ then the CFT on the boundary is known to be  a ${\cal N}=4$ superconformal gauge theory. This theory is strongly coupled and in many aspects resembles the QCD.
Thus utilizing the correspondence one, in particular,  may gain some information on how the theories of this type behave (for a review on the correspondence  and its applications see \cite{Aharony:1999ti}).

One of the aspects of the AdS/CFT correspondence is geometrical. The boundary theory provides  certain boundary conditions for the gravitational theory in the bulk so that
one may decode the {\it hologram}: reconstruct the bulk spacetime from the boundary data. As was analyzed  in  \cite{deHaro:2000xn}   for this reconstruction the boundary data one has to specify consist on the boundary metric and the vacuum expectation of the stress-energy tensor of the boundary CFT.  The details are presented in \cite{deHaro:2000xn}, see also review  \cite{Skenderis:2000in}.

Entanglement entropy is one of the fundamental quantities which characterize the boundary theory. One would think that it  should have an interpretation within the AdS/CFT correspondence. This interpretation was suggested in 2006 by Ryu and Takayanagi  \cite{Ryu:2006bv}, \cite{Ryu:2006ef} (for a review on this proposal see \cite{Nishioka:2009un}).
This proposal is very interesting since it allows one to compute the entanglement entropy in a purely geometrical way (see also \cite{Fursaev:2006ih}).

\subsection{Holographic proposal for entanglement entropy}

Let $M$ be a $(d+1)$-dimensional asymptotically anti-de Sitter spacetime. Its conformal boundary is a $d$-dimensional spacetime $\partial M$. On a slice of constant time $t$ in $\partial M$ one picks a closed $(d-2)$-dimensional surface $\Sigma$ and defines the entropy of entanglement with respect to $\Sigma$. Now, on the constant $t$ slice of  the $(d+1)$-dimensional anti-de Sitter spacetime consider a $(d-1)$-dimensional minimal surface $\gamma$ such that its boundary in $\partial M$ is the surface $\Sigma$, $\partial\gamma=\Sigma$. According to the proposal of \cite{Ryu:2006bv}, \cite{Ryu:2006ef}
the following quantity
\be
S={A(\gamma)\over 4 G_{d+1}}\, ,
\lb{holographic proposal}
\ee
where $A(\gamma)$ is the area of minimal surface $\gamma$ and $G_{d+1}$ is Newton's constant in the $(d+1)$-dimensional gravitational theory, is equal to the entanglement entropy one has calculated in the boundary conformal field theory. This holographic proposal has passed many tests and never failed. It has reproduced correctly the entropy in all those cases when it is known explicitly. In particular,  in two spacetime dimensions ($d=2$) it correctly reproduced (\ref{entropy-circle}) and (\ref{entropy-T}) for the entropy at finite size
and at finite entropy respectively. In higher dimensions ($d>2)$ the area of minimal surface $\gamma$ diverges when it is extended till $\partial M$. This is an important feature, typical of the AdS/CFT correspondence. In fact, instead of the conformal boundary $\partial M$ one has to consider a regularized boundary $\partial M_\epsilon$ located  at a small distance $\epsilon$ (measured in 
terms of some radial coordinate $\rho$). In the AdS/CFT correspondence the divergence in $\epsilon$ has the interpretation of a UV divergence in the boundary quantum field theory. 
Considering the regularized surface $\gamma$ which extends to $\partial M_\epsilon$, one finds that its area, to leading order in $\epsilon$, behaves as $A(\gamma)\sim A(\Sigma)/\epsilon^{d-2}$.
Taking this behavior of the area one sees that 
the proposal (\ref{holographic proposal}) correctly reproduces the proportionality  of the entropy 
to the area of surface $\Sigma$ and its dependence on the UV cutoff $\epsilon$.

We note however that in certain situations the choice of the minimal surface $\Sigma$ may not be  unique. In particular, if the dividing surface $\Sigma$ has several components or 
if the quantum field resides inside a cavity instead of being defined on an infinite space,  there is more than one natural choice of the minimal surface 
$\gamma$. Different choices may correspond to  different phases in the boundary theory \cite{Klebanov:2007ws}.

\subsection{Proposals for holographic entanglement entropy of black hole}

If one wants to generalize the proposal of Ryu and Takayanagi to black holes, the first step would be to find a $(d+1)$-dimensional asymptotically AdS metric which solves the Einstein equations in the bulk and whose conformal boundary describes a $d$-dimensional black hole. To find such a metric explicitly may be a difficult task although some exact solutions are known. First of all, it is easy to construct an AdS metric which gives a de Sitter spacetime on the boundary. A de Sitter horizon is in many aspects similar to a black hole horizon.
Entanglement entropy associated to a de Sitter horizon \cite{Hawking:2000da}, \cite{Iwashita:2006zj} has the same properties as the entropy of any other Killing horizon. In four dimensions ($d=3$) an exact solution to the Einstein equations has been found in \cite{Emparan:1999wa}   that describes a $3d$-black hole on the boundary. In three dimensions $(d=2)$ an exact solution which describes a generic
static two-dimensional black hole on the boundary has been found in \cite{Skenderis:1999nb}. On the other hand, the results of \cite{deHaro:2000xn} show that for any chosen metric on the boundary $\partial M$ one can find, at least in a small region  close to the boundary, an exact solution to Einstein equations with negative cosmological constant. Exact formulas are given in
\cite{deHaro:2000xn}. Thus, at least principally, it is not a problem to find an asymptotically AdS metric which describes a black hole on the boundary.

The next question is how to choose the minimal surface $\gamma$. A proposal of Emparan \cite{Emparan:2006ni} consists of the following. Suppose the metric on the boundary of asymptotically AdS spacetime describes a black hole with a Killing horizon at surface $\Sigma$. Presumably, the horizon on the boundary $\partial M$ is extended to the bulk.
The bulk horizon is characterized by vanishing extrinsic curvature and is a minimal $(d-1)$-dimensional  surface. Thus, one can choose the bulk horizon to be that  minimal surface $\gamma$,
the area of which should appear in the holographic formula (\ref{holographic proposal}). In this construction the Killing horizon $\Sigma$ is the only boundary of  the
minimal surface $\gamma$. This prescription is perfectly eligible if one computes the  entanglement entropy of a black hole in infinite spacetime. In \cite{Emparan:2006ni} it was applied to entropy of a black hole residing on the 2-brane in the 4d solution
of ref.\cite{Emparan:1999wa}.  In ref.\cite{Iwashita:2006zj} a similar prescription is used to compute entanglement entropy of de Sitter horizon.

On the other hand, in certain situations it is interesting to consider a black hole residing inside a cavity, the so-called ``black hole in a box''. Then, as we have learned  in two-dimensional case, the entanglement entropy will depend on the size $L$ of the box so that, in the limit of large $L$, the entropy will have a thermal contribution proportional to volume of the box. This contribution is additional to the purely entanglement part which is due to presence of the horizon $\Sigma$. In order to reproduce this dependence on the size of the box one should use a different proposal. A relevant proposal was suggested  by Solodukhin in \cite{Solodukhin:2006xv}. 

Let a $d$-dimensional spherically symmetric static black
hole with horizon $\Sigma$
 lie on the regularized boundary (with regularization parameter $\epsilon$)
of asymptotically anti-de Sitter space-time  $AdS_{d+1}$ inside a
spherical cavity $\Sigma_L$ of radius $L$. Consider a minimal
d-surface $\Gamma$ whose boundary is the union of $\Sigma$ and
$\Sigma_L$. $\Gamma$ has saddle points where the radial AdS
coordinate  has an extremum. By spherical symmetry the saddle points
form a (d-2)-surface ${\cal C}$ with the geometry of a sphere. Consider
the subset $\Gamma_h$ of $\Gamma$ whose boundary is  the union of
$\Sigma$ and ${\cal C}$. According to prescription of \cite{Solodukhin:2006xv}, the quantity
\be S={{\tt Area}(\Gamma_h)\over 4 G_N^{d+1}}
 \lb{proposal SS} 
 \ee is
equal to the entanglement entropy of the black hole in the boundary of
AdS. In particular, it gives the expected dependence of the
entropy on the UV regulator $\epsilon$. The minimal surface $\Gamma_h$ ``knows'' about the existence of the other boundary $\Sigma_L$. That is why (\ref{proposal SS}) reproduces correctly the dependence of the entropy on the size of the ``box''. 
In \cite{Solodukhin:2006xv} this proposal has been verified for $d=2$ and $d=4$.

It should be noted that what as far as the UV divergent part of the entanglement entropy is concerned, the two proposals \cite{Emparan:2006ni}, \cite{Iwashita:2006zj} and \cite{Solodukhin:2006xv}
give the same result. This is due to the fact that the UV divergences come from that part of the minimal surface which approaches the boundary $\partial M$ of the AdS spacetime.
In both proposals this part of the surfaces $\gamma$ and $\Gamma_h$  is the same. The difference thus is in the finite terms which are due to global properties of the minimal surface.

From  geometrical point of view  the holographic calculation of the logarithmic term in the entanglement entropy
 is related to the surface anomalies studied
by Graham and Witten \cite{Graham:1999pm}  (this point is discussed in \cite{Schwimmer:2008yh}).

\subsection{Holographic entanglement entropy of 2d black holes}
In order to check the  proposal (\ref{proposal SS}) one needs  a solution to the bulk
Einstein equations that describes a black hole on the boundary of
AdS. In three dimensions a solution of this type is known
explicitly \cite{Skenderis:1999nb}, \be ds^2 = {d\rho^2\over
4\rho^2} +{1 \over \r} \left[f(x)\left(1 + {1\over 16}{f'{}^2- b^2
\over f}\r\right)^2
 d\tau^2  +{1 \over f(x)}\left( 1 +{1\over 4} f'' \r - {1 \over
16}{f^{' 2}- b^2 \over f}\r\right)^2 dx^2 \right],\label{3d metric}
\ee where   the AdS radius is set to unity. At asymptotic infinity
($\r=0$) of  the metric (\ref{3d metric}) one has the 2d black hole
metric 
\be
ds^2_{2d}=f(x)d\tau^2+f^{-1}(x)dx^2\, ,
\lb{2d bh}
\ee
where $f(x)$ has simple zero in $x=x_+$.  The cavity $\Sigma_L$ is placed at $x=L$ so that
$x_+\leq x\leq L$. The regularity of the metric (\ref{2d bh}) at the horizon $x=x_+$ requires that
$0\leq \tau\leq \beta_H$, $\beta_H=4\pi/f'(x_+)$.
 Note, that (\ref{3d metric}) is a vacuum solution
of the Einstein equations for any function $f(x)$.
 The regularity of the 3d metric (\ref{3d metric}) requires that the constant
$b$ should be related to the Hawking temperature of the
two-dimensional horizon by $b=f'(x_+)$. The geodesic $\Gamma$ lies
in the hypersurface of constant time $\tau$. The induced metric on
the hypersurface $(\rho , x)$  has a constant curvature equal $-2$
for any function $f(x)$ and is, thus, related by a coordinate
transformation to the metric 
\be ds_\tau^2={dr^2\over 4r^2}+{1\over
r}dw^2. 
\lb{ads2} 
\ee 
The exact relation between the two
coordinate systems is 
\be
w={1\over 8b}e^{z(x)}\left({16f(x)-(b^2-f'^2_x)\r\over 16f(x)+(b-f'_x)^2\r}\right)\, ,\nonumber \\
r=f(x){e^{2z(x)}}{\r\over (16f(x)+(b-f'_x)^2\r)^2}, 
\lb{trans} \ee
where $z(x)={b\over 2}\int_L^x {dx\over f(x)}$. The equation for
the geodesic in metric (\ref{ads2}) is $r={1\over C^2}-(w-w_0)^2$.
The geodesic length between two points lying on the geodesic  with radial coordinates
$r_1$ and $r_2$ is 
$$\gamma(1,2)={1\over 2}\left(\ln
\left[{1-\sqrt{1-C^2r}\over 1+\sqrt{1-C^2r}}\right]\right)_{r_1}^{r_2}\, .
$$
The saddle point of the geodesic is at $r_m=1/C^2$. 
 The constant $C$ is determined from the condition that
geodesic $\Gamma$ joins points $(x=x_+,\rho=\epsilon^2)$ and
$(x=L,\rho=\epsilon^2)$ lying on the regularized (with
regularization parameter $\epsilon$) boundary. In the limit of small $\epsilon$ one finds that
$$
{C^2 r_+\over 4}=\epsilon^2{b^2\over
f(x_+)}\exp(-b\int^L_{x_+}{dx\over f(x)}) \, .
$$
The length of
the geodesic $\Gamma_h$ joining point $r_+$ corresponding to
$(x=x_+,\rho=\epsilon)$ and the saddle point is thus  $\gamma
(\Gamma_h)=-{1\over 2}\ln{C^2r_+\over 4}. $

Now, one has to take  into account that, in the AdS/CFT correspondence, the value of  Newton's constant in the bulk is related to the number of quantum fields
living in the boundary $\partial M$. In the $AdS_3/CFT_2$ case one has that
${1\over G_N}={2c\over 3}$, where $c$ is the central charge of
boundary CFT.  One thus  finds  the holographic entropy (\ref{proposal SS}) 
\be
&&S={1\over 4G_N}\gamma(\Gamma_h)  \\
&&={c\over 6}\ln{1\over \epsilon}+ {c\over 12}\left[
\int_{x_+}^L{dx\over f(x)}(b-f')+\ln f(L)-\ln b^2\right]\, ,\nonumber
\lb{Sgeom} \ee
where $b=f'(x_+)$,  indeed coincides with the expression (\ref{S})  for
the holographic entanglement entropy of the 2d black hole in conformal field
theory.  In particular, for large values of $L$ the holographic formula for the entropy correctly reproduces
the entropy of thermal gas $S_{th}={c\pi\o 3}T_HL$ (we remind that $T_H=f'(x_+)/4\pi$). This is a consequence of the choice of the minimal surface $\Gamma_h$ made  in the proposal (\ref{proposal SS}).

\subsection{ Holographic entanglement entropy of higher dimensional black holes} 

In higher dimensions there is no known
exact solution similar to (\ref{3d metric}). However, a solution in
the form of  a series expansion in $\rho$  is available.  In
the rest of this section we focus on the case of boundary dimension 4.
Then one finds \cite{Henningson:1998gx} 
\be
&&ds^2={d\r^2\over 4\r^2}+{1\over \r}g_{ij}(x,\rho)dx^idx^j \lb{exp1} \\
&&g_{}(x,\rho)=g_{(0)}(x)+g_{(2)}\rho+g_{(4)}\r^2+h_{(4)}\r^2\ln\r
+..,\nonumber 
\ee 
where $g_{(0)ij}(x)$ is the boundary metric,
coefficient \cite{Henningson:1998gx}
\be
g_{(2)ij}=-{1\over 2}(R_{ij}-{1\over 6}Rg_{(0)ij})
\lb{g2 coefficient}
\ee
is the local covariant function of boundary metric. Coefficient $g_{(4)}$ is not expressed
as a local function of the boundary metric and is related to the
stress-energy tensor of the boundary CFT \cite{deHaro:2000xn},
which has an essentially nonlocal nature. $h_{(4)}$ is a local,
covariant, function of the boundary metric and is obtained as a
variation of the integrated conformal anomaly with respect to the
boundary metric \cite{deHaro:2000xn}. Its explicit form  was computed
in \cite{deHaro:2000xn}. 
$h_{(4)}$ is a traceless tensor and in mathematics literature  it is
known as the ``obstruction" tensor \cite{GH}. The explicit form of $h_{(4)}$ or 
$g_{(4)}$ is not important if one wants to compute the UV divergence terms in the entropy.

One may choose the
boundary metric  describing
 a static spherically symmetric black hole to take the form
\be ds^2=f(r)d\tau^2+f^{-1}(r)dr^2+r^2(d\theta^2+\sin^2\theta
d\phi^2) \lb{spher} 
\ee 
The minimal surface $\Gamma$ lies in the
hypersurface of the constant $\tau$ of 5-dimensional space-time
(\ref{exp1}).
 The induced metric on the hypersurface takes the form
\be ds^2_{\tau}={d\r^2\over 4\r^2}+{1\over \r}\left[F{dr^2\over
f(r)} +R^2(d\theta^2+\sin^2\theta d\phi^2)\right], \lb{hyper} 
\ee
where functions $F(r,\rho)=g^{rr}_{(0)}g_{rr}$ and
$R^2(r,\rho)=g_{\theta\theta}$ have $\rho$-expansion due to
(\ref{exp1}). The minimal surface $\Gamma$ can be parameterized by
$(\rho, \theta, \phi)$. Instead of the radial coordinate $r$ it is
convenient to introduce the coordinate $y=\int dr/\sqrt{f}$ so that,
near the horizon, one has $r=r_++by^2/4+O(y^4)$. The coordinate $y$ measures the
invariant distance along the radial direction. By spherical
symmetry, the area to be minimized is 
\be {\tt Area}(\Gamma)=4\pi
\int {d\r \over \r}R^2 \sqrt{{1\over 4\r^2}+{F\over \r} ({dy\over
d\r})^2}, \lb{*} 
\ee 
where $\r_m$ is  the saddle
point. In the  vicinity of the horizon ($y\ll 1$), we can
neglect the dependence of the functions $F(y,\r)$ and $R^2(y,\rho)$ on the 
coordinate $y$. The minimization of the area of the surface gives
the equation 
\be {FR^2{dy\over d\r}\over \r^2\sqrt{{1\over
4\r^2}+{F\over \r} ({dy\over d\r})^2}}=C={\tt const}. \lb{**} 
\ee
The area of the minimal surface $\Gamma_h$  is then given by the
integral 
\be {\tt Area}(\Gamma_h)=2\pi
\int_{\epsilon^2}^{\r_m}d\r{\cal A} (\r),~{\cal A} ={R^2\over
\r^2\sqrt{1-{C^2\r^3\over FR^4}}} \lb{Ar} 
\ee 
Using (\ref{exp1})
we find that $ {\cal A} (\r)=[{r^2_+\over
\r^2}+{g_{(2)\theta\theta}(r_+)\over \r}+..]. $ Substituting this
expansion into (\ref{Ar}) we find that the first two terms produce
divergences (when $\epsilon$ goes to zero)  which, according to
our proposal, are to be interpreted as UV divergences of the
entanglement entropy. At the black hole horizon, one has the
relation $2R_{\theta\theta}|_{r_+}=r_+^2(R-R_{aa})$. Putting
everything together and applying  proposal (\ref{proposal SS}), one finds for the divergent part 
\be S_{\tt div}={A(\Sigma)\over 4\pi
\epsilon^2}N^2-{N^2\over 2\pi}\int_\Sigma({1\over 4}R_{aa}-{1\over
6}R)\ln\epsilon, \lb{Sdiv} 
\ee 
where $A(\Sigma)=4\pi r_+^2$ is the
horizon area.

The logarithmic term in (\ref{Sdiv}) is 
is related to the logarithmic  divergence (calculated holographically in  \cite{Henningson:1998gx})
$$
W_{log}={N^2\over 4\pi^2}\int ({1\over 8}R^2_{\mu\nu} -{1\over
24}R^2) \ln\epsilon$$
  in the quantum effective action of boundary
CFT. This relation is a particular manifestation of the general
formula (\ref{d=4 log term}) that relates the logarithmic term in the entropy to the conformal anomalies of type A and B. One notes that  in the ${\cal N}=4$ superconformal $SU(N)$ gauge theory 
one has that $A=B={N^2\o \pi^2}$.

It should be noted that the UV finite terms and their dependence  on the size $L_{\tt inv}$ of the box can be computed in the limit of small $L_{inv}$. This calculation is given in \cite{Solodukhin:2006xv}.  In particular, in any even dimension $d$ one finds an universal
 term in the entropy that takes  the form (up to numerical factor) $S\sim
r_+^{d-2}h_{({d})\theta\theta}(r_+) L^2_{\tt inv}\ln L_{\tt inv}$ and is proportional to the value of the ``obstruction tensor'' on the black hole horizon.
The direct  calculation of such terms in the  entanglement entropy on
the CFT side is not yet available.

\section{Can entanglement entropy explain the Bekenstein-Hawking entropy of black holes?}

Entanglement entropy of a black hole is naturally proportional to the area of the black hole horizon. This property makes it very similar to the
Bekenstein-Hawking entropy assigned to the horizon. This apparent similarity between the two entropies is the main motivation to raise the question, whether
the Bekenstein-Hawking entropy is in fact entirely the entropy of entanglement. In this section we discuss problems which  this interpretation has to face,
different  approaches to solve them and difficulties which still remain unsolved.

\subsection{Problems of interpretation of the Bekenstein-Hawking entropy as entanglement entropy} 

Any approach that wants to treat the Bekenstein-Hawking entropy as an entanglement entropy has to answer the following questions:

\medskip

\noindent (i) The entanglement entropy is a UV divergent quantity, while the Bekenstein-Hawking entropy is a finite quantity, defined with respect to Newton's constant, measured in experiments. How can these two quantities  be equal?

\medskip

\noindent (ii) The entanglement entropy is proportional to the number of different field species which exist in Nature. On the other hand, the Bekenstein-Hawking entropy does not seem to depend 
on any number of fields. This problem is known as the  ``species puzzle''.

\medskip

\noindent (iii) We have seen that entanglement entropy due to fields which are non-minimally coupled to gravity, the gauge bosons and gravitons, behave differently from the entropy due to minimally coupled fields.  Since the gauge bosons and gravitons are  fields that are clearly present in Nature  and thus should contribute to the entropy, does this contribution  spoil 
the possibility of interpreting  the black hole entropy as an entanglement entropy?

\subsection{Entanglement entropy in induced gravity}

One, possibly very natural, way, originally proposed by Jacobson \cite{Jacobson:1994iw},  to attack these problems is to consider gravity as an induced phenomenon, in the spirit of Sakharov's ideas \cite{Sakharov:1967pk} (for a review on a modern touch on these ideas see \cite{Visser:2002ew}). 
In this approach the gravitational field is not fundamental but arises as a mean field approximation of the underlying quantum field theory of fundamental particles (the constituents). 
This is based on the fact, that even if there is no gravitational action at tree level, it will appear at one-loop.  The details of this mechanism will, of course, depend on the
concrete model.

\paragraph*{Model with minimally coupled fields.} 

To start with, let us consider a simple model in which the constituents are minimally coupled fields: we consider $N_0$ scalars and $N_{1/2}$ Dirac fermions. The induced gravitational action in this model, to lowest order in curvature, is 
\be
W_{ind}=-{1\over 16\pi G_{ind}}\int_E R\sqrt{g} d^4x\, , 
\lb{induced action}
\ee
where the induced Newton's constant is
\be
 {1\o G_{ind}}={N\o 12\pi\epsilon^2}\, , \ \ N=N_0+2N_{1/2}\, ,
\lb{induced Newton constant}
\ee
$N$ is the number of field species in this model.  The renormalization statement which is valid for the minimally coupled fields guarantees that the there is a precise balance between
the induced Newton's constant and the entanglement entropy, so that 
\be
S_{ent}={N\over 48\pi\epsilon^2}A(\Sigma)={1\over 4 G_{ind}}A(\Sigma)=S_{BH}\, ,
\lb{entanglement=BH}
\ee
i.e. the entanglement entropy of the constituents is precisely equal to the Bekenstein-Hawking entropy, expressed in terms of the induced Newton's constant (\ref{induced Newton constant}).
Thus, if at a fundamental level the constituents in Nature were only minimal fields, the Bekenstein-Hawking entropy, as this example shows, would be explained as the entropy of entanglement. Of course, this example ignores the fact that there are elementary particles, namely the gauge bosons, which are non-minimally coupled.

\paragraph*{Models with non-minimally coupled fields.}  In the model with minimal fields the induced Newton's constant (\ref{induced Newton constant}) is set by the UV cutoff $\epsilon$. If one wants to deal with the UV finite quantities one has to add fields which contribute negatively to Newton's constant. Excluding  non-physical fields with wrong statistics, the only possibility is to include  non-minimally coupled fields, scalars or vectors.   Models of this type have been considered by Frolov, Fursaev and Zelnikov  \cite{Frolov:1996wd},  \cite{Frolov:1996aj},  \cite{Frolov:1997up}. One considers a multiplet of scalar fields of mass $m_s$ and non-minimal coupling $\xi_s$ and a set of massive Dirac fields with mass $m_d$.
The number of fields and their parameters are  fine tuned so that the ultra-violet divergences in the cosmological constant and in Newton's constant are canceled. The induced Newton's constant then
\be
{1\o G_{ind}}={1\o 12 \pi}\left(\sum_s (1-6\xi_s)m^2_s\ln m^2_s+2\sum_d m^2\ln m_d^2\right) 
\lb{induced newton constant in FFZ}
\ee
is dominated by  the mass of the heaviest constituents. 
However, as soon as we include the non-minimally coupled fields the precise  balance between Newton's constant and the entanglement entropy is violated,  so that
the Bekenstein-Hawking entropy $S_{BH}=A(\Sigma)/4G_{ind}$,  defined with respect to induced Newton's constant,  is no longer equal to the entanglement entropy. In the model considered in
\cite{Frolov:1996wd},  \cite{Frolov:1996aj},  \cite{Frolov:1997up}
 (various models of a similar nature are considered in  \cite{Frolov:1997xd}, \cite{Frolov:1996qh}, \cite{Frolov:1999my}, \cite{Frolov:1999hu},  \cite{Frolov:1999gy},    \cite{Frolov:1998ea}, \cite{Fursaev:1999jq}, \cite{Fursaev:1998hr}) the exact relation between two entropies is 
\be
S_{BH}=S_{ent}-Q\, ,
\lb{FFZ relation}
\ee
where the quantity $Q$ is determined by the expectation value of the non-minimally coupled scalar fields on the horizon $\Sigma$
\be
Q=2\pi \sum_s \xi_s\int_\Sigma <\phi_s^2>\, .
\lb{Q in FFZ}
\ee
This quantity is UV divergent. For a single field it is similar to the quantity (\ref{S-non-minimal}).
Thus the  sharp difference between the entanglement entropy and the Bekenstein-Hawking entropy in this model can be summarized as follows: even though the induced Newton's constant is made UV finite,  the entanglement entropy still (and, in fact, always)
remains UV divergent.   
Thus,  we conclude that, in the model of Frolov, Fursaev and Zelnikov, the entanglement entropy is clearly different from the Bekenstein-Hawking entropy.

\subsection{Entropy in brane-world}

An interesting example where the Bekenstein-Hawking entropy is apparently induced in the correct way is given in  \cite{Hawking:2000da}. This example is closely related to the AdS/CFT correspondence discussed in section \ref{section: Holographic description}.  In the Randall-Sundrum set-up \cite{Randall:1999vf} one may consider the regularized boundary, which appeared in our discussion of section \ref{section: Holographic description}, as a
3-brane with $Z_2$ symmetry in an anti-de Sitter space-time. In the framework of the
AdS/CFT correspondence this brane has a description in terms of CFT coupled to gravity at a UV cutoff \cite{Gubser:1999vj}. If the
brane is placed at the distance $\rho=\epsilon^2$ from
the Anti-de Sitter boundary, one obtains that there is a dynamical gravity
induced on the brane, with the induced Newton's constant
\be
1/G_N={2N^2/(\pi \epsilon^2)}\, ,
\lb{Randall-Sundrum}
\ee
where $N$ is the number of colors in the superconformal $SU(N)$ Yang-Mills theory. $N^2$ in this case plays the role of the number of species.  We notice that, according to the AdS/CFT dictionary, the parameter $\epsilon$ which is an infra-red cut-off on the anti-de Sitter side, is in fact, a UV cut-off on the CFT side.
Consider now a black hole on the 3-brane. The Bekenstein-Hawking entropy  can then be represented as follows
\be
S_{BH}={A(\Sigma)\o 4G_N}={N^2 A(\Sigma)\o 2\pi \epsilon^2}=S_{ent}\, .
\lb{brane entropy}
\ee
As  Hawking, Maldacena and Strominger  \cite{Hawking:2000da} suggested , the right hand side of (\ref{brane entropy}) can be interpreted as the entanglement entropy of $N^2$ fields. This interpretation
turns out to be the right one, if one uses the holographic entanglement entropy discussed in section \ref{section: Holographic description}. Indeed, taking the leading divergent term in (\ref{Sdiv})  and noting that, in a $Z_2$ brane configuration, this result should be multiplied by factor of 2, we get exactly the right hand side of (\ref{brane entropy}). In ref.\cite{Hawking:2000da} one considers
 de Sitter spacetime (so that (\ref{brane entropy}) is the entropy of the de Sitter horizon in this case)  on the brane since it is the simplest brane configuration one can construct in anti-de Sitter spacetime. In \cite{Emparan:2006ni}  this proposal was extended to the holographic entanglement entropy of black hole on 2-brane solution found in \cite{Emparan:1999wa}.
 The two-dimensional black hole is considered in \cite{Fursaev:2000ym}.
 Entropy of a generic black hole on the 3-brane was considered in \cite{Solodukhin:2006xv}.
 
 There are, however, certain open questions regarding this example. First of all we should note 
that the weakly coupled ${\cal N}=4$  $SU(N)$ supermultiplet contains the Yang-Mills fields (gauge bosons), conformally coupled scalars and the Weyl fermions \cite{Maldacena:1997re}.
Thus it is a bit of a mystery how the entanglement entropy of these, mostly non-minimally coupled, fields (gauge bosons and scalars) has managed to become equal to the Bekenstein-Hawking entropy, when recalling   the problems with the non-minimal coupling we have discussed in section \ref{section: non-minimal coupling}. A part of this mystery is the fact that the holographic regularization 
(which corresponds to infra-red cut-off on the anti-de Sitter side) does not have a clear analog on the boundary side. Indeed, if we take for example  a standard heat kernel regularization, we find that the term linear in the scalar curvature $R$ does not appear at all in the effective action produced by the weakly coupled ${\cal N}=4$ superconformal gauge multiplet.
 
 \subsection{Gravity  cut-off} 
If we compare the two examples when the Bekenstein-Hawking entropy is correctly reproduced, the model of induced gravity with minimally coupled constituents and the brane world model,
we find that the success of the two models is strongly based on a precise relation between Newton's constant, the number of species and the UV cut-off. This relation can be reformulated in terms of the Planck mass $M_{PL}$ ($1/G_N\sim M^2_{PL}$) and  energy cutoff $\Lambda\sim 1/\epsilon$,
\be
\Lambda=M_{PL}/\sqrt{N}\, ,
\lb{Gia formula}
\ee
where the precise numerical pre-factor depends on the concrete model. It is amazing to note that exactly this relation (\ref{Gia formula}) was proposed by Dvali \cite{Dvali:2007hz}, \cite{Dvali:2007wp}, \cite{Dvali:2008fd} to hold in general in a theory of Quantum Gravity coupled to a large number of matter species. 
The arguments which were used  to get this relation, although  they include some thought experiments with  black holes, are, in fact,  unrelated to (and are thus independent of) 
the entropy.
However, it is clear that relation (\ref{Gia formula}), provided it is correct,  helps to reproduce precisely  the Bekenstein-Hawking entropy as entropy of entanglement and 
automatically solves the species puzzle \cite{Dvali:2008jb}.

\subsection{Kaluza-Klein example} 
One example when relation (\ref{Gia formula}) holds is the Kaluza-Klein model. In this model one starts with $(4+n)$-dimensional theory of gravity which then is compactified so that one has $n$ compact directions, forming, for example, an  $n$-tours with characteristic size $R$, and $4$ non-compact directions which form our 4-dimensional geometry. The higher-dimensional Planck scale $\Lambda$ is considered fundamental in this model and plays the role of the UV cutoff while the 4-dimensional Planck scale $M_{PL}$ (or 4-dimensional Newton's constant) is derived,
\be
M^2_{PL}=\Lambda^2(R\Lambda)^n\, .
\lb{Kaluza-Klein}
\ee
Suppose that in higher dimensions there is only one particle -  the massless graviton. From the four-dimensional point of view one has, additionally to a single massless graviton,   a theory of the tower of spin-2 massive Kaluza-Klein (KK) particles.  Truncating  the tower at the cut-off $\Lambda$ one finds that $N=(R\Lambda)^n$ is precisely the number of these Kaluza-Klein species. Thus, as was noted in \cite{Dvali:2007hz}, \cite{Dvali:2007wp}, the relation (\ref{Kaluza-Klein}) is a particular example of relation (\ref{Gia formula})
in which $N$ should be understood as the number of the KK species.

In the Kaluza-Klein example the entanglement entropy is equal to the Bekenstein-Hawking entropy as demonstrate in Dvali and Solodukhin \cite{Dvali:2008jb}. Consider now a large black hole with horizon size $r_g\gg R$. This large black hole fills all  compact directions so that, from the higher-dimensional point of view, the black hole  horizon is a product of a 2-sphere of radius $r_g$ and an $n$-dimensional torus of size $R$.  The Bekenstein-Hawking entropy in the  $(4+n)$-dimensional theory is
\be
S^{(4+n)}_{BH}=4\pi\Lambda^{n+2}r_g^2 R^n\, ,
\lb{4+n-entropy}
\ee
where $4\pi r^2_g R^n$ is the area of $(4+n)$-dimensional horizon. From the 4-dimensional point of view this horizon is a 2-sphere of radius $r_g$ and the Bekenstein-Hawking entropy
in the 4-dimensional theory is
\be
S^{(4)}_{BH}=4\pi M^2_{PL}r^2_g\, .
\lb{4-entropy}
\ee
We remark observe  that these two entropies are equal so that the two pictures, the higher dimensional and 4-dimensional one, are consistent.
Let us discuss now the entanglement entropy. In the $(4+n)$-dimensional theory there is only one field, the massless graviton. Its entropy is
\be
S_{ent}^{(4+n)}=4\pi r_g^2R^n\Lambda^{n+2}\, ,
\lb{4+n -entanglement}
\ee
where the cut-off is the higher dimensional Planck scale $\Lambda$. On the other hand, in the 4-dimensional theory one computes the entanglement entropy of $N$ KK fields
\be
S^{(4)}_{ent}=N (4\pi r_g^2\Lambda^2)=4\pi r_g^2 \Lambda^2 (R\Lambda )^ñ\, .
\lb{4-entanglement}
\ee
These two entropies are equal to each other so that the two ways to compute the entanglement entropy  agree. Moreover the entanglement entropy (\ref{4-entanglement}) and             
  (\ref{4+n -entanglement}) exactly reproduces  the Bekenstein-Hawking entropy (\ref{4+n-entropy}) and  (\ref{4-entropy}).   
  
Discussing this result we should, however, note that the massless and massive gravitons are non-minimally coupled particles.  It remains to be understood how the problem of the non-minimal coupling   is overcome   in this example.

\section{Other directions of research}
In this section we briefly mention some other interesting directions of  on-going research.

\subsection{Entanglement entropy in string theory}

It is generally believed that the problem of  entanglement entropy of black hole can and should be resolved
in string theory. This was originally suggested by Susskind and Uglum \cite{Susskind:1994sm}. Indeed, assuming that string theory is UV finite, 
the corresponding entropy calculation should result in a finite quantity. More specifically, the effective action of a closed string can be decomposed
in powers of string coupling $g$ as $g^{2(n-1)}$, where $n$ is the genus of the string world sheet. The string configurations with spherical topology, $n=0$,
give  a $1/g^2$ contribution. In a low energy approximation this is exactly the contribution to Newton's constant $G\sim g^2$. Thus one may expect that taking into account just $n=0$  closed string configurations, one will correctly reproduce both the entanglement entropy and Newton's constant. In the Euclidean formulation the prescription of \cite{Susskind:1994sm} is to look at the zero genus string world sheet which intersects the Killing horizon. In the Lorentzian picture this corresponds to an open string with both ends attached to the horizon.  The higher genus configurations should give some corrections to the $n=0$ result. This is a very attractive idea. However a very little 
progress has been made in the literature to actually calculate the entanglement entropy directly in string theory. The reason is of course the technical complexity of the problem. 
Some support to the idea of Sussking and Uglum was found in work of Kabat,  Shenker and Strassler \cite{Kabat:1995jq},  where the entropy in a two-dimensional $O(N)$ invariant  $\sigma$-model was considered. In particular, it was found that the state counting of the entropy in the UV regime may be lost if considered in the low energy (IR) regime. This  type of behavior models the situation with the classical Bekenstein-Hawking entropy. Presumably this analysis could be generalized to the string theory $\sigma$-model considered either in optical target metric \cite{Barbon:1994ej}, \cite{Barbon:1994wa} or in the Euclidean metric with a conical singularity at horizon (as suggested in  \cite{Callan:1994py}). Possibly in the latter case the results obtained for strings on orbifolds \cite{Dixon:1985jw}
can be useful (see \cite{Dabholkar:1994ai}, \cite{Dabholkar:1994gg}, \cite{Dabholkar:2001if} for earlier works in this direction).

  Another  promising approach to attack the problem is to use some indirect methods based on dualities. For example, the AdS/CFT correspondence has been used in \cite{Brustein:2005vx}  to relate the entanglement entropy of a string propagating on gravitational AdS background with a Killing horizon 
to the thermal entropy of field theory defined on a boundary of AdS and then,  eventually, the thermal entropy  to the Bekenstein-Hawking entropy of the horizon.

An interesting approach to the entanglement entropy of extremal black holes via AdS$_2$/CFT$_1$ duality is considered in \cite{Azeyanagi:2007bj}, \cite{Sen:2011cn}, where, in particular, one can identify the entanglement entropy and the microcanonical statistical entropy. This approach is based on the earlier work of Maldacena  \cite{Maldacena:2001kr} in which the Hartle-Hawking state is identified with an entangled state of two copies of CFT, defined on two boundaries of the maximally extended BTZ spacetime. In the accurately taken zero temperature limit the reduced density matrix, obtained by tracing over the states of one copy of CFT,  of the extremal black hole is shown to take the form 
\be
\rho={1\o d(N)}\sum_{k=1}^{d(N)}|k><k|\, ,
\lb{maximally entangled}
\ee
which describes the maximally entangled  state in the two copies of the CFT$_1$ living on the two boundaries of global AdS$_2$. 
$d(N)$ is the dimension of the Hilbert space of CFT$_1$, it can be expressed in terms of the charges carried by the black hole.
The corresponding entanglement entropy
$S=-\Tr\rho\ln\rho=\ln d(N)$ then is precisely equal to the microcanonical entropy in the familiar counting of BPS states  and thus is equal to the black hole entropy \cite{Callan:1996dv}.

\subsection{Entanglement entropy in loop quantum gravity}

Another approach to Quantum Gravity, sometimes considered as competing with  string theory, is Loop Quantum  Gravity.  In this theory one considers polymeric excitations of the gravitational field represented by the states of {\it spin networks}. A spin network is a graph, a network of points with links representing the relation between points.
Each link is labeled by a half-integer $j$ (the label stands for $SU(2)$ representations). To points, or vertices, of a spin network are attached a $SU(2)$ so-called 
intertwiner, a $SU(2)$ invariant tensor between the representations attached to all the edges linked to the considered vertex. A simpler and more familiar object in particle physics is the Wilson loop.  A surface $\Sigma$ is represented by vertices (punctures) which divide the spin network on two parts.  By tracing over states of just one part of the network one obtains a density matrix. The entanglement entropy then reduces to a sum over intersections of the spin network with the surface $\Sigma$ \cite{Dasgupta:2005yu}, \cite{Livine:2005mw}, \cite{Donnelly:2008vx},
\be
S(\Sigma)=\sum_{p=1}^P\ln (2j_p+1)\, ,
\lb{entropy LQG}
\ee
where $P$ is number of punctures representing $\Sigma$.  This quantity should be compared to eigenvalues of the operator of area,
\be
A(\Sigma)=8\pi G \gamma\sum_{p=1}^P\sqrt{j_p(j_p+1)}\, .
\lb{area operator}
\ee
Both quantities scale as $P$ for large $P$,  which indicates that the area law is correctly reproduced. The exact relation between the two quantities and  the classical entropy
$S_{BH}=A(\Sigma)/G$ is, however, not obvious due to   ambiguities present in the formalism. The Immirzi parameter $\gamma$ is one of them. The question, whether the Bekenstein-Hawking entropy is correctly reproduced in this approach, is eventually related to the continuum limit of the theory \cite{Jacobson:2007uj}. As discussed by Jacobson \cite{Jacobson:2007uj}, the answering this question may require  certain renormalization of Newton's constant as well as area renormalization.  Indeed, quantity (\ref{area operator}) represents a microscopic area
which may be related to the macroscopic quantity in a non-trivial way. These issues remain open.

\subsection{Entropy in non-commutative theories and in models with minimal length}

One might have hoped that the UV divergence of the entanglement entropy could be  cured  in a natural way were the structure of spacetime  modified on some fundamental level.
For example, if spacetime becomes non-commutative at short distances. This idea was tested in the case of simple fuzzy spaces in \cite{Dou:2006ni}, \cite{Dou:2009cw}. 
Although the area law has been verified, the entanglement entropy appears to be sensitive to the size of the ignored region, a phenomenon which may be understood as a 
UV-IR mixing typical for the non-commutative models. 

A holographic calculation of the entanglement entropy in non-commutative Yang-Mills theory was considered in  \cite{Barbon:2008ut}, \cite{Barbon:2008sr}.
This calculation for a strip of width $l$ shows that for for large values of $l\gg l_c$ compared to some characteristic length $l_c\sim {\theta {\lambda^{1/2}}/\epsilon}$, where $\theta$ is the parameter of non-commutativity and $\lambda=g^2_{YM}N$ is the 't Hooft coupling, then the short-distance contribution to the entanglement entropy shows an area law of the form
\be
S\sim N_{eff}{A(\Sigma)\o \epsilon^2}\, , \ N_{eff}=N^2({\theta\lambda^{1/2}\o \epsilon})\, ,
\lb{entropy non-comm Yang-Mills}
\ee
while for smaller values $l\sim l_c$ the entropy occurs to be proportional to the volume. As  seen from (\ref{entropy non-comm Yang-Mills}) the non-commutativity does not improve
the UV behavior of the entropy but leads to the renormalization of the effective number of degrees of freedom that may be interpreted as a manifestation of non-locality of the model.

The other related idea is to consider models in which the Heisenberg uncertainty relation is modified as $\Delta x\Delta p\geq {\hbar\o 2}(1+{\lambda^2}(\Delta p)^2)$, which shows that there exists a minimal length $\Delta x\geq \hbar {\lambda}$ (for a review on the models of this type   see \cite{Garay:1994en}). In a brick wall calculation  the presence of this minimal length 
will regularize the entropy as discussed in \cite{Brustein:2010ms}, \cite{Yoon:2007aj}, \cite{Sun:2004ct}, \cite{Kim:2006hk}, \cite{Kim:2007if}.

\subsection{Transplanckian physics and entanglement entropy}

One way to check whether the entanglement entropy is sensitive to  the way the conventional theory is completed in the UV regime is to study the possible modifications of the standard Lorentz invariant   dispersion relation $\omega^2={\bf k}^2$ at large values of momentum $\bf k$ (or at short distances). A typical modification is to break the Lorentz invariance
as follows $\omega^2={\bf k}^2+f({\bf k}^2)$. This issue was studied in \cite{Jacobson:2007jx} and  \cite{Chang:2003sa} in the context of the brick wall model. The conclusions made in these papers are however opposite.
According to ref.\cite{Jacobson:2007jx} the entropy is still UV divergent although the degree of divergence is modified in  way which depends on the form of function $f({\bf k}^2)$.
On the other hand, the paper \cite{Chang:2003sa} claims that the entropy can be made completely UV finite. In a similar claim  ref.\cite{Padmanabhan:2010wg} suggests that
the short-distance finiteness of the 2-point correlation function should imply the UV finiteness of the entanglement entropy.
The entanglement entropy in a wide class of theories characterized by 
modified (Lorentz invariant or not) field operators (so that the UV behavior of the modified propagator is improved compared to the standard one) was calculated in \cite{Nesterov:2010yi}. The conclusion reached in \cite{Nesterov:2010yi} (see also discussion in sections \ref{section: theories with a modified propagator} and \ref{section: theories with a modified propagator, covariant} of this review) agrees with that of ref.\cite{Jacobson:2007jx}: no matter how well is the UV behavior of the propagator, the entanglement entropy remains UV divergent. That the short-distance regularity of correlation functions does not necessarily imply that the entanglement entropy is UV finite was pointed out in \cite{Nesterov:2010jh}.

\subsection{Entropy of more general states} 

The quantum pure state which is  the starting point in the entanglement entropy calculation should not  necessarily be a vacuum state.
According to refs.\cite{Das:2010su}, \cite{Das:2007mj}, \cite{Das:2005ah}, \cite{Das:2008sy}, if one starts with a mixed state of the vacuum and an  excited state, the entanglement entropy receives some power law corrections,
\be
S\sim{A\o \epsilon^2}(1+c ({A\o \epsilon^2})^{-\beta})\, ,
\lb{power law correction}
\ee
whith $\beta$  always less than unity, and  the power law correction is due to the excited state.

\subsection{Non-unitary time evolution }  

An interesting issue discussed in the literature is the time evolution of entanglement entropy. It was suggested in 
\cite{Brustein:2006wp} that the eigenvalues of a reduced density matrix depend on time $t$. This is not possible if the time evolution
of the density matrix is described by an unitary operator. Thus the time evolution  should be nonunitary. In particular, the entanglement entropy 
should depend on time $t$. Similar conclusions  have been made in
\cite{Calabrese:2005in} and \cite{Koksma:2010zi}, \cite{Koksma:2011fx}, \cite{Koksma:2011dy}, where, in particular, it was shown that the entanglement entropy is an increasing function of time.
These observations may have interesting applications for black holes. As was proposed by Hawking \cite{Hawking:1976ra} the evolution in time of a black hole should be nonunitary, so that a pure initial state 
may evolve into a mixed state. From the entanglement point of view, this behavior appears to be not in contradiction
with the principles of quantum mechanics,  rather it is a simple
consequence of the entangled nature of the system. The irreversible loss of information due to entanglement is also seen from  the evolution of the  entropy under RG flow
\cite{Solodukhin:2006ic}, \cite{Latorre:2004pk}.

\section{Concluding remarks}

Since the inspiring work of Srednicki in 1993 we have come a long way in understanding the entanglement entropy.   Of course, the main motivation for this research  always was and still is the attempt to consider the Bekenstein-Hawking entropy  as entropy of entanglement. If successful this approach would give a universal explication for the entropy of black holes,
valid for black holes of any size and mass  and carrying any charges.  Unfortunately, we are not yet at that point. Many important  unresolved problems remain.
However, after 17 years of 	continuous progress it would not be too surprising if we were actually not that far from  the final answer. Speaking about  the future developments, I think that the further progress could be made in advancing in two main directions: 	understanding the entanglement entropy directly in string theory and  resolving the puzzle of the non-minimal coupling.  Hopefully,   at some not so distant point in time the efforts made in these  directions will  provide us with the still missing elements in the picture.

\section{Acknowledgements}
\label{section:acknowledgements}

I would like to thank my collaborators and friends A. Barvinsky, G. Dvali, G. 't Hooft, W. Israel, V. Frolov, D. Fursaev, R. Mann, R. Myers, S. Mukhanov,  D. Nesterov  and A. Zelnikov
for fruitful collaboration and many inspiring discussions.
I thank G. 't Hooft, R. Myers and R. Mann for reading the review and valuable remarks.
I am indebted to my colleagues N. Mohammedi and, especially, to S. Nicolis for reading the review and  many comments which helped to improve  the presentation. 
As always, I wish to thank my family for support and encouragement. 

\newpage



\bibliography{Myreview}

\end{document}